\documentclass[aps,prb,showpacs,twocolumn,amsmath,amssymb,revsymb,a4paper,10]{revtex4-1}
\pdfoutput=1
\usepackage{graphicx}
\usepackage{color} 
\usepackage{dcolumn}
\usepackage{bm}
\usepackage{setspace}
\usepackage{sidecap}
\usepackage{natbib}
\usepackage{hyperref}

\providecommand{\e}{\varepsilon}
\providecommand{\h}{\hbar}

\newcommand{\kb}{k_{\textsc{b}}^{}}
\newcommand{\om}{\omega}
\newcommand{\kk}{\mathbf{k}}
\newcommand{\rr}{\mathbf{r}}
\newcommand{\RR}{\mathbf{R}}
\newcommand{\Ss}{\mathbf{S}}
\newcommand{\II}{\mathbf{I}}
\newcommand{\op}{\uparrow}
\newcommand{\ned}{\downarrow}
\newcommand{\tl}{\textrm{L}}
\newcommand{\tr}{\textrm{R}}
\newcommand{\PP}{\mathfrak{p}}
\newcommand{\Tct}{T_{c,\textsc{tt}}}
\newcommand{\Tcs}{T_{c,\textsc{st}}}

\begin{document}

\title{Temperature dependent dynamical nuclear polarization bistabilities in double quantum dots in the spin-blockade regime}

\author{Anders Mathias Lunde,$^1$ Carlos L\'opez-Mon\'is,$^{1,2}$ Ioanna A. Vasiliadou,$^{3,4}$ Luis L. Bonilla,$^{3,5}$ and Gloria Platero$^1$}
\affiliation{$^1$Instituto de Ciencia de Materiales de Madrid, CSIC, 28049 Cantoblanco, Madrid, Spain\\
$^2$Institute for Theoretical Physics, University of Regensburg, D-93040 Regensburg, Germany\\
$^3$G. Millan Institute, Fluid Dynamics, Nanoscience and Industrial
Mathematics, Universidad Carlos III de Madrid, 
Spain\\ 
$^4$Materials Science and Engineering and Chemical Engineering Department, Universidad Carlos III de Madrid, 
Spain\\
$^5$Unidad Asociada al Instituto de Ciencia de Materiales de Madrid, CSIC, 28049 Cantoblanco, Madrid, Spain}

\date{\today}
 
\begin{abstract}
The interplay of dynamical nuclear polarization (DNP) and leakage current through a double quantum dot in the spin-blockade regime is analyzed. A finite DNP is built up due to a competition between hyperfine (HF) spin-flip transitions and another inelastic escape mechanism from the triplets, which block transport. We focus on the temperature dependence of the DNP for zero energy-detuning (i.e.~equal electrostatic energy of one electron in each dot and a singlet in the right dot). Our main result is the existence of a \emph{transition temperature}, below which the DNP is \emph{bistable}, so a \emph{hysteretic leakage current} versus external magnetic field $B$ appears. This is studied in two cases: (i) Close to the crossing of the three triplet energy levels near $B=0$, where spin-blockade is lifted due to the inhomogeneity of the effective magnetic field from the nuclei. (ii) At higher $B$-fields, where the two spin-polarized triplets simultaneously cross two different singlet energy levels. We develop simplified models leading to \emph{different} transition temperatures $\Tct$ and $\Tcs$ for the crossing of the triplet levels and the singlet-triplet level crossings, respectively. We find $\Tct$ analytically to be given solely by the HF couplings, whereas $\Tcs$ depends on various parameters and $\Tcs>\Tct$. The key idea behind the existence of the transition temperatures at zero energy-detuning is the suppression of energy absorption compared to emission in the inelastic HF transitions. Finally, by comparing the rate equation results with Monte Carlo simulations, we discuss the importance of having both HF interaction and another escape mechanism from the triplets to induce a finite DNP.
\end{abstract}


\maketitle

\section{Introduction}

The high degree of experimental control in modern quantum dot systems allows detailed  manipulation of electrons and their spin in confined states.\cite{Hanson-Kouwenhoven-Petta-Tarucha-Vandersypen-RMP-2007,van-der-Wiel-et-al-RMP-2002,Reimann-Manninen-RMP-2002,Kouwenhoven-Austing-Tarucha-review-2001} A particularly intriguing example was investigated by Ono \emph{et al}.\cite{Ono-Science-2002,Ono-Tarucha-PRL-2004} in a series of experiments. These revealed that, not only the charge, but also the Pauli exclusion principle for spin states can block the electronic transport through a double quantum dot (DQD) coupled in series. To observe this phenomenon -- dubbed \emph{spin-blockade}\cite{Weinmann-Hausler-Kramer-PRL-1995,Weinmann-Hausler-Kramer-Annalen-der-Physik-1996,Ciorga-Sachrajda-Could-et-al-PRB-2000,Ono-Science-2002} (SB) -- the energy levels of the two dots are tuned asymmetrically, so an electron with a definite spin is trapped in -say- the right dot. Now, only electrons with the opposite spin (to the trapped one) can pass through the DQD, since two electrons with equal spins in the right dot is tuned to be energetically forbidden. Therefore, once an electron with the same spin (as the trapped one) tunnels into the left dot, then transport through the DQD is blocked. SB requires non-linear bias and due to the asymmetric energy level tuning of the dots, current is only blocked in one direction leading to the observed current rectification.\cite{Ono-Science-2002}

An electron can escape from the states blocking transport by a spin relaxation process, which leads to a small leakage current in the SB regime. Analyzing the leakage current is therefore an excellent tool to obtain information about the spin relaxation processes from a transport experiment.\cite{Ono-Tarucha-PRL-2004,Koppens-et-al-Science-2005,Johnson-et-al-nature-2005} There are several ways to escape from the blocking states: via co-tunneling processes,\cite{Ono-Science-2002,Johnson-Petta-Marcus-PRB-2005,Vorontsov-Vavilov-PRL-2008,Saito-et-al-Physica-E-2008,Qassemi-Coish-Wilhelm-PRL-2009,Coish-Qassemi-PRB-2011} spin-orbit meditated spin relaxation\cite{Khaetskii-Nazarov-PRB-2000,Golovach-Khaetskii-Loss-PRL-2004,Pfund-Shorubalko-Ensslin-Leturcq-PRL-2007,Danon-Nazarov-PRB-2009,Weiss-Rashba-Kuemmeth-Churchill-Flensberg-PRB-2010,Frolov-Danon-Nazarov-Kouwenhoven-et-al-PRB-2010,Stepanenko-Rudner-Bertrand-Loss-PRB-2012,Stano-Fabian-PRL-2006,Destefani-Ulloa-PRB-2005} and/or by hyperfine interaction\cite{Coish-Baugh-review-2009} (HFI) between the electronic spins and the nuclear spins of the host material.\cite{Erlingsson-Nazarov-Falko-PRB-2001,Ono-Tarucha-PRL-2004,Eto-et-al-JPSJ-2004,Koppens-et-al-Science-2005,Johnson-et-al-nature-2005,Tarucha-Ono-et-al-physica-status-solidi-2006,Jouravlev-Nazarov-PRL-2006,Baugh-Kitamura-Ono-Tarucha-PRL-2007,Baugh-Kitamura-Ono-Tarucha-physica-status-solidi-2008,Rudner-Levitov-PRL-2007,Rudner-Levitov-Nanotechnology-2010,Rudner-Rashba-PRB-2011,Rudner-Koppens-Folk-Vandersypen-Levitov-PRB-2011,Inarrea-Platero-MacDonald-PRB-2007,Inarrea-Carlos-MacDonald-Platero-APL-2007,Inarrea-Carlos-Platero-APL-2009,Tarucha-Baugh-JPSJ-2008,Rudner-Levitov-PRB-2010,Lopez-Monis-et-al-NJP-2011,Gullans-et-al-PRL-2010,Takahashi-Kono-Tarucha-Ono-PRL-2011,Kobayashi-et-al-PRL-2011,Takahashi-Kono-Tarucha-Ono-APEX-2012,Lopez-et-al-PRB-2012} The relative importance of these mechanisms depends on the material and the external parameters. For instance, a specific co-tunneling process can become important by tuning the gate-voltages such that the virtual energy exchange in the co-tunneling process becomes low.\cite{Johnson-Petta-Marcus-PRB-2005}

Ever since the experiments by Ono \emph{et al.}\cite{Ono-Science-2002,Ono-Tarucha-PRL-2004} in vertical GaAs DQDs, several geometries and materials have been used to further study the leakage current in the SB regime due to different relaxation mechanisms.\cite{Koppens-et-al-Science-2005,Johnson-et-al-nature-2005,Koppens-et-al-Science-2005,Liu-et-al-PRB-2005,Pfund-Shorubalko-Ensslin-Leturcq-PRL-2007,Buitelaar-et-al-PRB-2008,Shaji-Nature-2008,Churchill-Marcos-Marcus-et-al-nature-phys-2009,Lai-Coish-Dzurak-Scientific-Reports-2011} For instance, Churchill \emph{et al}.\cite{Churchill-Marcos-Marcus-et-al-nature-phys-2009} analyzed experimentally the leakage current in carbon nanotube DQDs varying the amount of $^{13}$C --- the only stable carbon isotope with a non-zero spin. This amounts to varying the spin relaxation due to HFI from very important (high $^{13}$C concentration) to not important (almost no $^{13}$C present). This shows how different the leakage current can be with and without nuclear spins.\cite{Churchill-Marcos-Marcus-et-al-nature-phys-2009} Nowadays, spin-orbit coupling is also thought to play a role in carbon nanotubes.\cite{Kuemmeth-Ilani-Ralph-McEuen-Nature-2008,Churchill-Flensberg-et-al-PRL-2009,Jespersen-et-al-nature-phys-2011}

SB in Silicon DQDs has also been studied.\cite{Shaji-Nature-2008,Liu-Fujisawa-Ono-et-al-PRB-2008,Prada-Blick-Joynt-PRB-2008,Simmons-et-al-PRB-2010,Lai-Coish-Dzurak-Scientific-Reports-2011,Raith-Stano-Fabian-PRB-2012} In a recent work, Lai \emph{et al.}\cite{Lai-Coish-Dzurak-Scientific-Reports-2011} eliminated the HFI in Silicon DQDs by isotopic purification -- along the same lines as Churchill \emph{et al.}\cite{Churchill-Marcos-Marcus-et-al-nature-phys-2009} In this case, co-tunneling processes caused the leakage current in the SB regime (in good agreement with recent theories\cite{Qassemi-Coish-Wilhelm-PRL-2009,Coish-Qassemi-PRB-2011}), since the spin-orbit coupling is expected to be weak in Silicon.\cite{Lai-Coish-Dzurak-Scientific-Reports-2011} In contrast, spin-orbit interaction is generally believed to be strong in InAs. This enabled Pfund \emph{et al.}\cite{Pfund-Shorubalko-Ensslin-Leturcq-PRL-2007} to investigate its importance on the leakage current in InAs nanowire DQDs. Finally, recent studies show bipolar SB triple dots\cite{Busl-nature-2013} and valley-spin blockade in carbon nanotube DQDs.\cite{Palyi-Burkard-PRB-2009,Palyi-Burkard-PRB-2010,Pei-Laird-Steele-Kouwenhoven-Nature-nanotech-2012}

\begin{figure}
\includegraphics[width=0.45\textwidth,angle=0]{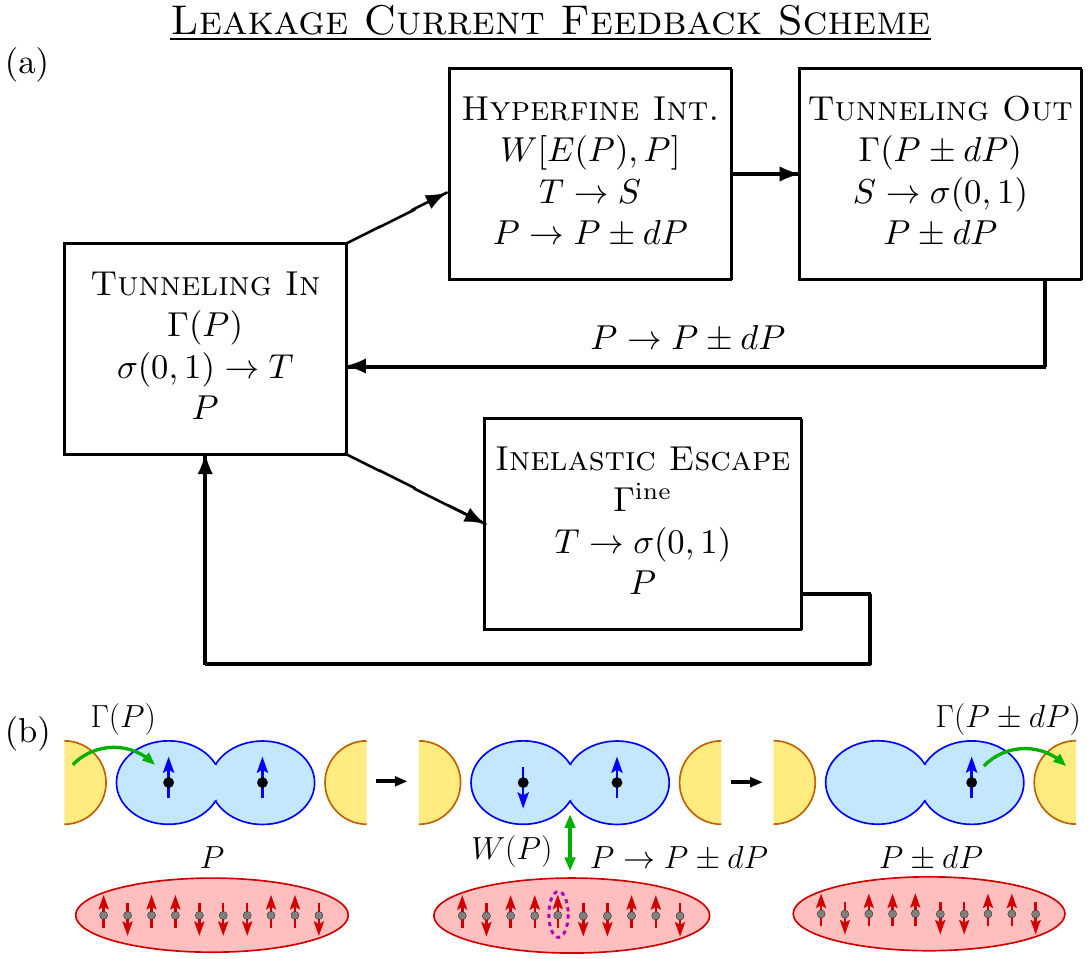}
\caption{(Color online) 
(a) Illustration of the two escape paths from the blocking triplets $T$, which compete to create a non-zero DNP in the SB regime. The cycle of transporting an electron through a blocking triplet state $T$ begins and ends with a single electron trapped in the right dot $\sigma(0,1)$. During a single transport cycle, the DNP $P$ is changed by a small amount $\pm dP$ (depending on the specific transition), if the escape from $T$ is HF mediated (upper branch). The DNP acts back on the energy levels $E(P)$ via the Overhauser fields and, in turn, on both the HF spin-flip rates $W$ and the tunneling rates $\Gamma$. In contrast, the weak inelastic escape mechanism\cite{Rudner-Levitov-PRL-2007,Rudner-Levitov-Nanotechnology-2010,Rudner-Rashba-PRB-2011,Rudner-Koppens-Folk-Vandersypen-Levitov-PRB-2011} (lower branch) leaves the DNP unchanged. (b) A real-space example of a HF-induced escape process from a triplet to the right contact through a singlet $S$. This changes the DNP.}
\label{fig:feedback-illustration}
\end{figure}

A HF-induced spin relaxation process from a blocking state will flip the electronic and nuclear spin in opposite directions, Fig.~\ref{fig:feedback-illustration}(b). The electronic spin relaxation in the SB regime can therefore change the average occupations of the nuclear spin states, since the nuclear spin relaxation time is very long compared to the electronic tunneling timescales.\cite{Koppens-et-al-Science-2005,Churchill-Marcos-Marcus-et-al-nature-phys-2009,Blick-et-al-PRB-2004} The repeated electronic spin-flip due to the leakage current can therefore produce a \emph{dynamical nuclear polarization} (DNP). The DNP acts back on the electronic states of the DQD as an effective magnetic field, the so-called Overhauser field.\cite{Overhauser-PR-1953} The Overhauser field is generally \emph{inhomogeneous} and therefore often different in the two dots. DNP is also studied in optical\cite{Brown-et-al-PRB-1996,Gammon-et-al-PRL-2001,Bracker-et-al-PRL-2005,Lai-et-al-PRL-2006,Maletinsky-Badolato-Imamoglu-PRL-2007,Rudner-Levitov-optics-PRL-2007,Tartakovskii-et-al-PRL-2007,Danon-Nazarov-PRL-2008,Xu-nature-optics-2009,Danon-Vandersypen-Nazarov-et-al-PRL-2009} and quantum Hall\cite{Dobers-Klitzing-et-al-PRL-1988,Wald-Kouwenhoven-McEuen-et-al-PRL-1994,Kim-Vagner-Xing-PRB-1994,Dixon-McEuen-et-al-PRB-1997,Machida-et-al-PRB-2002,Hashimoto-et-al-PRL-2002,Deviatov-et-al-PRB-2004,Wurtz-et-al-PRL-2005,Kou-Marcus-et-al-PRL-2010,Nakajima-Kobayashi-Komiyama-PRB-2010,Nakajima-Komiyama-PRB-2012} systems. 

It is important to emphasize that even though the HF-induced spin-relaxation do flip a nuclear spin, it will not \emph{always} change the \emph{average} nuclear polarization in the steady state of the SB regime. For instance, if HFI is the \emph{only} spin relaxation mechanism causing the leakage current, then the nuclear spins remain unpolarized.\cite{Jouravlev-Nazarov-PRL-2006,Rudner-Levitov-PRL-2007} Essentially, this is because tunneling into one of the two blocking states consisting of two spin-up electrons, $|\!\!\op,\op\rangle$, or two spin-down electrons, $|\!\!\ned,\ned\rangle$, are equal. Escape from these two states will polarize the nuclei in opposite directions and therefore on average the polarization does not change. This is so, even though the escape rates from the blocking states might be very different.\cite{Pfund-Shorubalko-Ensslin-Leturcq-PRL-2007,Rudner-Levitov-PRL-2007} In this case of HFI being the \emph{only} cause of leakage current, the nuclear spins can be modeled as an effective magnetic field with \emph{zero} mean value and non-zero statistical deviation\cite{Merkulov-Efros-Rosen-PRB-2002,Erlingsson-Nazarov-PRB-2002,Erlingsson-Nazarov-PRB-2004,Coish-Loss-PRB-2005,Jouravlev-Nazarov-PRL-2006} as has also been used to fit experimental data.\cite{Johnson-et-al-nature-2005} Nevertheless, if \emph{more than one} spin relaxation mechanism contribute to the leakage current in the SB regime, then the nuclei can indeed obtain a \emph{non-zero} DNP.\cite{Rudner-Levitov-PRL-2007,Inarrea-Platero-MacDonald-PRB-2007,Rudner-Levitov-Nanotechnology-2010,Inarrea-Carlos-MacDonald-Platero-APL-2007,Inarrea-Carlos-Platero-APL-2009} 

The finite  DNP leads to experimentally measurable signatures in the leakage current.\cite{Ono-Tarucha-PRL-2004,Koppens-et-al-Science-2005,Pfund-Shorubalko-Ensslin-Leturcq-PRL-2007,Baugh-Kitamura-Ono-Tarucha-PRL-2007,Baugh-Kitamura-Ono-Tarucha-physica-status-solidi-2008,Churchill-Marcos-Marcus-et-al-nature-phys-2009,Takahashi-Kono-Tarucha-Ono-PRL-2011,Kobayashi-et-al-PRL-2011,Takahashi-Kono-Tarucha-Ono-APEX-2012,Petersen-Ludwig-et-al-PRL-2013} Perhaps the most fascinating of these signatures is that of a \emph{hysteretic} leakage current versus external parameters like the magnetic field or a gate-voltage as has been observed.\cite{Ono-Tarucha-PRL-2004,Koppens-et-al-Science-2005,Pfund-Shorubalko-Ensslin-Leturcq-PRL-2007,Churchill-Marcos-Marcus-et-al-nature-phys-2009,Kobayashi-et-al-PRL-2011,Frolov-Danon-Kouwenhoven-et-al-PRL-2012} For instance, Pfund \emph{et al.}\cite{Pfund-Shorubalko-Ensslin-Leturcq-PRL-2007} found hysteresis due to a competition between HF and spin-orbit induced escape from the blocking states. The hysteresis signals a \emph{bistability} in the DNP: For a certain range of parameters, there exists two stable values of the DNP leading to two values of the current. For other parameters the nuclear spins might be polarized, but the DNP is single-valued and hence also the current.\cite{footnote-more-hys}

Very high polarizations\cite{footnote-experimental-polarizations-estimates} of about \emph{half} of the nuclei have been found experimentally in the SB regime,\cite{Baugh-Kitamura-Ono-Tarucha-PRL-2007,Churchill-Marcos-Marcus-et-al-nature-phys-2009} and even higher DNPs were not excluded. Spin diffusion from the DQD to the environment and dipole-dipole interactions are very weak, but nevertheless expected to reduce the polarizations somewhat.\cite{Koppens-et-al-Science-2005,Churchill-Marcos-Marcus-et-al-nature-phys-2009,Blick-et-al-PRB-2004} 

Other experimental findings like long-lived current oscillations in time\cite{Ono-Tarucha-PRL-2004,Koppens-et-al-Science-2005} and transient phenomena in the leakage current\cite{Takahashi-Kono-Tarucha-Ono-PRL-2011} have also been attributed to the nuclear spin environment.\cite{Erlingsson-Jouravlev-Nazarov-PRB-2005,Hu-Wang-PRB-2012,Rudner-Levitov-PRL-2012} Furthermore, it has been shown that DNP can build up in DQDs by cycles in gate-voltage space -- without transport through the DQD.\cite{Ramon-Hu-PRB-2007,Petta-Taylor-PRL-2008,Foletti-Bluhm-Yacoby-et-al-nature-phys-2009} 

The HFI is most effective to lift SB close to the crossing of the electronic energy levels between e.g.~a triplet and the singlet state such that energy is conserved in the spin-flip process.\cite{footnote-dE-nuclear-levels-is-negligible} In order to get close to a level crossings, the local gate voltages on the dots, the inter-dot tunneling or the external magnetic field can be varied experimentally. The local gate voltages change the energy levels of the individual dots and thereby the so-called \emph{energy detuning} (i.e.~the electrostatic energy difference between one electron in each dot and a singlet in the right dot). The barrier between the dots controls the wavefunction overlap and therefore the quantum mechanical \emph{exchange energy} between the singlet and triplet states. Moreover, the external magnetic field splits up the triplet levels.    

\subsection{Main ideas of this work and comparison to previous works}\label{subsec:Main-ideas-of-this-work}

In this work, we analyze the leakage current and the DNP in the SB regime. Finite DNP occurs due to a competition between (i) the HF-induced escape from the blocking states and (ii) another weaker inelastic escape mechanism such as co-tunneling or spin-orbit interaction -- as in the works by Rudner \emph{et al}.\cite{Rudner-Levitov-PRL-2007,Rudner-Levitov-Nanotechnology-2010,Rudner-Rashba-PRB-2011,Rudner-Koppens-Folk-Vandersypen-Levitov-PRB-2011} The induced DNP acts back on the electronic energy levels of the DQD, which in turn change the transition rates until the steady state is reached. Hence, we are dealing with a non-linear system with feedback present as illustrated in Fig.~\ref{fig:feedback-illustration}. 

We consider the gate-voltage configuration with \emph{zero energy detuning}, i.e.~the \emph{electrostatic} energy of one electron in each dot and a singlet in the right dot is the same. Thus, the external magnetic field is varied to get close to level crossings. The Overhauser magnetic field from the DNP is different in each dot, which mixes the triplet and singlet states with zero total angular momentum projection.\cite{Koppens-et-al-Science-2005,Kobayashi-et-al-PRL-2011,Rudner-Koppens-Folk-Vandersypen-Levitov-PRB-2011}  Thus, we study two kinds of level crossings in detail: The crossing of the triplet (like) levels and the crossing between singlet and triplet levels. What we name the crossing of triplet levels is in fact the crossing of the pure spin-polarized triplets and the triplet that has a small mixing with the singlet subspace. Hence, escape from the mixed state is possible. 

The main focus of this work is the presence and description of a \emph{transition temperature} $T_c$ for the DNP. For temperatures $T$ below $T_c$ the leakage current shows hysteresis versus the external magnetic field $B$, while for $T>T_c$ the hysteresis disappears even though the system can still have a non-zero DNP. The transition temperature is related to a bistability of the DNP for $T<T_c$, which is the reason for the current hysteresis. Interestingly, we find that the transition temperatures for the crossing of the triplet levels near $B=0$, $\Tct$, and the singlet-triplet crossing at finite $B$-field, $\Tcs$, respectively,  are in general \emph{different} and $\Tct<\Tcs$ for typical parameters. Thus, for $\Tct\!<\!T\!<\!\Tcs$ current hysteresis is expected near the singlet-triplet level crossings at finite $B$-field, but \emph{not} near the crossing of the triplet levels.   

The DNP in the SB regime is current induced and, hence, a result of a \emph{non-equilibrium} situation. Remarkably, \emph{spontaneous order} of the nuclear spins in \emph{equilibrium} generally happens at orders of magnitude lower temperatures than $\Tct$ and $\Tcs$ due to the weakness of dipole-dipole interaction among the nuclear spins.\cite{kold-finn-2001,Rudner-Levitov-PRL-2007,Coish-Baugh-review-2009}

We find the transition temperature $\Tct$ analytically to be given \emph{only}  by the strength of the HF couplings in the DQD. This is derived from a simplified model valid in the limit of the singlets being far away in energy from the triplets, i.e.~at large exchange energy. To describe the singlet-triplet crossing, we also derive a simplified model leading to an implicit equation for the DNP. In contrast to $\Tct$, we find the transition temperature $\Tcs$ for the singlet-triplet crossing to depend on various parameters. 

The possibility of not conserving energy in the HF transitions is present in this work. Rudner \emph{et al.}\cite{Rudner-Levitov-PRL-2007,Rudner-Levitov-Nanotechnology-2010,Rudner-Rashba-PRB-2011,Rudner-Koppens-Folk-Vandersypen-Levitov-PRB-2011} include this effect as level broadening, whereas we allow for energy emission and absorption e.g.~by phonons in the HF rates. Hence, energy absorption and emission in a HF process is \emph{equally likely} in Refs.[\onlinecite{Rudner-Levitov-PRL-2007,Rudner-Levitov-Nanotechnology-2010,Rudner-Rashba-PRB-2011,Rudner-Koppens-Folk-Vandersypen-Levitov-PRB-2011}]. In contrast, here the probability for energy absorption is exponentially suppressed compared to energy emission.\cite{Fujisawa-Science-1998,Brandes-Kramer-PRL-1999} We show that this is indeed an \emph{essential difference} between this study and the previous ones,\cite{Rudner-Levitov-PRL-2007,Rudner-Levitov-Nanotechnology-2010,Rudner-Rashba-PRB-2011,Rudner-Koppens-Folk-Vandersypen-Levitov-PRB-2011} since the presence of both transition temperatures exactly stem from this asymmetry between energy emission and absorption. 

In previous works by some of us,\cite{Inarrea-Platero-MacDonald-PRB-2007,Inarrea-Carlos-MacDonald-Platero-APL-2007,Inarrea-Carlos-Platero-APL-2009} non-zero DNP in the SB regime arise due to the competition between HF-induced spin-flips and escape from the blocking states by tunneling through excited states in the right dot. In contrast, such excited states are assumed to be far away in energy in this work and, hence, do not play a role. Moreover, we work with coherently coupled dots such that the inter-dot tunneling is not treated as a perturbation as in Refs.[\onlinecite{Inarrea-Platero-MacDonald-PRB-2007,Inarrea-Carlos-MacDonald-Platero-APL-2007,Inarrea-Carlos-Platero-APL-2009}]. This approach, for instance, includes the expected triplet with zero angular momentum projection. Moreover, previously\cite{Inarrea-Platero-MacDonald-PRB-2007,Inarrea-Carlos-MacDonald-Platero-APL-2007,Inarrea-Carlos-Platero-APL-2009} phonon absorption processes were neglected, so the physics treated here regarding the transition temperature was missed. 

Some of us have numerically studied a similar approach recently.\cite{Lopez-Monis-et-al-NJP-2011} However, in this case, the rate equation for the DNP turned out  to be inappropriate, because HFI was taken to be the \emph{only} escape mechanism from the blocking states. In contrast, here we find DNP to appear due to a competition between HF-induced escape and another inelastic escape mechanism. Here we put our results from the rate equation approach on a firm basis by comparing to Monte Carlo simulations. Furthermore, we point out in detail how the rate equation approach becomes sensitive to some initial occupations in the case without an inelastic escape path, and therefore become unable to describe the physical setup. Moreover, we emphasize that this work contains many new insights and results compared to Ref.[\onlinecite{Lopez-Monis-et-al-NJP-2011}]. For instance, the simple analytical models describing the various level crossings, which lead to the transition temperatures described above. 

The paper is organized as follows: Sec.~\ref{sec:model} describes the model of the DQD energy levels and their interplay with the DNP. Then we address the crossing of the triplets (Sec.~\ref{sec:triplet-cross}) and the singlet-triplet crossings (Sec.~\ref{sec:ST-cross}). Finally, the Monte Carlo simulations are discussed (Sec.~\ref{sec:rate-eq-breakdown-and-MC}).

\section{The model}\label{sec:model}

The model used below aims at describing the basic physics of a DQD coupled to a nuclear environment in the SB regime -- instead of focusing on a specific material. 

\subsection{The states of the DQD, the Hamiltonian and the Overhauser field}

The three triplet states blocking the transport in the SB regime are
\begin{subequations}
\label{eq:pure-triplet-states}
\begin{align}
|T_+\rangle&=d^\dag_{\tl\op}d^\dag_{\tr\op}|0\rangle,
\qquad
|T_-\rangle=d^\dag_{\tl\ned}d^\dag_{\tr\ned}|0\rangle,
\\
|T_0\rangle&=
\frac{1}{\sqrt{2}}
\left(d^\dag_{\tl\op}d^\dag_{\tr\ned}+d^\dag_{\tl\ned}d^\dag_{\tr\op}\right)
|0\rangle,
\end{align}
\end{subequations}
where the indices $0$ and $\pm$ represent the total angular momentum projection, $m=0,\pm1$. The singlet states with one electron in each dot, $|S(1,1)\rangle$, and two electrons in the right dot, $|S(0,2)\rangle$, respectively, are
\begin{subequations}
\label{eq:pure-singlet-states}
\begin{align}
|S(1,1)\rangle&=
\frac{1}{\sqrt{2}}
\left(d^\dag_{\tl\op}d^\dag_{\tr\ned} 
-d^\dag_{\tl\ned}d^\dag_{\tr\op}\right)
|0\rangle,
\\
|S(0,2)\rangle&=d^\dag_{\tr\ned}d^\dag_{\tr\op}|0\rangle.
\end{align}
\end{subequations}
A single electron with spin $\sigma$ trapped in the right dot is described by the one-electron state $|\sigma(0,1)\rangle=d^\dag_{\tr\sigma}|0\rangle$ for $\sigma=\op,\ned$. Here we only include a single spin-degenerate state created (annihilated) by $d^\dag_{\alpha\sigma}$ ($d^{}_{\alpha\sigma}$) in the right ($\alpha=\tr$) or left ($\alpha=\tl$) dot. The empty state is $|0\rangle$.

The entire system is described by the Hamiltonian $H=H_{\textrm{DQD}}+H_{\textrm{leads}}+H_{\textrm{T}}+H_{\textrm{HF}}$, where $H_{\textrm{DQD}}$, $H_{\textrm{leads}}$ and $H_{\textrm{T}}$ describe the two dots in series, the electronic leads and the tunneling coupling between them, respectively. The HFI, $H_{\textrm{HF}}$, is between the electrons in the DQD and the nuclear spins. The DQD is described by an Anderson-type Hamiltonian $H_{\textrm{DQD}}=\sum_{\alpha=\tl,\tr} \big(\e_{\alpha} n_{\alpha}+U_{\alpha}n_{\alpha\op}n_{\alpha\ned}+g\mu_BBS_{z,\alpha}\big)+U^{}_{\tr\tl}n^{}_{\tr}n^{}_{\tl}+t\sum_{\sigma}\big(d^\dag_{\tl\sigma}d^{}_{\tr\sigma}+d^\dag_{\tr\sigma}d^{}_{\tl\sigma}\big)$. Here $n_{\alpha}=n_{\alpha\op}+n_{\alpha\ned}$ is the number operator on each dot, $n_{\alpha\sigma}=d^\dag_{\alpha\sigma}d^{}_{\alpha\sigma}$, and $S_{z,\alpha}=\frac{1}{2}\left(n_{\alpha\op}-n_{\alpha\ned}\right)$ is the spin operators $z$-component on dot $\alpha$. Both inter-dot, $U_{\tr\tl}$, and intra-dot,  $U_{\alpha}$, Coulomb interaction are included. The onsite energies $\e_\alpha$ and the inter-dot tunneling $t$ can be tuned experimentally by gates. An external magnetic field $B$ causes a Zeeman splitting, $g\mu_BB$, where $g$ is the $g$-factor and $\mu_B$ is the Bohr magneton. The $B$-field is taken to be in-plane, so orbital effects can be safely neglected.\cite{Ono-Tarucha-PRL-2004,Coish-Qassemi-PRB-2011} For $t=0$, the triplets (\ref{eq:pure-triplet-states}) and singlets (\ref{eq:pure-singlet-states}) are eigenstates of $H_{\textrm{DQD}}$, whereas the singlets (\ref{eq:pure-singlet-states}) mix for $t\neq0$. The leads are described by $H_{\textrm{leads}}=\sum_{\alpha \kk \sigma} \e_{k\sigma\alpha} c^{\dag}_{\kk\sigma\alpha}c^{}_{\kk\sigma\alpha}$, where $c^{\dag}_{\kk\sigma\alpha}$ ($c^{}_{\kk\sigma\alpha}$) creates (annihilates) the orbital state $\kk$ with spin $\sigma$ in lead $\alpha$ of energy $\e_{k\sigma\alpha}$. The coupling of the leads to the DQD is $H_{\textrm{T}}=\sum_{\alpha \kk \sigma} (t^{}_{\alpha \kk}c^{\dag}_{\kk\sigma\alpha}d^{}_{\alpha\sigma}+t^\ast_{\alpha \kk}d^{\dag}_{\alpha\sigma} c^{}_{\kk\sigma\alpha})$ with the lead-DQD tunneling couplings $t^{}_{\alpha \kk}$.

The contact HFI relevant for $S$-like orbital states\cite{Schliemann-Khaetskii-Loss-JoPCM-Review-2003,Coish-Baugh-review-2009,Fischer-Coish-Bulaev-Loss-PRB-2008,Fischer-Trauzettel-Loss-PRB-2009,Fischer-Loss-PRL-2010,Lunde-Platero-HF-HgTe} $H_{\textrm{HF}}=\sum_{\alpha,n}A_\alpha(\RR_n)\Ss_{\alpha}\cdot\II_n$ connects the electronic spin $\Ss_{\alpha}=(S_{x,\alpha},S_{y,\alpha},S_{z,\alpha})$ on dot $\alpha=\tl,\tr$ with the nuclear spin $\II_{n}=(\II_{x,n},\II_{y,n},\II_{z,n})$ at position $\RR_n$. The coupling between the spins depends on the value of the electronic envelope wavefunction $\Psi_{\alpha}(\rr)$  of dot $\alpha=\tl,\tr$ at the position of the nuclear spin $\RR_n$, i.e.~$A_\alpha(\RR_n)=\nu \mathcal{A}|\Psi_{\alpha}(\RR_n)|^2$. Here $\nu$ is the unit cell volume and $\mathcal{A}$ is the atomic HF constant. For simplicity, the nuclear spins are taken to be spin--$1/2$ and we model the HFI using homogeneous HF constants,\cite{Khaetskii-Loss-Glazman-PRB-2003,Erbe-Schliemann-PRB-2010,Inarrea-Platero-MacDonald-PRB-2007} i.e. 
\begin{align}
H_{\textrm{HF}}=
\sum_{\alpha}\frac{A_\alpha}{N}\sum_n\Ss_{\alpha}\cdot\II_n, 
\end{align}
where $N$ is the total number of nuclear spins in the DQD. The \emph{effective} HF constants, $A_\alpha\simeq \mathcal{A}N/(2N_\alpha)$, for the two dots are of the same order of magnitude, but are not necessarily the same\cite{Coish-Loss-PRB-2005,Erbe-Schliemann-PRB-2010,Lopez-Monis-et-al-NJP-2011} for realistic dots containing different numbers of nuclear spins, $N_{\tl}\neq N_{\tr}$.

The polarization of the nuclei acts back on the electronic states as an effective Overhauser magnetic field.\cite{Overhauser-PR-1953} To include this, we divide the HF Hamiltonian into a mean-field part $H_{\textrm{HF}}^{\textrm{MF}}$, the Overhauser field, and a spin-flip part $H_{\textrm{HF}}^{\textrm{sf}}$, which leads to the HF-induced spin-flips necessary for \emph{dynamically} polarizing the nuclei. The external magnetic field provides a direction along which the nuclei can polarize,\cite{Takahashi-Kono-Tarucha-Ono-PRL-2011,Coish-Baugh-review-2009} such that the rotational symmetry is broken as in the experimental situation.\cite{Takahashi-Kono-Tarucha-Ono-PRL-2011} This is a customary approach for mean-field theories describing phenomena such as magnetization of a ferromagnet.\cite{Flensberg-BOOK} Hence, the mean-field from the nuclei is taken to be along the $z$-direction as the external magnetic field and given in terms of the average number of spin up and down $N_{\sigma}$ ($\sigma=\op,\ned$) in the nuclear environment,\cite{Baugh-Kitamura-Ono-Tarucha-PRL-2007,Lopez-Monis-et-al-NJP-2011,Rudner-Koppens-Folk-Vandersypen-Levitov-PRB-2011} i.e.~$\sum_n\langle\II_n\rangle=\sum_n\langle I_{z,n}\rangle=(N_\op-N_\ned)/2$. Thus,   
\begin{align}
H_{\textrm{HF}}^{\textrm{MF}}=
\sum_{\alpha=\tl,\tr}\frac{1}{2}A_\alpha P S_{z,\alpha},
\label{eq:H-HF-MF} 
\end{align}
where the nuclear spin polarization $P\equiv(N_\op-N_\ned)/N$ was introduced and $N=N_\op+N_\ned$ is the total number of nuclear spins. The number of nuclear spin-up (down) $N_{\op(\ned)}$ -- and thus $P$ -- change dynamically according to the external conditions of the current. The spin-flip part of the HFI is  
\begin{align}
H_{\textrm{HF}}^{\textrm{sf}}=
\sum_{\alpha,n}\frac{A_\alpha}{2N}
\left(S_{-,\alpha}I_{+,n}+S_{+,\alpha}I_{-,n}\right),
\label{eq:H-HF-fluc}
\end{align}
where $S_{\pm,\alpha}=S_{x,\alpha}\pm iS_{y,\alpha}$ and $I_{\pm,n}=I_{x,n}\pm iI_{y,n}$ are the raising and lowering operators of the electronic and nuclear spins, respectively. The electronic spin-flips induced by $H_{\textrm{HF}}^{\textrm{sf}}$ are included perturbatively below (see Sec.~\ref{subsec:rate-eq}).

From $H_{\textrm{HF}}^{\textrm{MF}}$, we can read off the Overhauser magnetic field in each dot:\cite{Inarrea-Platero-MacDonald-PRB-2007} $B^{\alpha}_{\textrm{nuc}}=\frac{A_\alpha P}{2g\mu_B}$. Importantly, the Overhauser fields in the two dots are \emph{different}, $B^{\tr}_{\textrm{nuc}}\neq B^{\tl}_{\textrm{nuc}}$, which is crucial for lifting the SB\cite{Koppens-et-al-Science-2005,Kobayashi-et-al-PRL-2011}  (as will be clear below). Here we introduce this difference by having $A_\tl\neq A_\tr$, but keep the DNP $P$ as a common quantity for both dots.\cite{Rudner-Levitov-PRL-2007,Rudner-Levitov-Nanotechnology-2010,Rudner-Rashba-PRB-2011,Lopez-Monis-et-al-NJP-2011} In principle, the DNP can be spatially inhomogeneous, which is challenging to model in detail. A step on that way, is having different  -- but homogeneous -- polarizations in the two dots and $A_\tl=A_\tr$.\cite{Inarrea-Platero-MacDonald-PRB-2007,Rudner-Koppens-Folk-Vandersypen-Levitov-PRB-2011} However, for coherently coupled dots, the overlap between the envelope functions is sizeable and thereby also the amount of nuclei under both envelope functions. This makes it less clear how to separate the nuclei into forming two independent homogeneous polarizations.\cite{Dominguez-Platero-PRB-2009} For simplicity, we therefore use a single DNP for both dots.

\begin{figure}
\includegraphics[width=0.37\textwidth,angle=0]{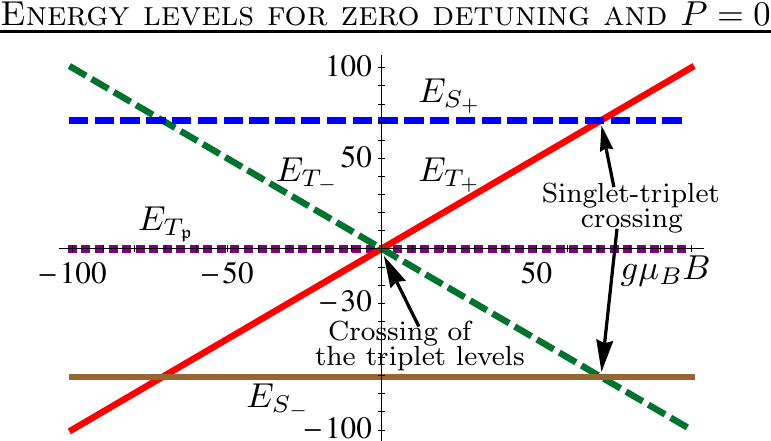}
\caption{(Color online) The energy levels (\ref{eq:energy-levels-def}) versus magnetic field in energy units $g\mu_BB$ for the five two-electron states $T_{+}$ (red full line), $T_-$ (green dashed line), $T_\PP$ (purple dotted line), $S_+$ (blue dashed line) and $S_-$ (brown full line) for zero detuning, $P\!=\!0$ and $t\!=\!50\mu$eV. The crossings of the triplets and the triplet-singlet crossing points (for $B>0$) are indicated.}
\label{fig:energy-levels}
\end{figure}
 
We find the basis states of the DQD including the Overhauser fields and the inter-dot tunneling by diagonalizing $H_{\textrm{DQD}}+H_{\textrm{HF}}^{\textrm{MF}}$ within the space of triplets $|T_{m}\rangle$ (\ref{eq:pure-triplet-states}), singlets (\ref{eq:pure-singlet-states}) and one-electron states $|\sigma(0,1)\rangle$, since all other states are not energetically relevant in the SB regime.\cite{footnote-degenerate-per} We specialize to the \emph{zero detuning} limit such that the electrostatic energy of one electron in each dot, $\e_{\tl}+\e_{\tr}+U_{\text{RL}}$, and of a singlet in the right dot, $2\e_{\tr}+U_{\tr}$, are the same: $\e_{\tl}+\e_{\tr}+U_{\text{RL}}=2\e_{\tr}+U_{\tr}=0$ (chosen as the zero of energy). For zero detuning, the diagonalization gives the following particularly simple basis states 
\begin{subequations}
\label{eq:2-electron-states}
\begin{align}
|T_+\rangle&=d^\dag_{\tl\op}d^\dag_{\tr\op}|0\rangle,
\qquad
\quad
|T_-\rangle=d^\dag_{\tl\ned}d^\dag_{\tr\ned}|0\rangle,
\\
|T_\PP\rangle&=
\frac{1}{\mathcal{N}}
\Big[
|T_0\rangle-\PP|S(0,2)\rangle
\Big],
\label{eq:T-p-state-def}
\\
|S_\pm\rangle&=
\frac{1}{\sqrt{2}}
\left[
|S(1,1)\rangle
\pm
\frac{1}{\mathcal{N}}
\Big(|S(0,2)\rangle+\PP|T_0\rangle\Big)
\right],
\end{align}
\end{subequations}
and the one-electron states remain the same, $|\sigma(0,1)\rangle=d^\dag_{\tr\sigma}|0\rangle$ and we set $\hbar=1$. 
Here we introduced 
\begin{align}
\label{eq:def-p-N-Apm}
\PP\equiv&\frac{A_-}{2\sqrt{2}t}P,\quad
\mathcal{N}\equiv\sqrt{1+\PP^2},\quad
A_{\pm}\equiv\frac{A_\tl\pm A_\tr}{2}
\end{align}
and the energies are found to be (see Fig.~\ref{fig:energy-levels})
\begin{subequations}
\label{eq:energy-levels-def}
\begin{align}
E_{T_{\PP}}&=0,
\qquad
E_{T_{\pm}}=\pm 
\left(g\mu_BB
+\frac{1}{2}A_+P\right),\\
E_{S_{\pm}}&=\pm\sqrt{2}t\mathcal{N}. 
\end{align}
\end{subequations}
Here the inter-dot tunneling mixes $|S(0,2)\rangle$ and $|S(1,1)\rangle$. Moreover, these two singlets mix with $T_0$ due to the \emph{difference} of the Overhauser fields between the two dots,\cite{Kobayashi-et-al-PRL-2011,Rudner-Koppens-Folk-Vandersypen-Levitov-PRB-2011,Lopez-Monis-et-al-NJP-2011} $g\mu_B(B^{\tl}_{\textrm{nuc}}-B^{\tr}_{\textrm{nuc}})=A_-P/2$. The triplet-singlet mixing is controlled by the dimensionless parameter $\PP$ in Eq.(\ref{eq:def-p-N-Apm}). The \emph{sum} of the Overhauser fields, $g\mu_B(B^{\tl}_{\textrm{nuc}}+B^{\tr}_{\textrm{nuc}})= A_+P$, splits the spin-polarized triplets $T_\pm$ as a magnetic field does. Hence, $E_{T_{\pm}}$ depend stronger on $P$ than the exchange energy splitting $|E_{S_\pm}-E_{T_\PP}|=\sqrt{2}t\mathcal{N}$ for $\PP\ll 1$. The singlet-triplet mixing $\PP$ is indeed small, since we are interested in the limit  $A_-\ll t$. Therefore, we keep calling the state $T_\PP$ for a \emph{triplet} and the states $S_\pm$ for \emph{singlets} (as indicated by the notation), even though \emph{strictly speaking} they are \emph{not}.

In the SB regime, the triplets $T_\pm$ and $T_0$ block transport. Due to the singlet-triplet mixing, the state $T_\PP$ is not a blocking state anymore, whereas $T_\pm$ still block transport. Therefore, if $P=0$ or $A_\tr=A_\tl$, then $\PP=0$ leading to transport blocking by all three triplets ($T_\PP\rightarrow T_0$).

In contrast to \emph{finite} detuning,\cite{Koppens-et-al-Science-2005,Rudner-Levitov-PRL-2007,Baugh-Kitamura-Ono-Tarucha-PRL-2007,Baugh-Kitamura-Ono-Tarucha-physica-status-solidi-2008,Kobayashi-et-al-PRL-2011} the crossing of the singlet and triplet energies (\ref{eq:energy-levels-def}) always happens in pairs for zero detuning, e.g.~$E_{T_+}$ and $E_{S_+}$ cross if and only if $E_{T_-}$ and $E_{S_-}$ cross. Moreover, here the energy levels relative differences have certain symmetries around $E_{T_\PP}$: $E_{T_+}\!-E_{T_\PP}\!=\!E_{T_\PP}\!-E_{T_-}$ and $E_{S_+}\!-E_{T_+}\!=\!E_{T_-}\!-E_{S_-}$.

\subsection{The dynamical coupling of the nuclear spins to the DQD energy levels and the leakage current}\label{subsec:rate-eq}

\begin{figure*}
\includegraphics[width=0.84\textwidth,angle=0]{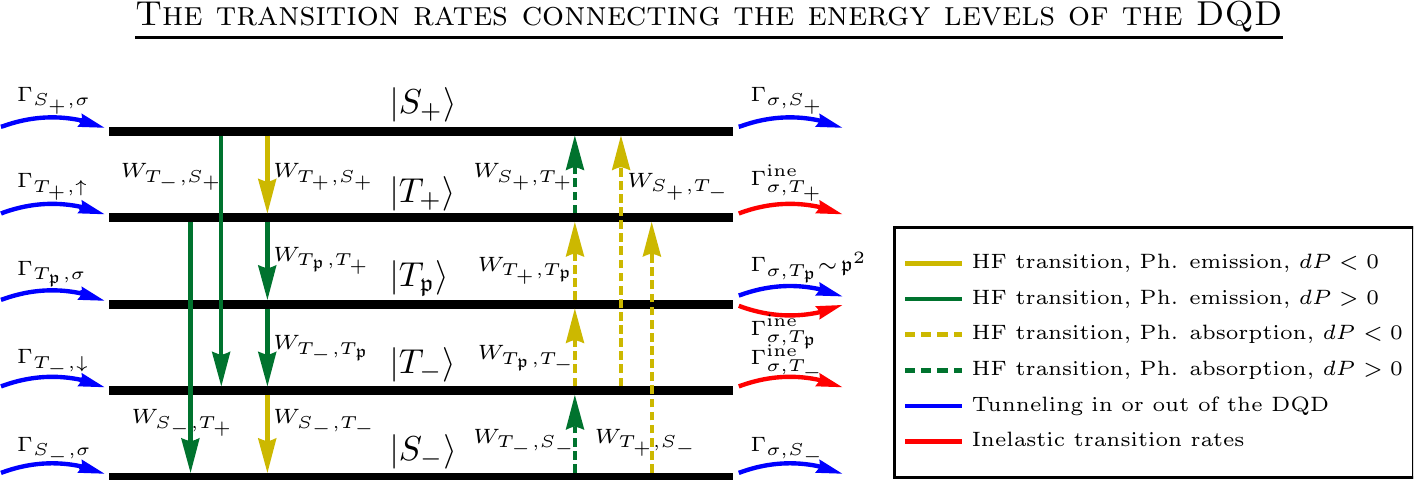}
\caption{(Color online) Illustration of the transitions connecting the DQD states. Each HF transition  $W_{f,i}$ flip the electronic and nuclear spins in opposite directions and change the DNP by $dP=+2/N$ or $dP=-2/N$ depending on the transition -- as specified in the label of the figure. We allow for emission or absorption of energy e.g. as a phonon (ph.) in the HF processes. Thus, transitions between misaligned energy levels are possible, but they become less probable the larger the difference between the energy levels. Moreover, as discussed in the main text, absorption of energy is suppressed by a factor of $e^{-\Delta E/\kb T}$ compared to emission, where $\Delta E>0$ is the difference between the levels. Additional inelastic escape rates $\Gamma^{\textrm{ine}}_{f,i}$ from the triplets without a nuclear spin-flip are also included (red arrows). The competition between these rates and the HF rates creates the possibility of finite DNP.\cite{Rudner-Levitov-PRL-2007,Pfund-Shorubalko-Ensslin-Leturcq-PRL-2007,Rudner-Levitov-Nanotechnology-2010} We work in the high bias limit, so transitions from one-electron to two-electron states is \emph{always} associated with tunneling \emph{into} the DQD from the \emph{left} lead and vice versa, as illustrated in the figure. Moreover, a particular order of the energy levels is chosen in the figure, which depends on the exchange energy $\sqrt{2}t\mathcal{N}$, the $B$-field and the DNP, see Eq.(\ref{eq:energy-levels-def}).}
\label{fig:full-rate-eq-illustration}
\end{figure*}

The dynamics is determined within the rate equation approach \cite{Inarrea-Platero-MacDonald-PRB-2007,Qassemi-Coish-Wilhelm-PRL-2009,Rudner-Levitov-Nanotechnology-2010,Rudner-Rashba-PRB-2011,Lopez-Monis-et-al-NJP-2011,Coish-Qassemi-PRB-2011} written compactly as 
\begin{align} 
\label{eq:full-rate-eqs}
\dot{n}^{}_{\nu}
=&  
\sum_{\nu^{\prime}}
\big(W^{}_{\nu,\nu^{\prime}} 
+\Gamma^{}_{\nu,\nu^{\prime}} 
+\Gamma_{\nu,\nu^{\prime}}^{\textrm{ine}}
\big)
n^{}_{\nu^{\prime}}
\nonumber\\
&
\hspace{1cm}
-n^{}_\nu
\sum_{\nu^{\prime}}
\big(
W^{}_{\nu^{\prime},\nu} 
+\Gamma^{}_{\nu^{\prime},\nu} 
+\Gamma_{\nu^{\prime},\nu}^{\textrm{ine}}
\big),
\end{align}
where $\dot{n}_\nu^{}$ denotes the time derivative of the average occupation $n_\nu^{}$ for $\nu=T_\pm,T_\PP,S_\pm,\op,\ned$ [using the short-hand notation $\sigma=\op,\ned$ for $|\sigma(0,1)\rangle$]. Moreover, the normalization condition is $\sum_\nu n_\nu=1$. We include three kinds of rates: (i) The HF spin-flip rates $W_{f,i}$ from the initial to the final two-electron state, $|i\rangle\rightarrow|f\rangle$. (ii) The tunneling rates $\Gamma_{f,i}$ from $|i\rangle$ to $|f\rangle$, which connect the leads to the DQD.  (iii) Finally, we include another inelastic escape rate $\Gamma_{f,i}^{\textrm{ine}}$ from the triplet states $T_\pm$ and $T_\PP$. Sec.~\ref{subsec:all-rates} gives the detailed rates. Fig.~\ref{fig:full-rate-eq-illustration} provides an illustration of the \emph{non-zero} rates in the rate equations (\ref{eq:full-rate-eqs}), which are given explicitly in Appendix \ref{appendix:full-rate-eqs} for completeness.

In order to obtain non-zero DNP, a preferred direction of angular momentum transfer from the electrons to the nuclei needs to exist. \emph{If} HFI is the \emph{only} source of SB lifting, then no such preferred direction exists. The reason is that HF-induced escape from the spin-polarized triplets $T_\pm$ changes the DNP in opposite directions and since the probabilities of loading $T_+$ and $T_-$ are equal ($\Gamma_{T_+,\op}=\Gamma_{T_-,\ned}$), no net DNP build up.\cite{Jouravlev-Nazarov-PRL-2006,Rudner-Levitov-PRL-2007,Pfund-Shorubalko-Ensslin-Leturcq-PRL-2007,Rudner-Levitov-Nanotechnology-2010} This is so, even though the HF rates from $T_\pm$ might be very different, but since only one way to escape from $T_\pm$ exists, it does not matter if escaping from $T_+$ or $T_-$ is the fastest.

Here we allow \emph{two} ways to escape from the triplets, both lifting SB. Either by the HF spin-flip transitions $W_{f,i}$ or by the inelastic escape rates $\Gamma_{f,i}^{\textrm{ine}}$. The additional inelastic rates can e.g. be provided by co-tunneling\cite{Johnson-Petta-Marcus-PRB-2005,Vorontsov-Vavilov-PRL-2008,Saito-et-al-Physica-E-2008,Qassemi-Coish-Wilhelm-PRL-2009,Coish-Qassemi-PRB-2011} or spin-orbit mediated\cite{Pfund-Shorubalko-Ensslin-Leturcq-PRL-2007,Danon-Nazarov-PRB-2009,Weiss-Rashba-Kuemmeth-Churchill-Flensberg-PRB-2010,Frolov-Danon-Nazarov-Kouwenhoven-et-al-PRB-2010,Stepanenko-Rudner-Bertrand-Loss-PRB-2012} spin relaxation processes. Importantly, the inelastic processes contained in $\Gamma_{f,i}^{\textrm{ine}}$ give an additional escape path from the triplets \emph{without} a nuclear spin-flip. Therefore, the two escape paths from the blocking states now \emph{compete}, such that it becomes important which of the HF-induced escape paths from $T_+$ or $T_-$ is the fastest. This competition is therefore crucial to obtain non-zero DNP.\cite{Rudner-Levitov-PRL-2007,Pfund-Shorubalko-Ensslin-Leturcq-PRL-2007,Rudner-Levitov-Nanotechnology-2010} We do not specify the inelastic escape rate further as Rudner \emph{et al.}\cite{Rudner-Levitov-PRL-2007,Rudner-Levitov-Nanotechnology-2010,Rudner-Koppens-Folk-Vandersypen-Levitov-PRB-2011,Rudner-Rashba-PRB-2011}

Each HF-induced electronic spin-flip will changes the DNP by $dP=\pm2/N$ depending on the transition, see Fig.~\ref{fig:full-rate-eq-illustration}. Therefore, in the case of competing escape rates, we describe the DNP, $P$, by the rate equation 
\begin{align} 
\dot{P}=&  
\frac{2}{N}\bigg[\big(W^{}_{T_-,S_+}-\!W^{}_{T_+,S_+}\big)n^{}_{S_+}
\nonumber \\ 
&+\big(W^{}_{T_-,S_-}\!-\!W^{}_{T_+,S_-}\big)n^{}_{S_-}
\!+\!\big(W^{}_{T_-,T_\PP}\!-\!W^{}_{T_+,T_\PP}\big)n^{}_{T_\PP}
\nonumber \\ 
&+\big(W^{}_{S_+,T_+}\!+\!W^{}_{S_-,T_+}\!+\!W^{}_{T_\PP,T_+}\big)n^{}_{T_+} 
\nonumber \\ 
&-\big(W^{}_{S_+,T_-}\!+\!W^{}_{S_-,T_-}\!+\!W^{}_{T_\PP,T_-}\big)n^{}_{T_-}
\bigg]
\label{eq:full-polarization-eq}
\end{align}
as used in several other studies.\cite{Rudner-Levitov-PRL-2007,Baugh-Kitamura-Ono-Tarucha-PRL-2007,Inarrea-Platero-MacDonald-PRB-2007,Petta-Taylor-PRL-2008,Rudner-Levitov-Nanotechnology-2010,Rudner-Koppens-Folk-Vandersypen-Levitov-PRB-2011,Rudner-Rashba-PRB-2011} The idea behind Eq.(\ref{eq:full-polarization-eq}) is to describe a competition between various rates that polarize the nuclear spins in opposite directions, which is different from the aforementioned competition between various escape paths. However, in order to \emph{have} a competition between various rates that polarize the nuclei in the first place, it is strictly necessary to have competing escape paths, i.e.~both $W_{f,i}$ and $\Gamma^{\textrm{ine}}_{f,i}$ non-zero. In Sec.~\ref{sec:rate-eq-breakdown-and-MC}, we show that the rate equation approach (\ref{eq:full-rate-eqs}-\ref{eq:full-polarization-eq}) fails for $\Gamma^{\textrm{ine}}_{f,i}=0$. Moreover, we validate the rate equation approach for $\Gamma^{\textrm{ine}}_{f,i}\neq0$, by showing that it leads to the same results as a Monte Carlo simulation. 

Furthermore, we neglect processes leading to depolarization of the nuclear bath in equilibrium,\cite{footnote-depol} since these are much slower than the HF spin-flip processes during transport through the DQD in the SB regime.\cite{Koppens-et-al-Science-2005,Churchill-Marcos-Marcus-et-al-nature-phys-2009,Blick-et-al-PRB-2004} 

In  this work, we use the \emph{high bias limit}, where electrons only enter the DQD from the left lead and leave it to the right lead. Hence, all the transitions from the two- to the one-electron states give the particle leakage current $I$ as
\begin{align}
I= 
\sum_{\sigma=\op,\ned}
\Big[
&
\Gamma_{\sigma,T_+}^{\textrm{ine}}n^{}_{T_+} 
+\Gamma_{\sigma,T_-}^{\textrm{ine}}n^{}_{T_-}
+(\Gamma_{\sigma,T_\PP}^{\textrm{ine}}+\Gamma^{}_{\sigma,T_\PP})n^{}_{T_\PP} 
\nonumber\\ 
&+\Gamma^{}_{\sigma,S_+}n^{}_{S_+}
+\Gamma^{}_{\sigma,S_-}n^{}_{S_-}
\Big].
\label{eq:current-def}
\end{align}
Experimentally, the high-bias limit and the zero energy detuning of the levels can be adjusted independently.

Below, the non-linear system of rate equations (\ref{eq:full-rate-eqs}-\ref{eq:full-polarization-eq}) is solved in the stationary limit by analytical and numerical means leading to the leakage current $I$ in Eq.(\ref{eq:current-def}).

\subsection{The transition rates}\label{subsec:all-rates} 

Now we give the rates used in the explicit calculations. 

\subsubsection{The inelastic escape rates}

For simplicity, we take the inelastic escape rates to be equal and constant following Refs.[\onlinecite{Rudner-Levitov-PRL-2007,Rudner-Levitov-Nanotechnology-2010,Rudner-Koppens-Folk-Vandersypen-Levitov-PRB-2011,Rudner-Rashba-PRB-2011}], i.e.
\begin{align}
\Gamma^{\textrm{ine}}_{\sigma,T_+}=
\Gamma^{\textrm{ine}}_{\sigma,T_-}=
\Gamma^{\textrm{ine}}_{\sigma,T_\PP}
\equiv
\Gamma^{\textrm{ine}}_{}
\quad\textrm{for}\quad
\sigma=\op,\ned. 
\label{eq:equal-Gamma-ine}
\end{align}
Since we consider the limit where the largest of the HF rates dominates over the inelastic escape rates close to the level crossings, then leaving out their energy dependence plays less of a role.\cite{Rudner-Levitov-PRL-2007,Rudner-Levitov-Nanotechnology-2010,Rudner-Koppens-Folk-Vandersypen-Levitov-PRB-2011,Rudner-Rashba-PRB-2011}
Experimentally, the co-tunneling rates can be decreased by tuning the energy levels compared to the Fermi levels of the leads,\cite{Johnson-Petta-Marcus-PRB-2005} while the spin-orbit strength e.g. can be decreased by material choice.\cite{Pfund-Shorubalko-Ensslin-Leturcq-PRL-2007} Appendix \ref{appendix:non-equal-inelastic-rates} discuss non-equal inelastic rates.

\subsubsection{The tunneling rates}\label{subsubsec:tunnel-rates}

The tunneling rates are found by Fermi golden rule. In general, the rates for tunneling into (out of) the DQD are proportional to (one minus) the Fermi distribution of the corresponding lead.\cite{Flensberg-BOOK} Due to the \emph{high bias limit}, we can leave out the Fermi functions from the explicit expressions of the tunneling rates below.

The non-zero rates for tunneling into the DQD are\cite{Lopez-Monis-et-al-NJP-2011} 
\begin{subequations}
\label{eq:tunnel-in-rates}
\begin{align}
\Gamma^{}_{T_+,\op}&=
\Gamma^{}_{T_-,\ned}=\Gamma^{}_\tl,
\\
\Gamma^{}_{T_\PP,\op}&=
\Gamma^{}_{T_\PP,\ned}=
\frac{\Gamma^{}_\tl}{2}\frac{1}{\mathcal{N}^2},
\\
\Gamma^{}_{S_+,\op}&=
\Gamma^{}_{S_-,\ned}=
\frac{\Gamma^{}_\tl}{4}\left(1-\frac{\PP}{\mathcal{N}}\right)^2,
\\
\Gamma^{}_{S_-,\op}&=
\Gamma^{}_{S_+,\ned}=
\frac{\Gamma^{}_\tl}{4}\left(1+\frac{\PP}{\mathcal{N}}\right)^2,
\end{align} 
\end{subequations}
and the non-zero rates for tunneling out of the DQD are
\begin{subequations}
\label{eq:tunnel-out-rates}
\begin{align} 
\Gamma_{\sigma,T_\PP}^{}&=
\Gamma_\tr^{}\frac{\PP^2}{\mathcal{N}^2},
\label{eq:tunnel-out-from-Tx}
\\
\Gamma^{}_{\sigma,S_+}&=
\Gamma^{}_{\sigma,S_-}=
\frac{\Gamma^{}_\tr}{2}\frac{1}{\mathcal{N}^2}
\quad\textrm{for}\quad \sigma=\op,\ned,
\end{align} 
\end{subequations}
i.e.~the probability of leaving behind a single electron in the right dot with spin up or down are equal. The rate of leaving $T_\PP$ goes to zero for $\PP\rightarrow0$ (i.e. $P\rightarrow0$ or $A_\tl\rightarrow A_\tr$), since the triplet-singlet mixing vanishes so $T_\PP$ becomes a blocking state. Here we use $\Gamma^{}_{\alpha}=2\pi |t_{\alpha k}|^2 D_{\alpha}$ in the standard wide-band approximation,\cite{Peskin-review-2010,Flensberg-BOOK} where $D_{\alpha}$ is the density of states of lead $\alpha$. 

For the calculations to follow, it is helpful to note that they are invariant under the simultaneous interchange of 
\begin{align} 
\label{eq:sym-for-tunnel-rates}
\op\leftrightarrow\ned,\quad
T_+\leftrightarrow T_- 
\quad\textrm{and}\quad 
S_+\leftrightarrow S_-.
\end{align}

\subsubsection{The hyperfine-induced spin-flip rates}\label{subsubsec:HF-rates}

The HF rates are found perturbatively in $H^{\textrm{sf}}_{\textrm{HF}}$ using the Fermi golden rule.\cite{Inarrea-Platero-MacDonald-PRB-2007,Rudner-Levitov-PRL-2007,Rudner-Levitov-Nanotechnology-2010,Lopez-Monis-et-al-NJP-2011,Flensberg-BOOK} The HF transition from, say, $T_+$ to $S_+$ implies a nuclear spin-flip from down to up, so the presence of a spin down among the nuclei is required. Thus, the rate $W_{S_+,T_+}$ is proportional to the probability of finding a random nuclear spin to be down: $W_{S_+,T_+}\propto N_{\ned}/N$, where $N_{\sigma}$ is the number of nuclei with spin $\sigma=\ned,\op$ and $N=N_\op+N_\ned$.\cite{Rudner-Levitov-PRL-2007,Inarrea-Platero-MacDonald-PRB-2007,Rudner-Levitov-Nanotechnology-2010,Lopez-Monis-et-al-NJP-2011,Lunde-Platero-PRB-2012} Likewise the other HF rates $W_{f,i}$ are proportional to either $\frac{N_{\ned}}{N}=\frac{1-P}{2}$ or  $\frac{N_{\op}}{N}=\frac{1+P}{2}$ depending on the direction of the nuclear spin-flip in the process, see Fig.~\ref{fig:full-rate-eq-illustration}. 

Here we allow the HF transitions to exchange energy with the environment e.g. by phonons.\cite{Kim-Vagner-Xing-PRB-1994,Erlingsson-Nazarov-Falko-PRB-2001,Erlingsson-Nazarov-PRB-2002,Abalmassov-Marquardt-PRB-2004,Meunier-Kouwenhoven-PRL-2007,Inarrea-Platero-MacDonald-PRB-2007,Prada-Blick-Joynt-PRB-2008,Golovach-Khaetskii-Loss-PRB-2008,Danon-arxiv-2013,Srinivasa-arxiv-2013}   Phonon emission has been shown to be significant even in low temperature transport experiments.\cite{Fujisawa-Science-1998,Meunier-Kouwenhoven-PRL-2007} In the transition, it is much easier to emit energy compared to absorbing energy by phonons,\cite{Fujisawa-Science-1998,Brandes-Kramer-PRL-1999} since the rate for \emph{absorbing} an energy of $\h\om$ is proportional to the phonon occupation factor $n_B(\h\om)$, while the rate for \emph{emitting} an energy of $\h\om$ is  $\propto n_B(\h\om)+1$. Here $n_B(E)=[e^{E/\kb T}-1]^{-1}$ is the Bose function.\cite{Fujisawa-Science-1998,Brandes-Kramer-PRL-1999,Erlingsson-Nazarov-PRB-2002,Abalmassov-Marquardt-PRB-2004} Thus, the absorption rate is suppressed by $n_B(\h\om)/[n_B(\h\om)+1]=e^{-\h\om/\kb T}$ compared to the emission rate, which we show below to be crucial for the DNP bistability at zero detuning. The asymmetry between emitting and absorbing energy is \emph{not} unique to phonons and can also appear from other ways of exchanging energy with a bath due to detailed balance.

Therefore, the two main physical ingredients in the HF rates $W_{f,i}$ are: (i) the asymmetry between absorbing and emitting energy and, (ii) including the number of the relevant nuclei spin species needed for the transition. A detailed derivation of the rates used here is given in Ref. \onlinecite{Lopez-Monis-et-al-NJP-2011}. The non-zero HF rates between the triplets are 
\begin{subequations}
\label{eq:TT-HF-rates}
\begin{align} 
W_{T_\PP,T_+}&=
W_{T_-,T_\PP}
\label{eq:WTmTx}\\
&=\frac{1}{2N}
\left[\frac{1-P}{2}\right]
\left[\frac{\sqrt{2}A_+}{\mathcal{N}}\right]^2
\!\mathcal{F}_{ph}(E_{T_+}-E_{T_\PP}),
\nonumber
\\
W_{T_+,T_\PP}&=
W_{T_\PP,T_-}
\label{eq:WTxTm}\\
&=\frac{1}{2N}
\left[\frac{1+P}{2}\right]
\left[\frac{\sqrt{2}A_+}{\mathcal{N}}\right]^2
\!\mathcal{F}_{ph}(E_{T_\PP}-E_{T_+}),
\nonumber
\end{align} 
\end{subequations}
where the phonon part of the rate $\mathcal{F}_{ph}(E_i-E_f)$ only depends on the difference between the initial and final energies.\cite{footnote-dE-nuclear-levels-is-negligible} Note that $E_{T_\PP}-E_{T_-}=E_{T_+}-E_{T_\PP}$ follows from Eq.(\ref{eq:energy-levels-def}). The non-zero singlet-triplet HF rates are
\begin{subequations}
\label{eq:TS-HF-rates}
\begin{align} 
W_{S_+,T_+}&=
W_{T_-,S_-}\\
&=
\frac{1}{2N} 
\left[\frac{1-P}{2}\right]
\left[A_--\frac{\PP A_+}{\mathcal{N}}\right]^2
\!\!\mathcal{F}_{ph}(E_{T_+}-E_{S_+}),
\nonumber
\\
W_{T_+,S_+}&=
W_{S_-,T_-}\\
&=
\frac{1}{2N} 
\left[\frac{1+P}{2}\right]
\left[A_--\frac{\PP A_+}{\mathcal{N}}\right]^2
\!\!\mathcal{F}_{ph}(E_{S_+}-E_{T_+}),
\nonumber
\\
W_{S_-,T_+}&=
W_{T_-,S_+}\\
&=
\frac{1}{2N}
\left[\frac{1-P}{2}\right]
\left[A_-+\frac{\PP A_+}{\mathcal{N}}\right]^2
\!\!\mathcal{F}_{ph}(E_{T_+}-E_{S_-}),
\nonumber\\
W_{T_+,S_-}&=
W_{S_+,T_-}\\
&=
\frac{1}{2N} 
\left[\frac{1+P}{2}\right]
\left[A_-+\frac{\PP A_+}{\mathcal{N}}\right]^2
\!\!\mathcal{F}_{ph}(E_{S_-}-E_{T_+}),
\nonumber
\end{align} 
\end{subequations}
where $E_{T_+}-E_{S_-}=E_{S_+}-E_{T_-}$, so the difference between initial and final energies is the same for e.g. $W_{S_-,T_+}$ and  $W_{T_-,S_+}$. 
In the explicit calculations, we use the function  
\begin{align} 
\mathcal{F}_{ph}(E_i-E_f)=\frac{\gamma_{ph}}{\gamma_{ph}^2+(E_i-E_f)^2}c(E_i-E_f)
\label{eq:ph-factor}
\end{align} 
to account for the phonon emission/absorption. Here $\gamma_{ph}$ is a characteristic phonon energy scale (e.g.\cite{Fujisawa-Science-1998} $\gamma_{ph}\sim\mu$eV) and $c(E_i-E_f)=\theta(E_i-E_f)+\theta(E_f-E_i)e^{-(E_f-E_i)/\kb T}$ is the crucial factor that exponentially suppresses absorbing compared to emitting energy. Here $\theta(E)$ is the unit step function. For simplicity, we disregard many details of the phonon description and only include two important aspects: (i) the asymmetry between absorbing and emitting energy and (ii) the further apart the energy levels are, the less probable a transition is -- included phenomenologically in the Lorentzian.\cite{Inarrea-Platero-MacDonald-PRB-2007,Inarrea-Carlos-MacDonald-Platero-APL-2007,Lopez-Monis-et-al-NJP-2011} Moreover, this form includes the limit of HF spin-flips without energy exchange. To get a more detailed phonon description in the rates, both $H^{\textrm{sf}}_{\textrm{HF}}$ and the electron-phonon interaction could be included as perturbations in a $T$-matrix approach,\cite{Flensberg-BOOK} which gives a description depending on more parameters e.g. the material.\cite{Erlingsson-Nazarov-PRB-2002,Abalmassov-Marquardt-PRB-2004,Prada-Blick-Joynt-PRB-2008}

An important difference between the triplet-triplet and singlet-triplet rates, is that the triplet-triplet rates (\ref{eq:TT-HF-rates}) are $\propto A_+^2$, whereas the singlet-triplet rates (\ref{eq:TS-HF-rates}) have a common prefactor of $A_-^2$ (remembering that $\PP\propto A_-$). Thus, the strength of the two kinds of rates near their respective crossings are very different, and the singlet-triplet rates are sensitive to the difference in the Overhauser field between the dots -- in contrast to the triplet-triplet rates. However, if $A_\tr=A_\tl$, then all three triplets block transport, since $\PP=0$ so $\Gamma_{\sigma,T_\PP}=0$, see Eq.(\ref{eq:tunnel-out-from-Tx}).

Finally, we observe that the HF rates Eqs.(\ref{eq:TT-HF-rates}-\ref{eq:TS-HF-rates}) are invariant under the interchange of
\begin{align} 
\label{eq:sym-for-HF-rates}
\op\leftrightarrow\ned,\
T_+\leftrightarrow T_-, \
S_+\leftrightarrow S_-
\ \textrm{and}\
\textrm{initial}\leftrightarrow\textrm{final}. 
\end{align} 
This is not the same as (\ref{eq:sym-for-tunnel-rates}) for the tunneling rates, since here the final and initial states are also interchanged.

\section{The crossing of the triplets}\label{sec:triplet-cross}

Next, we analyze the DNP and leakage current close to the crossing of the three triplet levels near $B=0$. Since $T_\PP$ Eq.(\ref{eq:T-p-state-def}) is not a pure triplet state, it allows for leakage current. As we shall see below, analytical insights -- such as the \emph{transition temperature} -- can be achieved from the rate equations (\ref{eq:full-rate-eqs}-\ref{eq:full-polarization-eq}) in this case.

\subsection{The implicit equation for the nuclear polarization and the simplified rate equations}\label{subsec:pol-eq-TT}

\begin{figure}
\includegraphics[width=0.43\textwidth,angle=0]{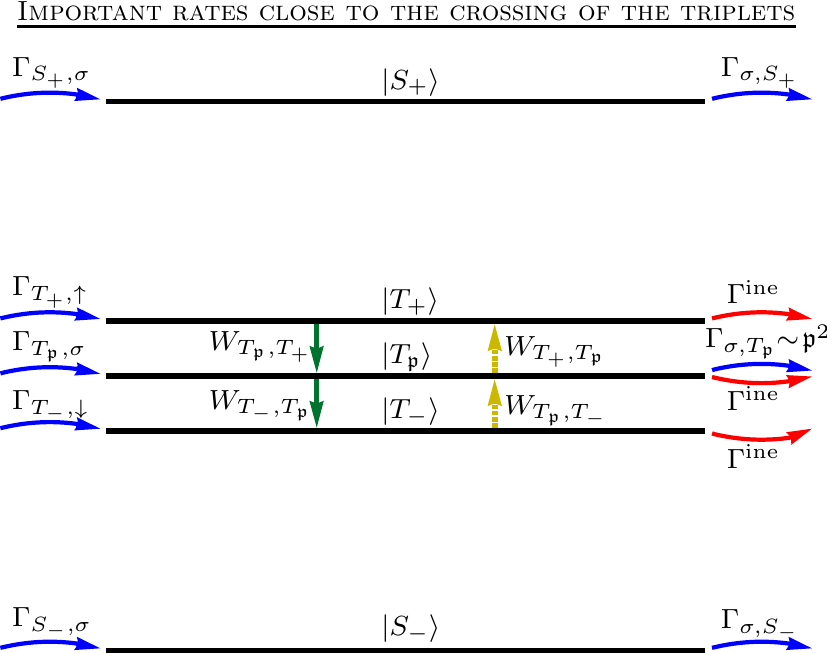}
\caption{(Color online) 
Illustration of the processes in the rate equations (\ref{eq:TT-rate-eqs}) relevant for large energy separation between singlets and triplets and close to the crossing of the three triplets. In this case, only HF rates between the triplets are effective: $W_{T_+,T_\PP}=W_{T_\PP,T_-}$ Eq.(\ref{eq:WTxTm}) (vertical yellow dashed arrows) polarizing the nuclei negatively ($dP<0$) and $W_{T_\PP,T_+}=W_{T_-,T_\PP}$ Eq.(\ref{eq:WTmTx}) (vertical green full arrows) polarizing the nuclei positively ($dP>0$). The tunneling in and out of the DQD $\Gamma_{f,i}$ (blue arrows) and the inelastic escape rates $\Gamma^{\textrm{ine}}_{}$ (red arrows) are the same as in Fig.~\ref{fig:full-rate-eq-illustration}. In the present case, we show that the stationary DNP is determined by the condition that all triplet-triplet rates are equal, see Eq.(\ref{eq:pol-condition-for-rates}).}
\label{fig:TT-rate-eq-illustration}
\end{figure}

Now we derive an implicit equation for the DNP from a simplified system of rate equations -- valid close to the crossing of the triplet levels. We consider the limit of large energy separation between the triplet and singlet levels compared to $\gamma_{ph}$. This can be obtained by large inter-dot tunneling $t\gg\gamma_{ph}$ and sweeping the magnetic field close to zero. In this limit, we can neglect the singlet-triplet HF rates (\ref{eq:TS-HF-rates}) compared to the triplet-triplet HF rates (\ref{eq:TT-HF-rates}), so the system of rate equations (\ref{eq:full-rate-eqs-appendix}) for equal inelastic escape rates $\Gamma_{}^{\textrm{ine}}$ Eq.(\ref{eq:equal-Gamma-ine}) simplifies to
\begin{widetext}
\begin{subequations}
\label{eq:TT-rate-eqs}
\begin{align} 
\dot{n}^{}_{T_+}
&=  
W^{}_{T_+,T_\PP}n^{}_{T_\PP}
+\Gamma^{}_{T_+,\op} n^{}_\op 
-[W^{}_{T_\PP,T_+}+2\Gamma_{}^{\textrm{ine}}]n^{}_{T_+}, 
\label{eq:TT-n-Tp-dot}
\\  
\dot{n}^{}_{T_-}
&=
W^{}_{T_-,T_\PP}n^{}_{T_\PP}
+\Gamma^{}_{T_-,\ned} n^{}_\ned 
-[W^{}_{T_\PP,T_-}+2\Gamma_{}^{\textrm{ine}}]n^{}_{T_-}, 
\label{eq:TT-n-Tm-dot}
\\  
\dot{n}_{T_\PP}
&=  
W_{T_\PP,T_+}n^{}_{T_+}+W_{T_\PP,T_-}n^{}_{T_-} 
+\Gamma_{T_\PP,\op}n_{\op}+\Gamma_{T_\PP,\ned}n^{}_{\ned}
-\big[W_{T_+,T_\PP} + W_{T_-,T_\PP}+\Gamma_{\op,T_\PP} 
+\Gamma_{\ned,T_\PP} 
+2\Gamma_{}^{\textrm{ine}}\big] n^{}_{T_\PP},
\label{eq:TT-n-Tx-dot}
\\  
\dot{n}^{}_{S_+}&=
\Gamma^{}_{S_+,\op}n^{}_{\op}+\Gamma^{}_{S_+,\ned}n^{}_{\ned}
-\big[\Gamma^{}_{\op,S_+} + \Gamma^{}_{\ned,S_+}\big]n^{}_{S_+},
\label{eq:TT-n-Sp-dot}
\\  
\dot{n}^{}_{S_-}&=
\Gamma^{}_{S_-,\op}n^{}_{\op}+\Gamma^{}_{S_-,\ned}n^{}_{\ned}
-\big[\Gamma^{}_{\op,S_-}+\Gamma^{}_{\ned,S_-}\big]n^{}_{S_-},
\label{eq:TT-n-Sm-dot}
\\
\dot{n}^{}_\op&=
\Gamma^{}_{\op,S_+}n^{}_{S_+}+\Gamma^{}_{\op,S_-}n^{}_{S_-} 
+(\Gamma^{}_{\op,T_\PP}+\Gamma_{}^{\textrm{ine}})n^{}_{T_\PP} 
+\Gamma_{}^{\textrm{ine}}(n^{}_{T_+}+n^{}_{T_-}) 
-\big[\Gamma^{}_{S_+,\op}+\Gamma^{}_{S_-,\op}+\Gamma^{}_{T_\PP,\op}+\Gamma^{}_{T_+,\op}\big]n^{}_\op,
\label{eq:TT-n-op-dot}
\\
\dot{n}^{}_\ned&=
\Gamma^{}_{\ned,S_+}n^{}_{S_+}+\Gamma^{}_{\ned,S_-}n^{}_{S_-} 
+(\Gamma^{}_{\ned,T_\PP}+ \Gamma_{}^{\textrm{ine}})n^{}_{T_\PP} 
+\Gamma_{}^{\textrm{ine}}(n^{}_{T_-} +n^{}_{T_+})
-\big[\Gamma^{}_{S_+,\ned}+\Gamma^{}_{S_-,\ned}
+\Gamma^{}_{T_\PP,\ned}+\Gamma^{}_{T_-,\ned}\big]n^{}_\ned,
\label{eq:TT-n-ned-dot} 
\end{align}
\end{subequations}
\end{widetext}
as illustrated in Fig.~\ref{fig:TT-rate-eq-illustration}. Similarly, Eq.(\ref{eq:full-polarization-eq}) simplifies to
\begin{align} 
\dot{P}=  
\frac{2}{N}\Big[
&
(W^{}_{T_-,T_\PP}{-}W^{}_{T_+,T_\PP})
n^{}_{T_\PP}
\nonumber \\ 
&
+W^{}_{T_\PP,T_+}n^{}_{T_+}
-W^{}_{T_\PP,T_-}n^{}_{T_-}
\Big].
\label{eq:TT-polarization-eq}
\end{align}
In this approximation, the rate equations for the triplet and singlet occupations \emph{only} couple indirectly through the one-electron  occupations. Moreover, we avoid very low temperatures, where the approximation could fail.\cite{footnote-low-T-fail}

To find the stationary DNP, we solve the system of rate equations (\ref{eq:TT-rate-eqs}-\ref{eq:TT-polarization-eq}), i.e. $\dot{P}=0$ and $\dot{n}_{\nu}=0$ for all $\nu$. To this end, we subtract $\dot{n}^{}_\ned$ Eq.(\ref{eq:TT-n-ned-dot}) from $\dot{n}^{}_\op$ Eq.(\ref{eq:TT-n-op-dot}) and use the symmetries under interchange of indices for the tunneling rates (see Eqs.(\ref{eq:tunnel-in-rates}-\ref{eq:sym-for-tunnel-rates})), i.e. 
$\dot{n}^{}_\op-\dot{n}^{}_\ned=(\Gamma_{T_+,\op}+\Gamma_{S_+,\ned}+\Gamma_{S+,\op}+\Gamma_{T_\PP,\sigma})(n_\ned-n_\op)=0$, so
\begin{align} 
n_\ned=n_\op
\label{eq:nned-lig-nop}
\end{align} 
in the stationary limit. Using the symmetries for the tunneling rates and the HF rates, the relation (\ref{eq:nned-lig-nop}) leads to $\dot{n}^{}_{T_+}-\dot{n}^{}_{T_-}+\frac{N}{2}\dot{P}=2 \Gamma^{\textrm{ine}} (n_{T_-}-n_{T_+})=0$, i.e.
\begin{align} 
n_{T_+}^{}=n_{T_-}^{}
\label{eq:nTm-lig-nTp}
\end{align} 
in the stationary limit. Hence, Eq.(\ref{eq:TT-polarization-eq}) becomes $\dot{P}=\frac{2}{N}\big(W^{}_{T_\PP,T_+}-W^{}_{T_+,T_\PP}\big)\big(n^{}_{T_\PP}+n^{}_{T_+}\big)=0$ using that $W_{T_\PP,T_-}=W_{T_+,T_\PP}$ and $W_{T_-,T_\PP}=W_{T_\PP,T_+}$, see Eq.(\ref{eq:TT-HF-rates}). Thus, since the occupations are positive, we arrive at  
\begin{align} 
W^{}_{T_\PP,T_+}=W^{}_{T_+,T_\PP},
\label{eq:pol-condition-for-rates}
\end{align}
which is \emph{the implicit equation for the steady state DNP} $P$ -- remembering that both rates depend on $P$ explicitly and through the Overhauser split energy levels $E_{T_{\pm}}$, see Eq.(\ref{eq:TT-HF-rates}). Equivalently, this can be written as  $W^{}_{T_\PP,T_-}=W^{}_{T_-,T_\PP}$ using Eq.(\ref{eq:TT-HF-rates}). Physically, the relation (\ref{eq:pol-condition-for-rates}) means that the DNP stabilizes at a value such that the phonon emission and absorption transitions between the two levels $T_\PP$ and $T_+$ ($T_-$) become equally probable. Inserting Eq.(\ref{eq:TT-HF-rates}), the relation (\ref{eq:pol-condition-for-rates}) can be rewritten as
\begin{align} 
\frac{1+P}{1-P}=
\frac{\mathcal{F}_{ph}(E_{T_+}-E_{T_\PP})}{\mathcal{F}_{ph}(E_{T_\PP}-E_{T_+})}.
\label{eq:pol-condition-for-rates-mellem}
\end{align}
This shows that the DNP is insensitive to the part of $\mathcal{F}_{ph}$, which is even in energy, and \emph{only} depends on the difference between emitting and absorbing energy. In other words, the even-energy part of the function $\mathcal{F}_{ph}$ cancels out on the right-hand side of Eq.(\ref{eq:pol-condition-for-rates-mellem}) and we are left with the ratio between absorbing and emitting energy in the transition. Hence, the DNP is largely independent of the way the phonons are modelled in the HF rates, as long as the important asymmetry between emitting and absorbing energy is included.\cite{footnote-general-TcTT} We observe a crucial difference to previous works,\cite{Rudner-Levitov-PRL-2007,Rudner-Levitov-Nanotechnology-2010,Rudner-Rashba-PRB-2011,Rudner-Koppens-Folk-Vandersypen-Levitov-PRB-2011} where non-zero DNP is induced only at finite detuning, since emitting and absorbing energy is modelled as being equally likely in these works, see Sec.~\ref{subsec:Main-ideas-of-this-work}. By inserting the energies Eq.(\ref{eq:energy-levels-def}) into Eq.(\ref{eq:pol-condition-for-rates-mellem}), we end up with the following implicit equation for $P$, 
\begin{align} 
P=
\tanh\left(\frac{g\mu_BB+\frac{1}{2}A_+P}{2\kb T}\right).
\label{eq:pol-condition-tanh}
\end{align}
From this equation, it is not possible to obtain a closed analytical expression for $P$. Remarkably, it shows that $P$ near the crossing of the triplet levels \emph{only} depends on the Zeeman splitting $g\mu_BB/A_+$ and the temperature $\kb T/A_+$ --- both measured in units of $A_+=(A_\tl+A_\tr)/2$.

\begin{figure}
\includegraphics[width=0.39\textwidth,angle=0]{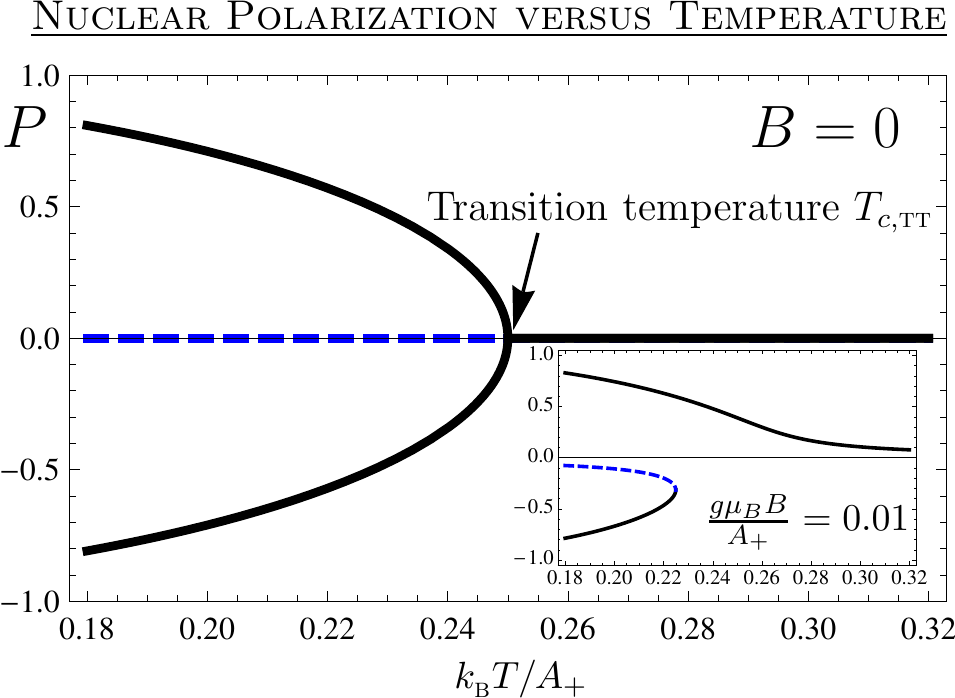}
\caption{(Color online) The DNP $P$ versus temperature in units of the average hyperfine coupling, $\kb T/A_+$, at zero (main figure) and finite magnetic field $g\mu_BB/A_+=0.01$ (inset). Both figures are obtained by solving numerically the implicit equation (\ref{eq:pol-condition-tanh}). For $B=0$ and $T<\Tct$, the DNP has two stable solutions (full black lines) and one unstable solution (dashed blue line). Above $\Tct$, only one stable DNP exist. The inset shows similar behavior for $B\neq0$, but here the multiple DNP solutions appear at a lower temperature.}
\label{fig:P-vs-T-triplet-crossing}
\end{figure}

The form of the implicit equation (\ref{eq:pol-condition-tanh}) resembles the one found by using mean-field theory to a Heisenberg spin model, which describes an equilibrium ferromagnetic phase transition driven by temperature.\cite{Flensberg-BOOK} In contrast, here the polarization is dynamically induced by the leakage current, i.e.~a \emph{non-equilibrium} situation.

The DNP versus $T$ and $B$ is easily found numerically from Eq.(\ref{eq:pol-condition-tanh}), see Fig.~\ref{fig:P-vs-T-triplet-crossing}.  For $B=0$, the DNP has a bifurcation at a certain transition temperature $\Tct$, where the system goes from one stable DNP for $T>\Tct$ to two stable DNPs and an unstable one for $T<\Tct$. The stability of the DNP is found by numerical iteration of the set of differential equations.\cite{Strogatz-BOOK} The \emph{transition temperature} for $B=0$ is readily found from Eq.(\ref{eq:pol-condition-tanh}) to be\cite{footnote-way-to-get-TcTT}
\begin{align} 
\kb \Tct =\frac{1}{4}A_+.
\label{eq:trans-temp-TT}
\end{align}
This is a remarkably simple and insightful result. It shows that the transition temperature $\Tct$ is given \emph{only} by the average HF constants $A_+=(A_\tl+A_\tr)/2$. For $T<\Tct$, the DNP can have two stable values and therefore so can the current. Hence, the current shows \emph{hysteresis} for $T<\Tct$, which disappears for $T\geq\Tct$. 

\begin{figure}
\includegraphics[width=0.39\textwidth,angle=0]{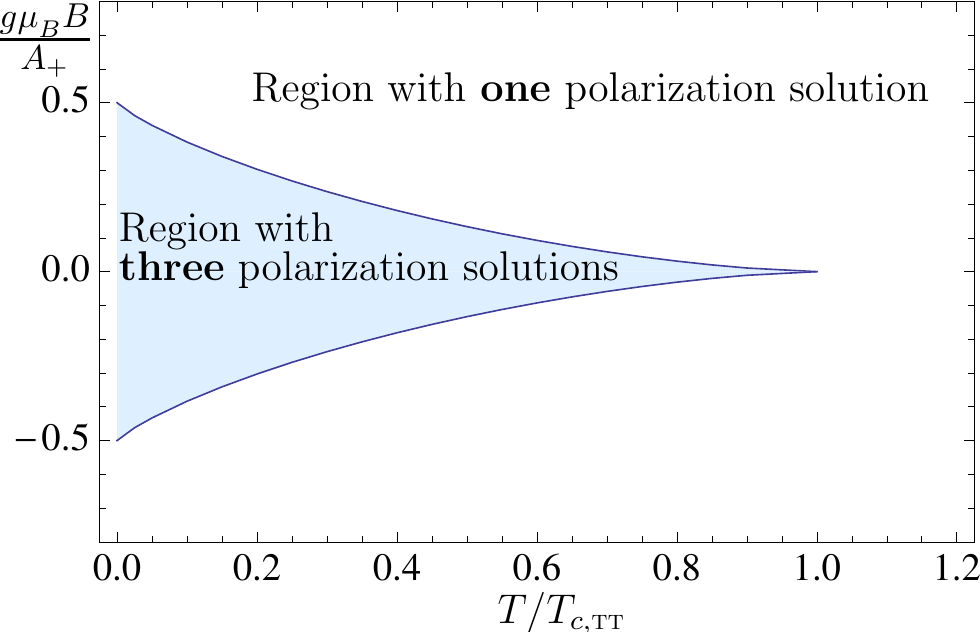}
\caption{(Color online) The regions in parameter space close to the crossing of the triplets with one (white region) or three (blue region) DNP solutions, respectively. These regions are found numerically from Eq.(\ref{eq:pol-condition-tanh}). In the blue region, only two out of the three DNP solutions are stable against small fluctuations. From Eq.(\ref{eq:pol-condition-tanh}), it is evident that the DNP \emph{only} depend on $g\mu_BB/A_+$ (vertical axis) and $T/\Tct$ (horizontal axis), where $\Tct\!=\!A_+/(4\kb)$. The DNP versus $B$-field for fixed $T$ in Fig.~\ref{fig:cross-triplets}(a,d,g),  corresponds to vertical sweeps in this figure.}
\label{fig:phase-diagram-triplet-crossing}
\end{figure}

For $B\neq0$, the DNP also has multiple solutions below a certain temperature, which is generally lower than the transition temperature $\Tct$ Eq.(\ref{eq:trans-temp-TT}) for $B=0$, see the inset of Fig.~\ref{fig:P-vs-T-triplet-crossing}. Figure \ref{fig:phase-diagram-triplet-crossing} shows the number of DNP solutions for a specific value of $g\mu_BB$ and $\kb T$. From Fig.~\ref{fig:phase-diagram-triplet-crossing}, it is evident that by sweeping the external magnetic field, the region of multiple solutions of DNP -- and therefore also current hysteresis -- becomes broader the lower the temperature. Especially, multiple solutions appear only for $T<\Tct$, which underlines the importance of $\Tct$ in connection to the current hysteresis.  The leakage current and its hysteresis is treated in greater detail in Sec.~\ref{subsec:current-TT-crossing}.

For $B=0$ and $T$ close to $\Tct$, we can expand the right-hand side of Eq.(\ref{eq:pol-condition-tanh}) in $|P|\ll1$, which gives that the DNP vanishes as  $P\propto\pm[(\Tct-T)/\Tct]^{1/2}$ for $0\leq(\Tct-T)/\Tct\ll 1$. This is typical behavior for mean-field theories\cite{Flensberg-BOOK} as the one used here. 

Above equal inelastic escape rates are used. If we instead use non-equal inelastic rates following the symmetries (\ref{eq:sym-for-tunnel-rates}) under interchange of indices, then the implicit equation (\ref{eq:pol-condition-for-rates}) for the DNP and the transition temperature (\ref{eq:trans-temp-TT}) remain \emph{unchanged}, see Appendix \ref{appendix:non-equal-inelastic-rates}. However, the polarization condition (\ref{eq:pol-condition-for-rates}) and/or transition temperature might change, if the inelastic escape rates follow e.g. other symmetries under index exchange.

\subsection{The leakage current and the occupations close to the crossing of the triplets}\label{subsec:current-TT-crossing}

\subsubsection{Analytical stationary occupation and current expressions in terms of the rates}

Next, we find the leakage current close to the crossing of the triplets using the simplified rate equations (\ref{eq:TT-rate-eqs}-\ref{eq:TT-polarization-eq}). First, we find the stationary occupations using $n_\ned=n_\op$ Eq.(\ref{eq:nned-lig-nop}), $n_{T_+}=n_{T_-}$ Eq.(\ref{eq:nTm-lig-nTp}) and the implicit equation (\ref{eq:pol-condition-for-rates}) for $P$. Now, subtracting $\dot{n}_{S_+}$(\ref{eq:TT-n-Sp-dot}) and $\dot{n}_{S_-}$ (\ref{eq:TT-n-Sm-dot}) using $n_\op=n_\ned$ and the index invariances of $\Gamma_{f,i}$ Eqs.(\ref{eq:tunnel-in-rates}-\ref{eq:sym-for-tunnel-rates}), we find $\dot{n}_{S_+}-\dot{n}_{S_-}=2 \Gamma_{\op,S_+} (n_{S_-} - n_{S_+})=0$, so 
\begin{align}
n_{S_-} = n_{S_+}
\label{eq:nSp-lig-nSm}
\end{align}
in the steady state. This simplifies the sum of $\dot{n}_{S_+}$ (\ref{eq:TT-n-Sp-dot}) and $\dot{n}_{S_-}$ (\ref{eq:TT-n-Sm-dot}), $\dot{n}_{S_+}+\dot{n}_{S_-}=0$, and leads to
\begin{align}
n_{S_+}=\frac{\Gamma_{S_+,\ned}+\Gamma_{S_+,\op}}{2\Gamma_{\op,S_+}} 
\ n_{\op}
\label{eq:nSp-propto-np}
\end{align}
by again using $n_\op\!\!=\!n_\ned$ and the index exchange symmetries of $\Gamma_{f,i}$ Eqs.(\ref{eq:tunnel-in-rates}-\ref{eq:sym-for-tunnel-rates}). Inserting these occupation relations into the normalization condition $\sum_{\nu}n_{\nu}{=}1$, we get 
\begin{align}
n_{\op}=\frac{\Gamma_{\op,S_+}}{\Upsilon}[1-n_{T_\PP}-2n_{T_+}],
\label{eq:np-via-norm-condition}
\end{align}
where $\Upsilon\equiv 2 \Gamma_{\op,S_+}+\Gamma_{S_+,\ned}+\Gamma_{S_+,\op}$. By inserting Eqs.(\ref{eq:nned-lig-nop}), (\ref{eq:nTm-lig-nTp}), (\ref{eq:pol-condition-for-rates}), (\ref{eq:nSp-lig-nSm}), (\ref{eq:nSp-propto-np}) and (\ref{eq:np-via-norm-condition}) into $\dot{n}_{T_\PP}=0$ (\ref{eq:TT-n-Tx-dot}) and $\dot{n}_{T_+}+\dot{n}_{T_-}=0$ (\ref{eq:TT-n-Tp-dot}-\ref{eq:TT-n-Tm-dot}), we obtain two coupled equations for the occupations $n_{T_\PP}$ and $n_{T_+}$ with the solution
\begin{subequations}
\label{eq:triplet-occupations}
\begin{align}
n_{T_\PP}=&
\frac{\Gamma_{\op,S_+} \!\big[2\Gamma_{}^{\textrm{ine}} \Gamma_{T_\PP,\op} 
+W_{T_+,T_\PP}
(\Gamma_{T_+,\op} {+} \Gamma_{T_\PP,\op})\big]}{\Lambda},
\label{eq:Tx-occupation}
\\
n_{T_+}=&
\frac{\Gamma_{\op,S_+}}{\Lambda}
\!\Big[\!
\Gamma_{T_+,\op}  (\Gamma_{}^{\textrm{ine}} {+} \Gamma_{\op,T_\PP})
+W_{T_+,T_\PP}(\Gamma_{T_+,\op} {+} \Gamma_{T_\PP,\op})\!\Big]\!,
\label{eq:Tp-occupation}
\end{align}
where
\begin{align}
\!\Lambda\equiv&
\Gamma_{\op,S_+} 
\!
\Big[2\Gamma_{T_+,\op}\Gamma_{\op,T_\PP} 
+
(2\Gamma_{}^{\textrm{ine}} {+} 3W_{T_+,T_\PP}) 
(\Gamma_{T_+,\op} {+} \Gamma_{T_\PP,\op}) \Big]
\nonumber\\
+&
\!\Upsilon\Big[2\Gamma_{}^{\textrm{ine}} (\Gamma_{}^{\textrm{ine}} {+} \Gamma_{\op,T_\PP}) 
+(3\Gamma_{}^{\textrm{ine}} {+} \Gamma_{\op,T_\PP}) W_{T_+,T_\PP}\!\Big].
\end{align}
\end{subequations}
The explicit expressions for $n_{\op}=n_{\ned}$ can easily be found by inserting Eq.(\ref{eq:triplet-occupations}) into Eq.(\ref{eq:np-via-norm-condition}). This in turn leads to the expression for $n_{S_+}=n_{S_-}$ via Eq.(\ref{eq:nSp-propto-np}). Thus, we now have \emph{all the stationary occupations} close to the crossing of the triplets in terms of the rates. 
 
The leakage current in the high bias limit is now obtained by inserting the occupations into Eq.(\ref{eq:current-def}), i.e.
\begin{subequations}
\label{eq:current-general-for-TTT-crossing}
\begin{align}
I=&
\frac{2\Gamma_{\op,S_+}}{\zeta}  
(\Gamma_{T_+,\op} + \Gamma_{S_+,\ned} + \Gamma_{S_+,\op} + \Gamma_{T_\PP,\op})
\\ 
&\times
\Big[2 \Gamma_{}^{\textrm{ine}} (\Gamma_{}^{\textrm{ine}} + \Gamma_{\op,T_\PP}) 
+ W_{T_+,T_\PP}
(3 \Gamma_{}^{\textrm{ine}} + \Gamma_{\op,T_\PP})\Big]
\nonumber
\end{align}
where we introduced
\begin{align}
\zeta&\equiv
(\Gamma_{}^{\textrm{ine}} + \Gamma_{\op,T_\PP}) 
(2\Gamma_{T_+,\op} \Gamma_{\op,S_+} +2 \Gamma_{}^{\textrm{ine}} \Upsilon)
\\
&+2 \Gamma_{}^{\textrm{ine}} \Gamma_{\op,S_+} \Gamma_{T_\PP,\op}
\nonumber\\
&+
\!\!W_{T_+,T_\PP}
\!\Big[3 \Gamma_{T_+,\op} \Gamma_{\op,S_+} 
{+} 3 \Gamma_{\op,S_+} \Gamma_{T_\PP,\op}
{+} \Upsilon(3 \Gamma_{}^{\textrm{ine}} {+} \Gamma_{\op,T_\PP}) 
\Big].
\nonumber
\end{align}
\end{subequations}
We emphasize that in the derivation of the occupations and the leakage current, we have only used the invariance of the rates under exchange of indices (see Secs.~\ref{subsubsec:tunnel-rates} and \ref{subsubsec:HF-rates}) and \emph{not} the explicit expressions for the rates. Thus, the above expressions are indeed rather general. 

Furthermore, equal inelastic escape rates $\Gamma^{\textrm{ine}}$ from the three triplets were used here. If we instead only assume that the inelastic rates follow the same symmetries under index exchange as $\Gamma_{f,i}$ in Eq.(\ref{eq:sym-for-tunnel-rates}), then the current expressions above only change slightly, see Appendix \ref{appendix:non-equal-inelastic-rates}.
 
Next, we focus on the case of the explicit tunneling rates in Eqs.(\ref{eq:tunnel-in-rates}-\ref{eq:tunnel-out-rates}), so the leakage current (\ref{eq:current-general-for-TTT-crossing}) becomes
\begin{subequations}
\label{eq:current-inserting-rates-for-TTT-crossing}
\begin{align}
\frac{I}{\Gamma_\tl}=
\frac{8\gamma_{\tr\tl}}{\xi} 
&\Big\{
\gamma_{\textrm{ine}} (2 \gamma_{\textrm{ine}} + 3 w) 
\\
&+\big[
2 \gamma_{\textrm{ine}} (\gamma_{\textrm{ine}} + \gamma_{\tr\tl}) 
+(3 \gamma_{\textrm{ine}} + \gamma_{\tr\tl})w
\big] \PP^2
\Big\}
\nonumber
\end{align}
which is overall proportional to $\Gamma_\tl$. Here we introduced
\begin{align}
\label{eq:def-w-gRL-gine}
\gamma_{\tr\tl}^{}&\equiv\frac{\Gamma_\tr}{\Gamma_\tl},
\quad
\gamma_{\textrm{ine}}^{}\equiv\frac{\Gamma^{\textrm{ine}}}{\Gamma_\tl},
\quad
w\equiv\frac{W^{}_{T_+,T_\PP}}{\Gamma_\tl},
\end{align}
and
\begin{align}
\xi\equiv&
(2 \gamma_{\textrm{ine}} + 3 w)
\big[
3\gamma_{\tr\tl} + 
2 \gamma_{\textrm{ine}} (1 + 2 \gamma_{\tr\tl})
\big]
\nonumber
\\
&+2 \Big[
4 \gamma_{\textrm{ine}} \gamma_{\tr\tl} (1 + \gamma_{\tr\tl}) 
+\gamma_{\textrm{ine}}^2 (6 + 4 \gamma_{\tr\tl})
\nonumber\\ 
&+3 \gamma_{\textrm{ine}} (3 + 2 \gamma_{\tr\tl}) w
+2 \gamma_{\tr\tl} (\gamma_{\tr\tl}w + \gamma_{\tr\tl} + 2 w)
\Big] \PP^2
\nonumber\\
&+4\Big[
2\gamma_{\textrm{ine}}(\gamma_{\textrm{ine}} + \gamma_{\tr\tl}) 
+ (3 \gamma_{\textrm{ine}} + \gamma_{\tr\tl})w
\Big] \PP^4.
\end{align}
\end{subequations}
Thus, the current is expressed in terms of the dimensionless triplet-singlet mixing parameter $\PP$ in Eq.(\ref{eq:def-p-N-Apm}) and the three rates, $\gamma_{\tr\tl}$, $\gamma_{\textrm{ine}}$ and $w$ -- all measured in units of the basic tunneling rate $\Gamma_\tl$. The asymmetry between the coupling of the DQD to the left and right lead is described by $\gamma_{\tr\tl}$, where $\gamma_{\tr\tl}=1$ for the symmetric case. Hence, $\gamma_{\tr\tl}$ is on the order of unity. In contrast, both the dimensionless HF triplet-triplet rate\cite{footnote-w-universal} $w$ and the dimensionless inelastic escape rate $\gamma_{\textrm{ine}}$ are much smaller than unity: $\gamma_{\textrm{ine}},\  w\ll\gamma_{\tr\tl}\sim 1$. Moreover, here we focus on the limit of the inelastic escape rate being much smaller than the HF rate close to the crossing of the triplets.

We note that \emph{without} singlet-triplet mixing, $\PP=0$ (i.e.~$P=0$ or $A_{\tl}=A_{\tr}$), the current (\ref{eq:current-inserting-rates-for-TTT-crossing}) reduces to 
\begin{align}
\frac{I(\PP=0)}{\Gamma_\tl}=
\frac{8\gamma_{\tr\tl}\gamma_{\textrm{ine}}}{3\gamma_{\tr\tl}+2 \gamma_{\textrm{ine}} (1 + 2 \gamma_{\tr\tl})},
\label{eq:current-for-zero-TS-mixing-parameter} 
\end{align}
which is independent of the HF rate $w$. Physically, this can be understood in the following way. For $\PP=0$, the escape channel from $T_\PP$ due to singlet-triplet mixing disappears. Thus, for $\PP=0$ only the inelastic escape channel $\Gamma^{\textrm{ine}}$ contributes to the current through the three triplet states (remembering that the $S_\pm$ singlet levels are far away in energy). Since we use equal inelastic escape rates $\Gamma^{\textrm{ine}}$ from the three triplet states here, then the current does not depend on from which triplet state the electron tunnels out. Thus, the current has to be independent of the HF transitions between the three triplets for $\PP=0$ as found in Eq.(\ref{eq:current-for-zero-TS-mixing-parameter}). In contrast, if the inelastic escape rates from the three triplets are \emph{not} equal, then the current can indeed depend on the HF rate even for $\PP=0$. An example of this, is given in Eq.(\ref{eq:current-non-equal-ine-rates-without-ST-mixing}) in Appendix \ref{appendix:non-equal-inelastic-rates}.

We stress that the rates still depend on the DNP, which in general is \emph{not} analytically known in terms of the external parameters. Thus, the occupations (\ref{eq:triplet-occupations}) and current expressions (\ref{eq:current-general-for-TTT-crossing}-\ref{eq:current-inserting-rates-for-TTT-crossing}) are also \emph{not} explicit functions of the external parameters. To obtain explicit expression versus external parameters, the DNP needs to be found from the implicit DNP equation (\ref{eq:pol-condition-tanh}). This will be done below.

\begin{figure*}
\includegraphics[width=0.93\textwidth,angle=0]{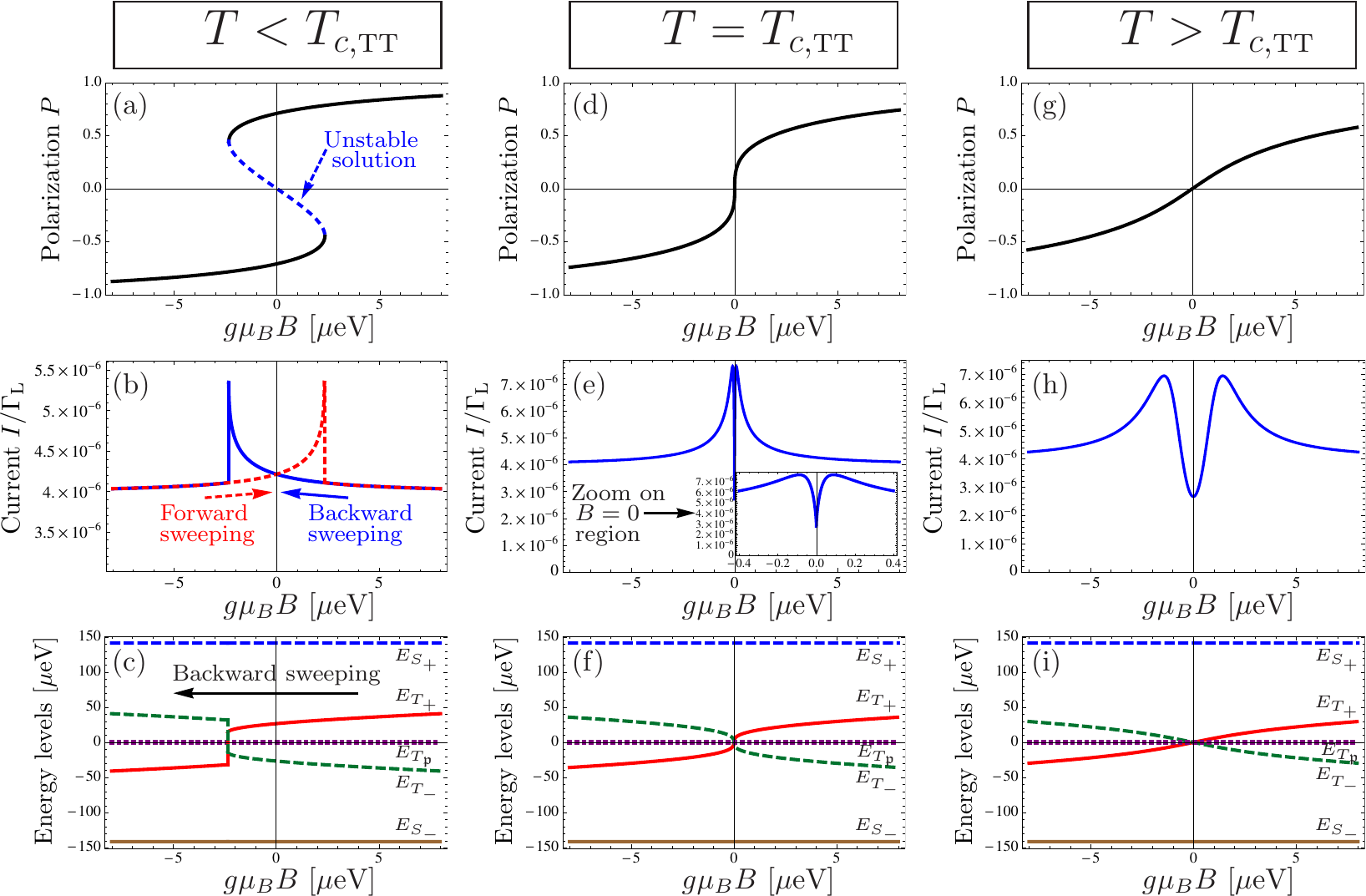}
\caption{(Color online) The DNP $P$, the leakage current $I/\Gamma_\tl$ (\ref{eq:current-inserting-rates-for-TTT-crossing}) and the energy levels versus external magnetic field $g\mu_BB$ (in energy units) close to the crossing of the triplet states $T_+$, $T_-$ and $T_\PP$ for temperatures $T=0.8\Tct<\Tct$ (a-c), $T=\Tct$ (d-f) and $T=1.2\Tct>\Tct$ (g-i). It is clearly seen that a hysteretic current appears only below the transition temperature $T<\Tct$. The reason is that for $T<\Tct$, a magnetic field region with two stable DNPs (black full lines) exists as seen in Fig.~(a). In this region, an unstable solution to the DNP (blue dashed line) also shows up. In contrast, for $T\geq \Tct$ only a single stable DNP is found in Figs.~(d) and (g) [see also Fig.~\ref{fig:phase-diagram-triplet-crossing}].  For clarity, only the backward sweeping of the magnetic field (from a $g\mu_BB$ higher than $2.34\mu$eV) is seen for the energy levels on Fig.~(c). Furthermore, note that the vertical scale in (b) does not include zero, in contrast to (e) and (h). The  inset in Fig.~(e) for $T=\Tct$ shows the sharp current dip at very low magnetic fields. The parameters used here are: $A_{\tl}=80\mu$eV, $A_{\tr}=70\mu$eV, $t=100\mu$eV, $\gamma_{ph}=1\mu$eV, $\Gamma_\tl=\Gamma_\tr$ (i.e.~$\gamma_{\tr\tl}=1$) and\cite{footnote-number-discussion} $\h\Gamma_\tl N=10^7\mu$eV. Moreover, we choose the dimensionless inelastic escape rate to be $\gamma_{\textrm{ine}}=\Gamma^{\textrm{ine}}/\Gamma_\tl=10^{-6}$, such that it is much smaller than the HF rate $w$ Eq.(\ref{eq:def-w-gRL-gine}) in the regions close to the crossing of the triplets (see Fig.~\ref{fig:HF-rates-close-to-TTT-crossing}).}
\label{fig:cross-triplets}
\end{figure*}

\subsubsection{The leakage current versus magnetic field:\\ Emergence of hysteresis below the transition temperature}

Next, we analyze the leakage current versus $B$-field as shown in Fig.~\ref{fig:cross-triplets} for $T<\Tct$, $T=\Tct$ and $T>\Tct$. To this end, the DNP $P$ is found numerically from the implicit equation (\ref{eq:pol-condition-tanh}) [Fig.~\ref{fig:cross-triplets}(a,d,g)] and then inserted into the current (\ref{eq:current-inserting-rates-for-TTT-crossing}) [Fig.~\ref{fig:cross-triplets}(b,e,h)]. Hence, if multiple stable DNP solutions exist, then there will also be multiple possible stable values of the current. The actually stationary leakage current and DNP in a concrete situation therefore depend on the initial value in time of the DNP as in other non-linear dynamical systems.\cite{Strogatz-BOOK} 

The hysteresis in the current comes about in the following way: Consider $T<\Tct$ and the magnetic field tuned so high that there is only a single DNP solution, e.g. $g\mu_BB=7\mu$eV in Fig.~\ref{fig:cross-triplets}(a). By decreasing $g\mu_BB$ one will enter the region of multiple possible DNPs [at $g\mu_BB\sim2.34\mu$eV in Fig.~\ref{fig:cross-triplets}(a)]. Since the DNP is a stable solution against small fluctuations, the system will remain on the upper stable branch ($P>0$) until the critical $B$-field, where the upper branch cease to exist [about $g\mu_BB\sim-2.34\mu$eV in Fig.~\ref{fig:cross-triplets}(a)]. At this critical field, the system has to go to the lower stable DNP branch (with $P<0$). Thus, the DNP change discontinuously versus $B$. This in turn leads to a \emph{jump} in the current (as seen in the blue full curve on Fig.~\ref{fig:cross-triplets}(b) for sweeping the field backwards from a high value). For $B$-fields lower than the critical one, the DNP is single valued again and so is the current. Now, if at this point the field is increased beyond the critical field [of $g\mu_BB\sim-2.34\mu$eV in Fig.~\ref{fig:cross-triplets}(a)] one will follow the \emph{lower} stable DNP branch with $P<0$, leading to the dashed red curve in Fig.~\ref{fig:cross-triplets}(b). This sweep direction also leads to a sharp jump once the lower stable DNP branch cease to exist [at $g\mu_BB\sim+2.34\mu$eV]. Thus, the hysteretic behavior of the leakage current for $T<\Tct$ is now evident. The discontinuity versus $g\mu_BB$ for $T<\Tct$ is also seen in the energy levels Fig.~\ref{fig:cross-triplets}(c), where only one sweeping direction is shown for clarity.

We observe that the width of the hysteresis loop increases with decreasing $T$, since this width is given by the vertical distance between the two (full) lines in Fig.~\ref{fig:phase-diagram-triplet-crossing}.

Identifying that the transition temperature $\Tct$ simply is given by the average HF constants  (\ref{eq:trans-temp-TT}), is an important result of this paper. Experimentally, the HF constants are of order\cite{Pfund-Shorubalko-Ensslin-Leturcq-PRL-2007,Churchill-Marcos-Marcus-et-al-nature-phys-2009} 100$\mu$eV, so $\Tct$ is on the order of 0.3K, which is within range of modern experiments.       

To test the results of the simplified model without HF triplet-singlet rates presented in Fig.~\ref{fig:cross-triplets},  we have numerically iterated the full set of rate equations (\ref{eq:full-rate-eqs-appendix},\ref{eq:full-polarization-eq}) including \emph{all} rates. For the parameters of Fig.~\ref{fig:cross-triplets} -- where the $E_{S_\pm}$ levels are far way from the triplet levels -- the two calculations give the same results (not shown in the figure), i.e.~the presented simplified model works well.

Now we give a better understanding of the form of the current versus $B$-field. In this work, we focus on the limit where the HF rates dominate the inelastic escape rate close to the level crossings as in Refs.[\onlinecite{Rudner-Levitov-PRL-2007,Rudner-Levitov-Nanotechnology-2010,Rudner-Koppens-Folk-Vandersypen-Levitov-PRB-2011,Rudner-Rashba-PRB-2011}]. Nevertheless, the inelastic escape rate plays an important role for the current in the following. First, we analyze in detail current versus $g\mu_BB$ for $T\geq \Tct$, where the DNP is single valued and therefore no current hysteresis is found [Fig.~\ref{fig:cross-triplets}(e,h)]. The DNP goes continuously through $P=0$ at $g\mu_BB=0$. As discussed above (see Eq.(\ref{eq:current-for-zero-TS-mixing-parameter})), the singlet-triplet mixing disappears at $P=0$, which in turn closes the escape path from $T_\PP$ as $\Gamma_{\sigma,T_\PP}\sim P^2$ for $|P|\ll1$, see Eq.(\ref{eq:tunnel-out-from-Tx}). Thus, the current decreases for $B\rightarrow0$ and $T\geq \Tct$ to a value only given by the inelastic escape rate -- even though it is weak. For $\gamma_{\tr\tl}=1$ Eq.(\ref{eq:current-for-zero-TS-mixing-parameter}) gives
\begin{align}
\frac{I(B=0,T\geq \Tct)}{\Gamma_\tl}=
\frac{8\gamma_{\textrm{ine}}}{3+6\gamma_{\textrm{ine}}},
\end{align}
which agrees perfectly with the value of $\sim2.7\times10^{-6}$ found in Fig.~\ref{fig:cross-triplets}(e,h) for $B=0$. The slope of $P$ at $g\mu_BB=0$ increases rapidly when approaching $T=\Tct$ from temperatures above $\Tct$, see Fig.~\ref{fig:cross-triplets}(d,g). Hence, the dip in the current at $g\mu_BB=0$ becomes increasingly sharper for $T$ approaching $\Tct$, see Fig.~\ref{fig:cross-triplets}(e,h) and the inset. 

\begin{figure}
\includegraphics[width=0.38\textwidth,angle=0]{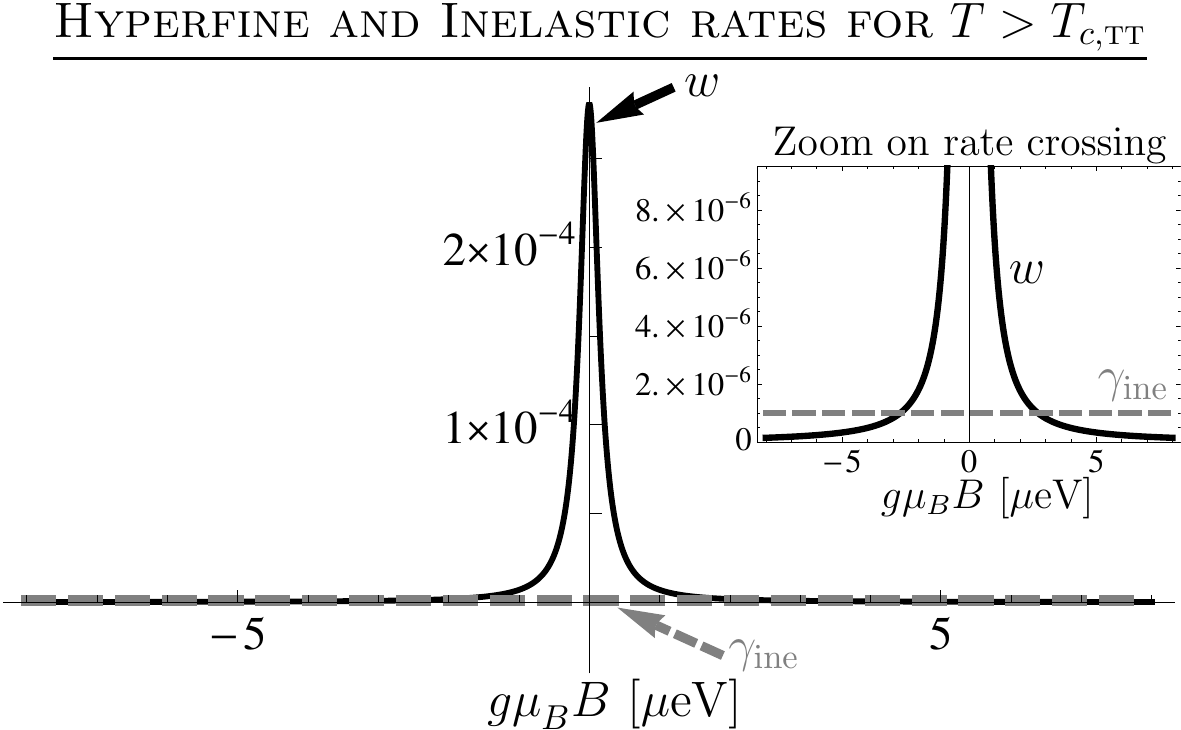}
\caption{(Color online) The dimensionless HF rate $w$ (black full line) and inelastic escape rate $\gamma_{\textrm{ine}}=\Gamma^{\textrm{ine}}/\Gamma_{\tl}$ (gray dashed line) versus magnetic field $g\mu_BB$ for $T=1.2\Tct$. The inset shows that $\gamma_{\textrm{ine}}\gtrsim w$ for $|g\mu_BB|\gtrsim2.7\mu$eV. In contrast, the HF rate $w$ dominates by orders of magnitude over $\gamma_{\textrm{ine}}$ around $B=0$. The parameters are the same as in Fig.~\ref{fig:cross-triplets}(g-i).}
\label{fig:HF-rates-close-to-TTT-crossing}
\end{figure} 

The HF rates $w$ goes to zero as the triplet energies move apart, i.e. $w\rightarrow0$ for increasing $|g\mu_BB|$, see Fig.~\ref{fig:HF-rates-close-to-TTT-crossing}. Thus, the inelastic escape rate $\gamma_{\textrm{ine}}$ will eventually become larger than $w$, since $\gamma_{\textrm{ine}}$ is constant. For the values used in Fig.~\ref{fig:cross-triplets}(g-i), we have $\gamma_{\textrm{ine}}>w$ for $|g\mu_BB|\gtrsim2.7\mu$eV as shown in the inset of Fig.~\ref{fig:HF-rates-close-to-TTT-crossing}. Thus, once the triplet levels move further apart, the triplet-triplet rate $w$ goes to zero and can be neglected in the current (\ref{eq:current-inserting-rates-for-TTT-crossing}). That corresponds to $|g\mu_BB|\gtrsim2.7\mu$eV for $T>\Tct$ in the numerical example of Fig.~\ref{fig:cross-triplets}(g-i). For $\gamma_{\tr\tl}=1$, we find\cite{footnote-lowest-order-same}   
\begin{align}
\frac{I_{w\rightarrow0}}{\Gamma_\tl}
&=
\frac{8 \gamma_{\textrm{ine}} \big[\gamma_{\textrm{ine}} + (1 + \gamma_{\textrm{ine}}) \PP^2\big]}
{2 \PP^2 + \gamma_{\textrm{ine}} (3 + 2 \PP^2) \big[1 + 2 \gamma_{\textrm{ine}} + 2 (1 + \gamma_{\textrm{ine}}) \PP^2\big]}
\nonumber\\
&=4 \gamma_{\textrm{ine}} 
+\mathcal{O}[(\gamma_{\textrm{ine}})^2].
\label{eq:current-far-from-TTT-crossing}
\end{align}
Here $\gamma_{\textrm{ine}}\ll |\PP|$ for $|P|\gtrsim0.1$ for $A_-$ and $t$ similar to those used in Fig.~\ref{fig:cross-triplets}. In Figs.~\ref{fig:cross-triplets}(b,e,h), we see that the current levels off to a constant value of $\sim4\times10^{-6}$ far away from the level crossing, which is in perfect agreement with the prediction (\ref{eq:current-far-from-TTT-crossing}). Therefore, we have found that 
\begin{align}
\frac{I(B=0,T\geq \Tct)}{I_{w\rightarrow0}}
\simeq\frac{2}{3} 
\quad\textrm{for}\quad 
\gamma_{\textrm{ine}}\ll1, 
\end{align}
so the value at the crossing for $B=0$ (for $T\geq \Tct$) is generically lower than the value that the current levels off to asymptotically [Fig.~\ref{fig:cross-triplets}(e,h)]. 

\begin{figure}
\includegraphics[width=0.38\textwidth,angle=0]{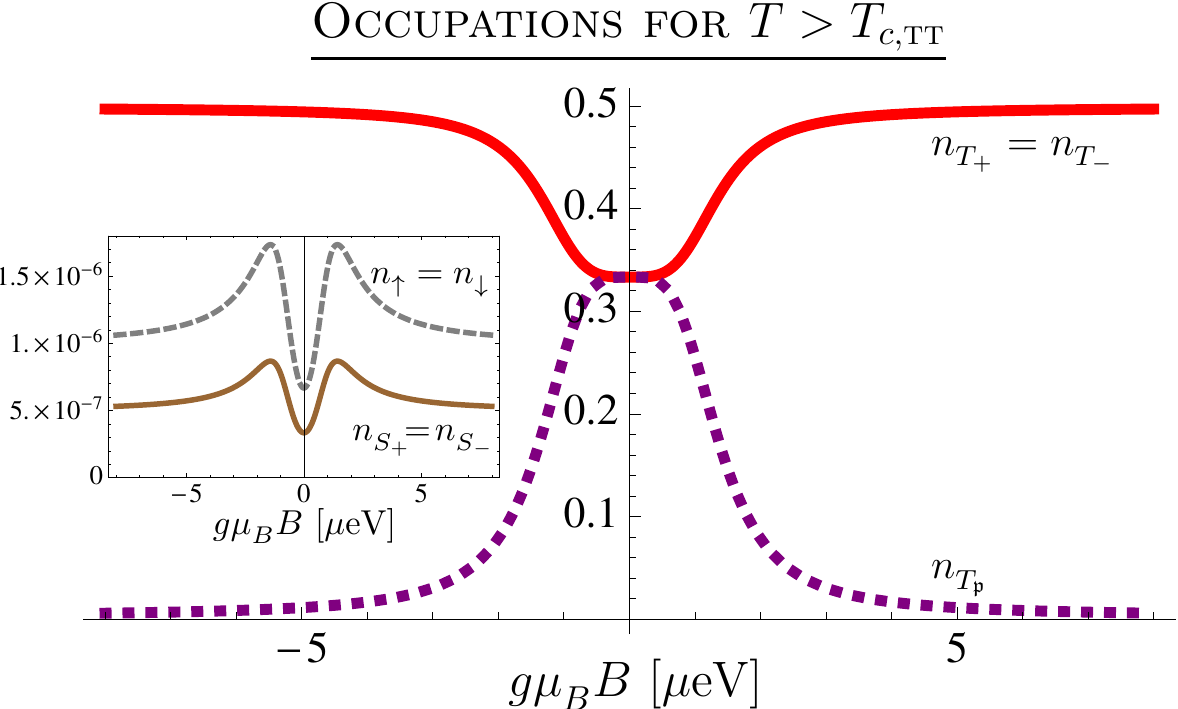}
\caption{(Color online) The occupations of the triplets $n_{T_+}\!=\!n_{T_-}$ (red full line) and $n_{T_\PP}$ (purple dotted line) Eq.(\ref{eq:triplet-occupations}) versus magnetic field $g\mu_BB$ for $T=1.2\Tct>\Tct$. Inset: The occupations of the singlets $n_{S_+}=n_{S_-}$ (brown full line) and the one-particle states $n_{\op}=n_{\ned}$ (gray dashed line). In the SB regime, the triplets are the bottleneck of the transport, so the system spends a long time in these states, leading to high occupations. In contrast, the singlet and one-electron states have orders of magnitude lower occupations. Furthermore, we remark that at the leakage current side-peak positions $g\mu_BB\simeq\pm1.4\mu$eV in Fig.~\ref{fig:cross-triplets}(h), the occupations $n_{T_+}=n_{T_-}$ dominate $n_{T_\PP}$. The parameters are the same as in Fig.~\ref{fig:cross-triplets}(g-i).}
\label{fig:occupations-close-to-TTT-crossing}
\end{figure}

Now we have shown that the current value both \emph{at} the crossing and \emph{far away} form the crossing of the triplets is determined by the inelastic escape rate. Albeit the HFI of course is essential in having a DNP in the first place. Next, we discuss how the HF triplet-triplet transitions can increase the leakage current close to -- but not exactly at -- the level crossing. At the crossing ($B=0$ for $T\geq \Tct$), all three triplet states form the bottleneck for transport through the DQD. For $P\neq0$,  the triplet-singlet mixing leads to the additional escape from $T_\PP$ such that the transport bottleneck (far away from the crossing) now only consist of $T_\pm$. For $T\geq \Tct$, we only get $P\neq0$ for $B\neq0$. Moreover, note that $n_{T_\PP}$ becomes negligible compared to $n_{T_\pm}$ far from the level crossing for $T\geq \Tct$ (see Fig.~\ref{fig:occupations-close-to-TTT-crossing}). The point is that in the region close to the level crossing, the HF triplet-triplet transitions leads to an escape path from $T_\pm$ via $T_\PP$. Since the HF rate $w$ is much larger than $\gamma_{\textrm{ine}}$ close to the crossing, this escape route is so effective that it creates the side-peaks of the current at finite $|g\mu_BB|$ seen in Fig.~\ref{fig:cross-triplets}(e,h). These current side-peaks are therefore sensitive to the value of the inelastic rate: if $\gamma_{\textrm{ine}}$ is increased by a factor of 5 or more, then the side-peaks in Fig.~\ref{fig:cross-triplets}(h) disappear. In contrast, if $\gamma_{\textrm{ine}}$ is decreased the side-peaks remain. Fig.~\ref{fig:occupations-close-to-TTT-crossing} shows that near the current side-peaks [with maxima at $g\mu_BB\simeq\pm1.4\mu$eV in Fig.~\ref{fig:cross-triplets}(h)], the occupations $n_{T_\pm}$ are much larger than $n_{T_\PP}$.

Finally, we point out that the form of the current for $T<\Tct$ can be understood by using the above considerations, but taking into account that the DNP jumps between rather high values [e.g.~$P\sim0.45$ to $P\sim-0.8$ at $g\mu_BB\simeq2.34\mu$eV in Fig.~\ref{fig:cross-triplets}(a)]. Therefore, the rich region around $P=0$ is simply skipped. [Note also the vertical scale change in Fig.~\ref{fig:cross-triplets}(b) compared to (e) and (h).] 

Therefore, we have now obtained an understanding of the leakage current versus magnetic field close to the crossing of the triplet levels under the assumption that the singlet  levels $E_{S_\pm}$ are far away in energy.

\subsubsection{The current in the high temperature and low $B$-field limit}

Next, we show that the leakage current can be given analytically in terms of the external parameters for low magnetic fields and $T>\Tct$. Specifically, if 
$\left|2g\mu_BB/A_++P\right|\ll T/\Tct$ then the hyperbolic tangent in the implicit DNP equation (\ref{eq:pol-condition-tanh}) can be expanded, so 
\begin{align} 
P\simeq\frac{\frac{1}{2}g\mu_BB/\kb}{ T- \Tct}.
\label{eq:DNP-for-B-low-T-high}
\end{align}
This is similar to a Curie-Weiss law for a ferromagnet in the paramagnetic region and the fact that $P\propto (T-\Tct)^{-1}$ is typical for the mean-field approach used here.\cite{Ashcroft-Mermin-BOOK} 
Expanding the current (\ref{eq:current-inserting-rates-for-TTT-crossing}) in $P$ and inserting Eq.(\ref{eq:DNP-for-B-low-T-high}), the current for low $B$-fields explicitly becomes:  
\begin{align}
&\frac{I(T>\Tct)}{\Gamma_\tl}
\simeq
\frac{8\gamma_{\tr\tl}\gamma_{\textrm{ine}}}{3\gamma_{\tr\tl}+2 \gamma_{\textrm{ine}} (1 + 2 \gamma_{\tr\tl})}
\label{eq:current-for-low-B-high-T} 
\\
&\hspace{8mm}
\!+\!\frac{A_-^2}{4t^2}
\frac{\gamma_{\tr\tl} (\gamma_{\tr\tl}^2+\gamma_{\textrm{ine}}\gamma_{\tr\tl}-4 \gamma_{\textrm{ine}}^2)}{\big[3 \gamma_{\tr\tl} +\gamma_{\textrm{ine}}(2 + 4 \gamma_{\tr\tl})\big]^2}
\!\bigg[\frac{g\mu_BB}{\kb [T- \Tct]}\bigg]^2.
\nonumber
\end{align}
This describes the current dip close to $B=0$ seen in Fig.~\ref{fig:cross-triplets}(h). It shows explicitly that the current increases by changing slightly $B$ away from $B=0$, since $\gamma_{\textrm{ine}}<(1+\sqrt{17}) \gamma_{\tr\tl}/8$ for reasonable parameters. We observe that the HF rate $w$ only appears beyond the second order term, however, already this term contains $A_-=(A_\tl-A_\tr)/2$. The lowest order term for $B=0$ coincides with Eq.(\ref{eq:current-for-zero-TS-mixing-parameter}) for no triplet-singlet mixing as expected.

\section{The singlet-triplet crossing}\label{sec:ST-cross}

In this section, we analyse the DNP and leakage current close to the crossing of the singlet levels $E_{S_{\pm}}$ and the pure triplet levels $E_{T_\pm}$ at finite magnetic field. Since we consider zero detuning, the crossing of the levels always happens in pairs, e.g. $E_{T_+}$ and $E_{S_+}$ cross at the same $B$-field as $E_{T_-}$ and $E_{S_-}$ do. Interestingly, here we find that the transition temperature $\Tcs$ for the singlet-triplet level crossing is enhanced compared to $\Tct$ Eq.(\ref{eq:trans-temp-TT}).

\subsection{A simplified model for the singlet-triplet crossing and its implicit polarization equation}\label{subsec:implicit-pol-eq-ST-cross}

Next, we develop a simplified set of rate equations valid close to the crossing of $E_{T_+}$ ($E_{T_-}$) and $E_{S_+}$ ($E_{S_-}$) for \emph{positive} $B$-field splitting, $g\mu_BB>0$. From these equations, we derive an implicit equation for the DNP. The level crossings for $g\mu_BB<0$ follow along similar lines.\cite{footnote-negative-B}

\begin{figure}
\includegraphics[width=0.43\textwidth,angle=0]{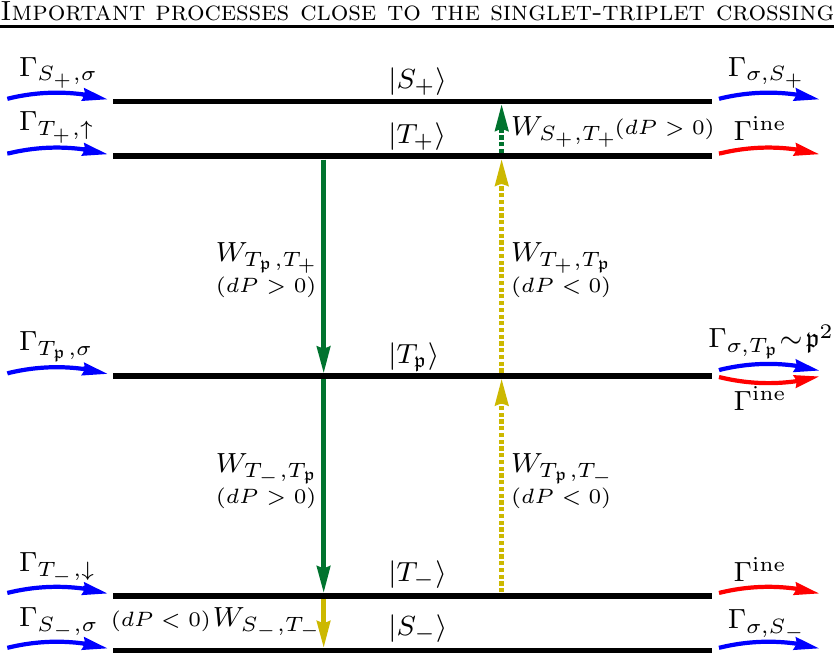}
\caption{(Color online) The transitions included in the simplified rate equations (\ref{eq:rate-eqs-ST-crossing}-\ref{eq:polarization-eq-ST-cross}) tailored to described the DNP and leakage current close to the singlet-triplet crossing for $g\mu_BB>0$. The HF transitions $S_\pm\rightarrow T_\pm$ are neglected, since they are much less probable than simply tunneling out of the singlets. Furthermore, the triplet-triplet transitions are included, since these can play a role even though the levels have a large energy separation (see the main text for further discussion). The HF phonon emission (absorption) processes are shown as full (dotted) vertical arrows. The processes indicated by green (yellow) arrows change the DNP positively (negatively). The inelastic escape rates $\Gamma^{\textrm{ine}}_{}$ (red arrows) and the tunneling rates $\Gamma_{f,i}$ (blue arrows) are the same as in Fig.~\ref{fig:full-rate-eq-illustration}.}
\label{fig:ST-rate-eq-illustration}
\end{figure}

As a first approach, one might intend to follow the same strategy as in Sec.~\ref{subsec:pol-eq-TT} for the crossing of the triplets: Keep only the HF rates between the levels, which are close in energy. Hence, we keep only the HF terms in Eqs.(\ref{eq:full-rate-eqs-appendix},\ref{eq:full-polarization-eq}) involving $W_{T_\pm,S_\pm}$ and $W_{S_\pm,T_\pm}$ in the present case. Such a simplification leads to $W_{S_+,T_+}=W_{T_+,S_+}$ as an implicit equation for the DNP --- much like in the case of the crossing of the triplets Eq.(\ref{eq:pol-condition-for-rates}). This leads to the same transition temperature as $\Tct$ Eq.(\ref{eq:trans-temp-TT}) to a very good approximation. \emph{However, for the singlet-triplet level crossing, this approach is actually not a good approximation}. Explicitly, we find that this approach does \emph{not} reproduce the DNP found by a numerical iteration of the rate equations (\ref{eq:full-rate-eqs-appendix},\ref{eq:full-polarization-eq}) including all rates (for an inter-dot coupling $t$ about two orders of magnitude larger than $\gamma_{ph}$).\cite{footnote-remark-on-approx} The approach fails for the following reasons: Firstly, the occupations $n_{S_\pm}$ are much smaller than the occupations $n_{T_\pm}$ and $n_{T_\PP}$, since escape from the singlets are much easier than from the triplets in the SB regime. Secondly, the triplet-triplet and the singlet-triplet HF rates have the overall prefactors of $A_+^2$ and $A_-^2$, respectively, such that the triplet-triplet rate is enhanced compared to the singlet-triplet rate (for comparable energy level splitting). Due to these two facts, the triplet-triplet HF terms can still be comparable in magnitude to the singlet-triplet terms in the rate equations (\ref{eq:full-rate-eqs-appendix},\ref{eq:full-polarization-eq}) -- even though $|E_{T_\pm}-E_{T_\PP}|\gg \gamma_{ph}$ close to the singlet-triplet crossing. In other words, we \emph{cannot} neglect terms like $W_{T_\PP,T_+}n_{T_+}$ compared to terms like $W_{S_+,T_+}n_{T_+}$.

Here, we have to adopt a different approach of simplifying the rate equations from the one used for the crossing of the triplets in Sec.~\ref{subsec:pol-eq-TT}. This is done in order to describe the regime of large singlet-triplet energy splitting $|E_{S_\pm}-E_{T_\PP}|$ compared to $\gamma_{ph}$, however, not so large that the triplet-triplet rates cannot still play a role. Our approach is to neglect two kinds of terms in the rate equations (\ref{eq:full-rate-eqs-appendix},\ref{eq:full-polarization-eq}). (i) The HF singlet-triplet terms between $T_+$ ($T_-$) and $S_-$ ($S_+$) can safely be neglected, because of large energy separation (for $g\mu_BB>0$) combined with an overall prefactor of $A_-^2$ in the rate. Thus, we neglect terms of the form $W^{}_{T_{\bar{\nu}},S_\nu}n^{}_{S_\nu}$ and  $W^{}_{S_{\bar{\nu}},T_\nu}n^{}_{T_\nu}$, where $\nu=\pm$ and $\bar{\nu}=-\nu$. (ii) Due to the SB regime, $n^{}_{S_\nu}\!\!\ll n^{}_{T_{\nu^{\prime}}}$ for $\nu^{\prime},\nu=\pm$, so we neglect the terms $W^{}_{T_\pm,S_\pm}n^{}_{S_\pm}$.  We thereby neglect HF transitions from $S_\pm$ to $T_\pm$, since tunneling out from $S_\pm$ are \emph{much} more probable processes. Using equal inelastic escape rates (\ref{eq:equal-Gamma-ine}), these two simplifications lead to the following rate equations for $g\mu_BB>0$ 
\begin{widetext}
\begin{subequations}
\label{eq:rate-eqs-ST-crossing}
\begin{align} 
\dot{n}^{}_{T_+}
&=  
W^{}_{T_+,T_\PP}n^{}_{T_\PP}
+\Gamma^{}_{T_+,\op} n^{}_\op 
-\big[W^{}_{S_+,T_+} + W^{}_{T_\PP,T_+} 
+2\Gamma_{}^{\textrm{ine}}\big]n^{}_{T_+}, 
\\  
\dot{n}^{}_{T_-}&=
W^{}_{T_-,T_\PP}n^{}_{T_\PP}
+\Gamma^{}_{T_-,\ned} n^{}_\ned 
-\big[
W_{S_-,T_-}^{} 
+W^{}_{T_\PP,T_-} 
+2\Gamma_{}^{\textrm{ine}}\big] n^{}_{T_-}, 
\\  
\dot{n}_{T_\PP}&=  
W_{T_\PP,T_+}n^{}_{T_+}
+W_{T_\PP,T_-}n^{}_{T_-} 
+\Gamma_{T_\PP,\op}n^{}_{\op}
+\Gamma_{T_\PP,\ned}n^{}_{\ned} 
-\big[W_{T_+,T_\PP} + W_{T_-,T_\PP}+\Gamma_{\op,T_\PP} 
+\Gamma_{\ned,T_\PP} 
+2\Gamma_{}^{\textrm{ine}} 
\big] n^{}_{T_\PP},
\\  
\dot{n}^{}_{S_+}&=
W^{}_{S_+,T_+}n^{}_{T_+}
+\Gamma^{}_{S_+,\op}n^{}_{\op}+\Gamma^{}_{S_+,\ned}n^{}_{\ned}
-\big[
\Gamma^{}_{\op,S_+} + \Gamma^{}_{\ned,S_+}\big]n^{}_{S_+},
\\  
\dot{n}^{}_{S_-}&=
W^{}_{S_-,T_-}n^{}_{T_-}
+\Gamma^{}_{S_-,\op}n^{}_{\op}+\Gamma^{}_{S_-,\ned}n^{}_{\ned}
-\big[
\Gamma^{}_{\op,S_-}+\Gamma^{}_{\ned,S_-}\big]n^{}_{S_-},
\\
\dot{n}^{}_\op&=
\Gamma^{}_{\op,S_+}n^{}_{S_+}
+\Gamma^{}_{\op,S_-}n^{}_{S_-} 
+(\Gamma^{}_{\op,T_\PP}+\Gamma_{}^{\textrm{ine}})n^{}_{T_\PP} 
+\Gamma_{}^{\textrm{ine}}(n^{}_{T_+}+n^{}_{T_-}) 
-\big[\Gamma^{}_{S_+,\op}+\Gamma^{}_{S_-,\op}+\Gamma^{}_{T_\PP,\op}+\Gamma^{}_{T_+,\op}\big]n^{}_\op,
\\
\dot{n}^{}_\ned&=
\Gamma^{}_{\ned,S_+}n^{}_{S_+}
+\Gamma^{}_{\ned,S_-}n^{}_{S_-} 
+(\Gamma^{}_{\ned,T_\PP}+\Gamma_{}^{\textrm{ine}})n^{}_{T_\PP} 
+\Gamma_{}^{\textrm{ine}}(n^{}_{T_-} +n^{}_{T_+}) 
-\big[\Gamma^{}_{S_+,\ned}+\Gamma^{}_{S_-,\ned}
+\Gamma^{}_{T_\PP,\ned}+\Gamma^{}_{T_-,\ned}\big]n^{}_\ned. 
\end{align}
\end{subequations}
\end{widetext}
Likewise, Eq. (\ref{eq:full-polarization-eq}) for the DNP simplifies to
\begin{align} 
\dot{P}=&  
\frac{2}{N}\Big[ 
(W^{}_{T_-,T_\PP}-W^{}_{T_+,T_\PP})n^{}_{T_\PP}
\label{eq:polarization-eq-ST-cross}
\\ 
&
{+}(W^{}_{S_+,T_+}{+}W^{}_{T_\PP,T_+})n^{}_{T_+} 
-(W^{}_{S_-,T_-}{+}W^{}_{T_\PP,T_-})n^{}_{T_-}
\Big].
\nonumber
\end{align}
The simplified model is illustrated on Fig.~\ref{fig:ST-rate-eq-illustration}. The only difference compared to the rate equations (\ref{eq:TT-rate-eqs}-\ref{eq:TT-polarization-eq}) for the crossing of the triplets, is the \emph{direct} coupling of the singlets and triplets via the terms $W^{}_{S_\pm,T_\pm}n^{}_{T_\pm}$. 

Now we derive the implicit equation for the stationary DNP from these rate equations. We do not use the explicit form of the rates, but only the invariances under index interchange (\ref{eq:sym-for-tunnel-rates},\ref{eq:sym-for-HF-rates}). We begin by noting that $\dot{n}^{}_\op-\dot{n}^{}_\ned=0$ leads to $n^{}_\op=n^{}_\ned$ in the stationary state. Inserting this into $\dot{n}^{}_{T_+}-\dot{n}^{}_{T_-}+(N/2)\dot{P}=0$ gives $n^{}_{T_+}=n^{}_{T_-}$ in the steady state. These two relations are the same as in the description of the triplet level crossing, see Eqs.(\ref{eq:nned-lig-nop}-\ref{eq:nTm-lig-nTp}). In fact, $n^{}_\op=n^{}_\ned$ and $n^{}_{T_+}=n^{}_{T_-}$ can be derived form the rate equations (\ref{eq:full-rate-eqs-appendix},\ref{eq:full-polarization-eq}) including \emph{all} rates and equal inelastic rates. However, at this point the two descriptions separate, since the stationary singlet occupations are no longer equal [as in Eq.(\ref{eq:nSp-lig-nSm})]. Instead, we find
\begin{align}
\label{eq:occ-singlets-ST-cross}
n^{}_{S_\pm}=&
\frac{(\Gamma_{S_+,\ned} + \Gamma_{S_+,\op}) n^{}_{\op} +  W_{S_\pm,T_\pm} n^{}_{T_+}}{2 \Gamma_{\op,S_+}},
\end{align}
by solving $\dot{n}^{}_{S_\pm}=0$ using $n^{}_\op=n^{}_\ned$ and $n^{}_{T_+}=n^{}_{T_-}$ and that $\Gamma_{\op,S_+}\neq0$. Hence, $n^{}_{S_+}\neq n^{}_{S_-}$ if and only if $W_{S_-,T_-}\neq W_{S_+,T_+}$. These relations are used to derive
\begin{subequations}
\begin{align}
n^{}_{T_+}=&
n_\op
\frac{2}{\kappa}
\Big[
2 \Gamma_{T_+,\op} (\Gamma^{\textrm{ine}}+\Gamma_{\op,T_\PP})
\\
&\hspace{8mm}
+ (\Gamma_{T_\PP,\op}+\Gamma_{T_+,\op}) (W_{T_+,T_\PP} + W_{T_\PP,T_+}) 
\Big],
\nonumber\\
n^{}_{T_\PP}=&
n_\op
\frac{2}{\kappa}
\Big[
\Gamma_{T_\PP,\op} (4 \Gamma^{\textrm{ine}} + W_{S_+,T_+} + W_{S_-,T_-}) 
\\
&\hspace{8mm}
+(\Gamma_{T_\PP,\op}+\Gamma_{T_+,\op})(W_{T_+,T_\PP} + W_{T_\PP,T_+}) 
\Big]
\nonumber
\end{align}
from $\dot{n}^{}_\op+\dot{n}^{}_\ned=0$ and $\dot{n}^{}_{T_\PP}=0$. Here we introduced the non-zero quantity 
\begin{align}
\kappa 
\equiv
&
8 (\Gamma^{\textrm{ine}})^2 
+ 2( \Gamma_{\op,T_\PP} + 3 \Gamma^{\textrm{ine}})
(W_{T_+,T_\PP} + W_{T_\PP,T_+}) 
\nonumber\\
&
+ 8 \Gamma^{\textrm{ine}} \Gamma_{\op,T_\PP} 
+ (W_{S_+,T_+} + W_{S_-,T_-}) 
\nonumber\\
&
\hspace{4mm}\times
\big[2 \Gamma^{\textrm{ine}} +2 \Gamma_{\op,T_\PP}+W_{T_+,T_\PP} + W_{T_\PP,T_+}\big].
\end{align}
\end{subequations}
Note that the triplet occupation expressions are proportional to the one-electron occupation. By inserting the occupation expressions into Eq.(\ref{eq:polarization-eq-ST-cross}), we find $\dot{P}=n_{\op}\frac{2}{N}\frac{\chi}{\kappa}$, where $\chi$ is a combination of rates (given below). In order to satisfy $\dot{P}=0$ in steady state, we have to require that $\chi=0$, since the occupation is positive. Thus, we arrive at the implicit DNP equation, $\chi=0$, which explicitly is
\begin{widetext}
\begin{align}
0=&
(W_{T_+,T_\PP} - W_{T_\PP,T_+})
\bigg\{
\Gamma_{T_+,\op}(\Gamma_{\op,T_\PP} {+} \Gamma^{\textrm{ine}})
+(W_{T_+,T_\PP} {+} W_{T_\PP,T_+}) (\Gamma_{T_+,\op} {+} \Gamma_{T_\PP,\op}) 
+\Gamma_{T_\PP,\op} 
\Big[\frac{1}{2}(W_{S_+,T_+} {+} W_{S_-,T_-}) {+} 2\Gamma^{\textrm{ine}}\Big]
\bigg\}
\nonumber\\ 
&+
(W_{S_-,T_-} - W_{S_+,T_+})
\bigg\{
\Gamma_{T_+,\op} (\Gamma_{\op,T_\PP} + \Gamma^{\textrm{ine}})
+\frac{1}{2} (W_{T_+,T_\PP} + W_{T_\PP,T_+})
(\Gamma_{T_+,\op} + \Gamma_{T_\PP,\op})
\bigg\}
=\chi.
\label{eq:implicit-eq-ST-cross}
\end{align}
\end{widetext}
This implicit equation for $P$ is more involved than the one describing the DNP around the crossing of the triplet levels Eq.(\ref{eq:pol-condition-for-rates}). Moreover, a simple formula  for the transition temperature $\Tcs$ is not immediately apparent. Nevertheless, the implicit equation (\ref{eq:implicit-eq-ST-cross}) does give some insights. For instance, it describes the crossing of the triplets as a special case: Close to the crossing of the triplets, $W_{S_\pm,T_\pm}$ is negligible such that Eq.(\ref{eq:implicit-eq-ST-cross}) simplifies to $W_{T_+,T_\PP}-W_{T_\PP,T_+}=0$, which is exactly Eq.(\ref{eq:pol-condition-for-rates}).

Furthermore, the implicit equation (\ref{eq:implicit-eq-ST-cross}) shows that $\Tcs$ stems from the asymmetry between energy emission and absorption in the HF process. We show this by assuming the opposite: absorbing or emitting an energy in the HF process is equally likely [i.e.~$\mathcal{F}_{ph}(E)$ is even in Eqs.(\ref{eq:TT-HF-rates},\ref{eq:TS-HF-rates})]. This assumption leads to $(W_{S_-,T_-} - W_{S_+,T_+})\propto P$ and $(W_{T_+,T_\PP} - W_{T_\PP,T_+})\propto P$ by using Eqs.(\ref{eq:TT-HF-rates},\ref{eq:TS-HF-rates}), such that the implicit equation (\ref{eq:implicit-eq-ST-cross}) can be written as $0=PG(P)$, where $G(P)$ is a strictly positive function.\cite{footnote-G-details} Thus, $P=0$ is the only DNP solution without the asymmetry between emission and absorption of energy such that no DNP bistability occurs.     

Here, we find the DNP from the implicit equation (\ref{eq:implicit-eq-ST-cross}) numerically. This in turn gives the leakage current and transition temperature $\Tcs$ as we will discuss next.

\begin{figure*}
\includegraphics[width=0.93\textwidth,angle=0]{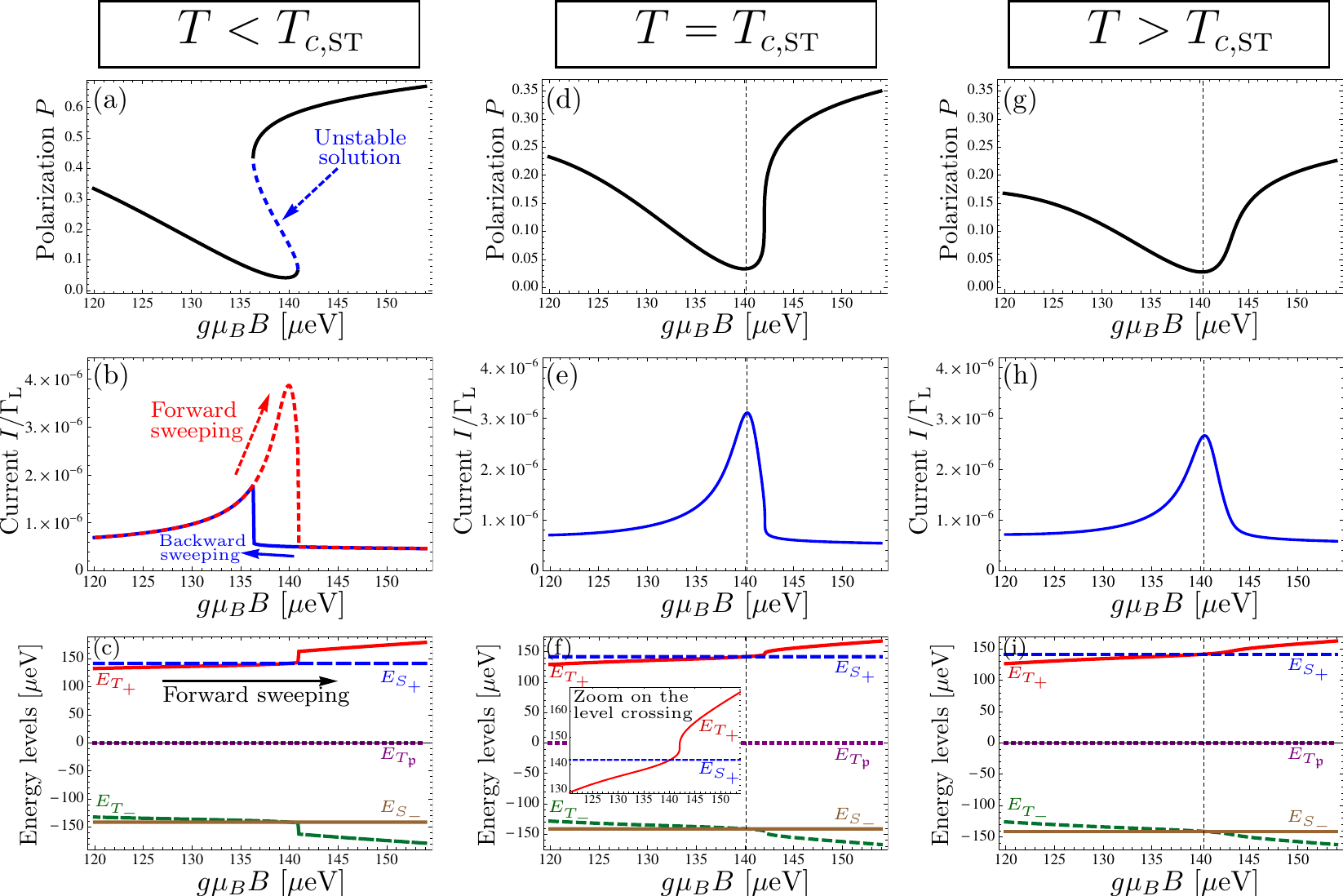}
\caption{(Color online) The nuclear polarization $P$, leakage current $I/\Gamma_\tl$ and energy levels versus positive external magnetic field $g\mu_BB$ (in energy units) close to the singlet-triplet crossing for temperatures $T=0.5\Tcs<\Tcs$ (a-c), $T=\Tcs$ (d-f) and $T=1.5\Tcs>\Tcs$ (g-i). For $T<\Tcs$, we observe two stable DNP values (black full lines) and an unstable one (blue dashed line) in Fig.~(a), which leads to the hysteretic leakage current as seen in Fig.~(b). The corresponding energy levels are seen in Fig.~(c), where only the case of sweeping the magnetic field forward is shown for clarity. For $T\geq\Tcs$, the DNP is single valued [Fig.~(d,g)] such that no hysteretic current appears, see Fig.~(e,h). Note the difference in the vertical scales between the DNP in Fig.~(a) and Fig.~(d,g). The vertical dashed black line indicates the simultaneous crossing of (i) the triplet energy $E_{T_+}$ (red full line) with the singlet energy $E_{S_+}$ (blue dashed line) and (ii) $E_{T_-}$ (green dashed line) with  $E_{S_-}$ (brown full line). The current is seen to peak at the level crossing -- essentially due to the enhanced HF singlet-triplet rate, which lifts the SB. The inset of Fig.~(f) shows the non-monotonous energy level variation close to the crossing of $E_{T_+}$ and $E_{S_+}$. In general, $\Tcs$ depends on several parameters of the system (see the main text). For the numerical example seen here, we find $\kb\Tcs\simeq 2.80A_+$, which is about \emph{one order of magnitude larger} than $\kb\Tct=A_+/4$. The parameters used here are: $A_{\tl}=80\mu$eV, $A_{\tr}=70\mu$eV, $t=100\mu$eV, $\gamma_{ph}=1\mu$eV, $\Gamma_\tl=\Gamma_\tr$ (i.e.~$\gamma_{\tr\tl}=1$), $\h\Gamma_\tl N=10^6\mu$eV  and the dimensionless inelastic escape rate is chosen to be $\gamma_{\textrm{ine}}=\Gamma^{\textrm{ine}}/\Gamma_\tl=10^{-7}$, such that the HF rates dominates close to the singlet-triplet crossing (see Fig.~\ref{fig:rates-ST-crossing}).} 
\label{fig:ST-cross}
\end{figure*}

\begin{figure}
\includegraphics[width=0.4\textwidth,angle=0]{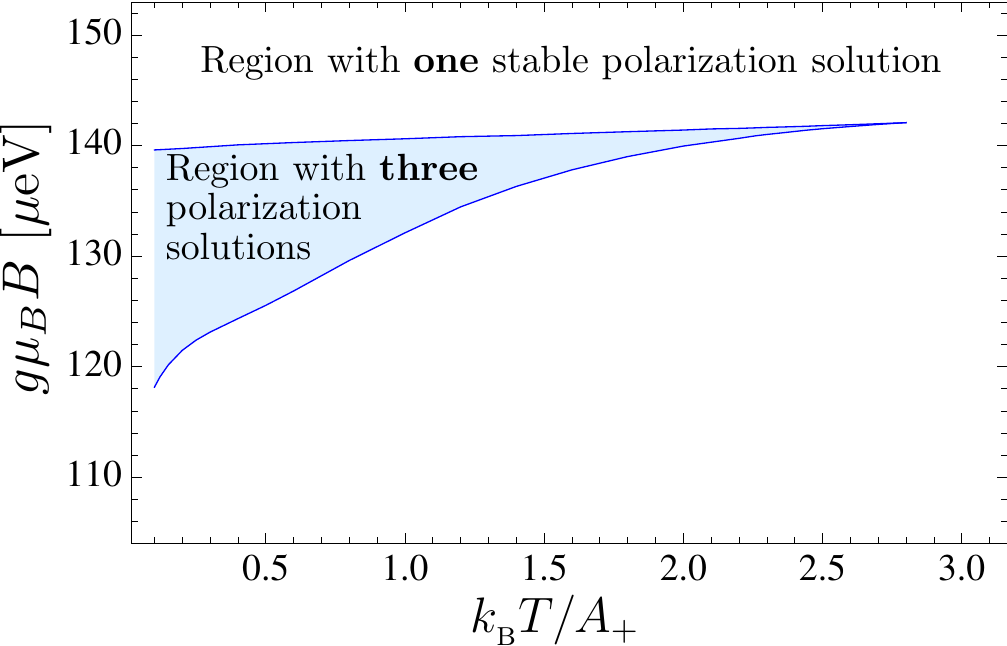}
\caption{(Color online) The regions in parameter space close to the singlet-triplet crossing with one (white region) or three (blue region) DNP solutions, respectively. Only two of the three solutions for $P$ in the blue region are stable against small fluctuations. Here we use the same parameters as in Fig.~\ref{fig:ST-cross} and find the transition temperature to be $\Tcs\simeq 2.80 A_+/\kb$.}
\label{fig:phase-diagram-ST-crossing}
\end{figure}

\subsection{The nuclear polarization, leakage current and the singlet-triplet crossing transition temperature}

We extract the DNP versus magnetic field for various temperatures numerically from the implicit equation (\ref{eq:implicit-eq-ST-cross}), see Fig.~\ref{fig:ST-cross}(a,d,g). In this way, we can pinpoint the region of temperature and $B$-field with one and three DNP solutions, respectively, as shown in Fig.~\ref{fig:phase-diagram-ST-crossing}. This in turn allows to determine the transition temperature $\Tcs$ for the triplet-singlet crossing, where the DNP becomes single-valued. In the specific case of parameters in Fig.~\ref{fig:ST-cross} and \ref{fig:phase-diagram-ST-crossing}, we find $\Tcs\simeq2.80A_+/\kb$, which is about one order of magnitude larger than $\Tct=A_+/(4\kb)$ Eq.(\ref{eq:trans-temp-TT}). 

\begin{table}
\centering
  \begin{tabular}{ l | c | c | c | c | }
     & $\gamma_{\textrm{ine}}\!=\!10^{-6}$ & $\gamma_{\textrm{ine}}\!=\!10^{-7}$ & $\gamma_{\textrm{ine}}\!=\!10^{-8}$ & $\gamma_{\textrm{ine}}\!=\!10^{-9}$ \\ 
    \cline{2-5}
     \hline
    {\footnotesize$t\!=\!50\mu$eV} & $2.64A_+$ & $2.65A_+$ & $2.65A_+$ & $2.65A_+$ \\ 
    \hline
    {\footnotesize$t\!=\!100\mu$eV} & $2.72A_+$ & $2.80A_+$ & $2.81A_+$ & $2.82A_+$ \\ 
    \hline
    {\footnotesize$t\!=\!150\mu$eV} & $2.50A_+$ & $2.62A_+$ & $2.66A_+$ & $2.67A_+$ \\
    \hline
   \end{tabular}
\caption{The transition temperatures $\kb\Tcs$ for the singlet-triplet crossing for various inter-dot couplings $t$ and inelastic escape rates $\gamma_{\textrm{ine}}=\Gamma^{\textrm{ine}}/\Gamma_\tl$. Here, $\gamma_{\textrm{ine}}$ are chosen smaller than the dominant hyperfine rate close to the level crossing. The fixed parameters here are: $A_{\tl}=80\mu$eV, $A_{\tr}=70\mu$eV, $\gamma_{ph}=1\mu$eV, $\Gamma_\tl=\Gamma_\tr$ and $\h\Gamma_\tl N=10^6\mu$eV.}
\label{Table:Tc-ST-cross-inelastic-rate-vs-t}
\end{table}

\begin{table}
\centering
  \begin{tabular}{ l | c | c | c | }
     & $A_-\!=\!1\mu$eV & $A_-\!=\!5\mu$eV  & $A_-\!=\!10\mu$eV  \\ 
     & {\scriptsize($A_\tl\!=\!76\mu$eV)} & {\scriptsize($A_\tl\!=\!80\mu$eV)} & {\scriptsize($A_\tl\!=\!85\mu$eV)}  \\ 
     \hline
    $\h\Gamma_\tl N\!=\!10^5\mu$eV & $6.65A_+$ & $2.38A_+$ & $1.32A_+$  \\ 
    \hline
    $\h\Gamma_\tl N\!=\!10^6\mu$eV & $6.94A_+$ & $2.80A_+$ & $1.39A_+$  \\ 
    \hline
  \end{tabular}
\caption{The transition temperatures $\kb\Tcs$ for the singlet-triplet crossing varying the difference between the effective HF constants $A_-=(A_\tl-A_\tr)/2$ and the number of nuclei $\h\Gamma_\tl N$ (measured with the rate $\Gamma_\tl$). By varying $A_-$, we change the overall strength of the singlet-triplet rates compared to the triplet-triplet rates. The fixed parameters here are: $A_{+}=75\mu$eV,  $\gamma_{\textrm{ine}}=10^{-7}$, $t=100\mu$eV,  $\gamma_{ph}=1\mu$eV and $\Gamma_\tl=\Gamma_\tr$.}
\label{Table:Tc-ST-cross-Am-og-N}
\end{table}

We have repeated this procedure to find the transition temperatures for different parameters as seen in tables \ref{Table:Tc-ST-cross-inelastic-rate-vs-t} and \ref{Table:Tc-ST-cross-Am-og-N}. We find that \emph{the transition temperature} $\Tcs$ \emph{depends on various external parameters} -- in contrast to the crossing of the triplets, where $\Tct=A_+/(4\kb)$. In table \ref{Table:Tc-ST-cross-inelastic-rate-vs-t}, we find that $\Tcs$ seems largely insensitive to decreasing the inelastic rate $\gamma_{\textrm{ine}}=\Gamma^{\textrm{ine}}/\Gamma_\tl$ as long as it is smaller than the dominant singlet-triplet rates close to the level crossing. This makes sense from the implicit equation (\ref{eq:implicit-eq-ST-cross}), since a small $\Gamma^{\textrm{ine}}$ is negligible compared to $W_{S_\pm,T_\pm}$ and $\Gamma_{\op,T_\PP}$.  Table \ref{Table:Tc-ST-cross-inelastic-rate-vs-t} also reveals a small non-monotonous dependence of $\Tcs$ on 
$t$, which controls the level splitting $|E_{S_\pm}-E_{T_\PP}|$ and, in turn, the size of the HF triplet-triplet rates close to the singlet-triplet crossing.

An effective way to change the relative magnitudes of the singlet-triplet and the triplet-triplet rates, is to change $A_-$, since the singlet-triplet rate have an overall prefactor of $A_-^2$ (whereas $W_{T,T}\propto A_+^2$). Table \ref{Table:Tc-ST-cross-Am-og-N} shows $\Tcs$ for varying the relative strength of the singlet-triplet and triplet-triplet rates. The largest transition temperature, $\kb \Tcs\simeq6.94A_+$, is found when the singlet-triplet, triplet-triplet and inelastic rates all are of the same order. In contrast, the smallest value,  $\kb \Tcs\simeq1.32 A_+$,  is found when the singlet-triplet rate dominates by more than two orders of magnitude over the triplet-triplet rate. Moreover, the number of nuclei change $\Tcs$ slightly. Finally, we remark that $\Tcs$ is not simply proportional to $A_+$. Nevertheless, we give $\kb \Tcs$ in units of $A_+$ in order to compare it with a typical energy scale of the problem. Altogether, a common feature for all the parameters considered here, is that $\Tcs$ is found to be larger than $\Tct$.

\begin{figure}
\includegraphics[width=0.4\textwidth,angle=0]{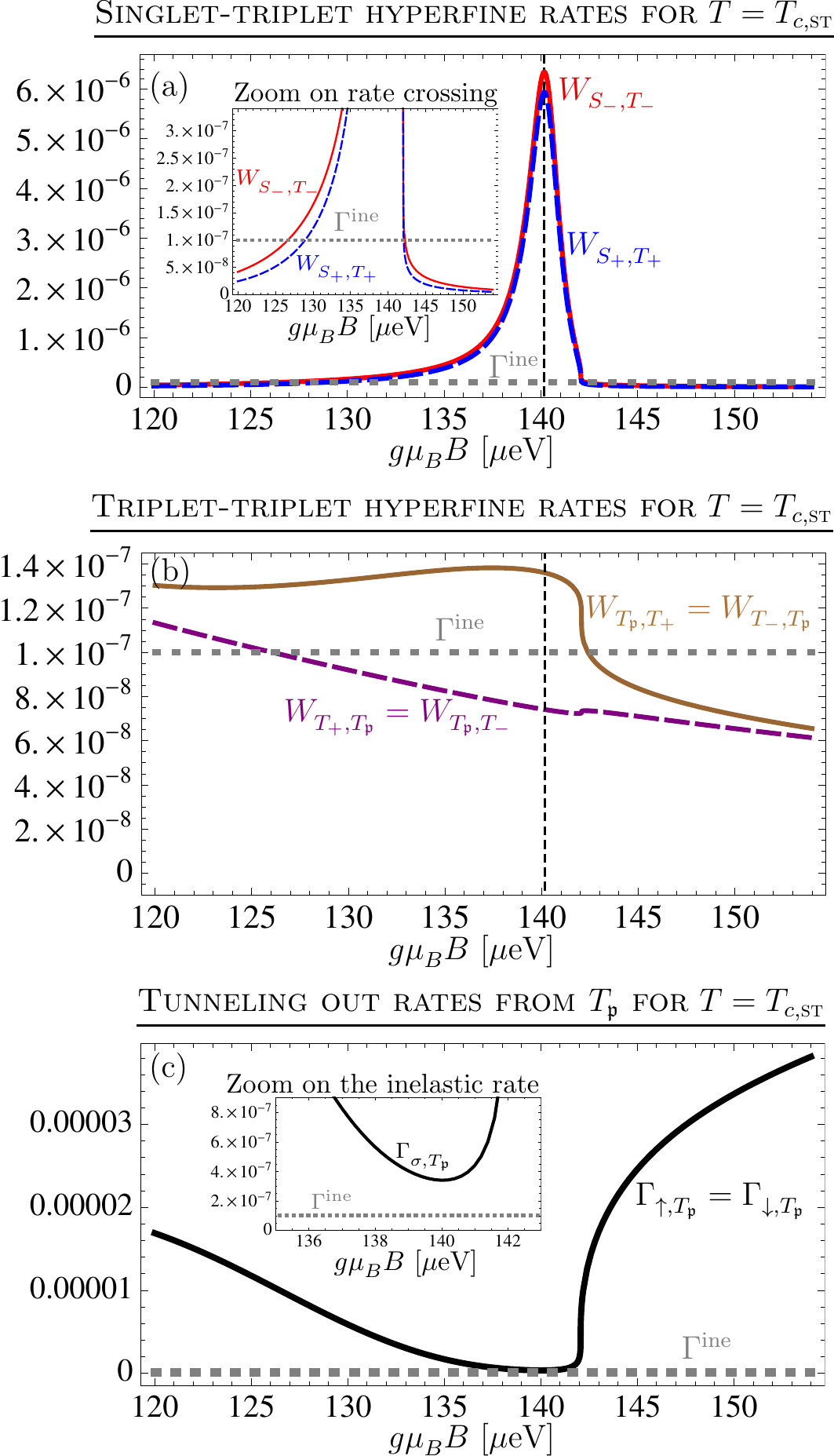}
\caption{(Color online) (a) The HF triplet-singlet rates $W_{S_-,T_-}$ (red full line) and $W_{S_+,T_+}$ (blue dashed line) become much larger than the inelastic escape rate $\Gamma^{\textrm{ine}}$ (gray dotted line) close to the singlet-triplet crossing (vertical dashed line). Inset: far away from the singlet-triplet crossing the HF rates goes to zero so $\Gamma^{\textrm{ine}}$ becomes larger. (b) The HF triplet-triplet rates $W_{T_\PP,T_+}=W_{T_-,T_\PP}$ (brown full line) and $W_{T_+,T_\PP}=W_{T_\PP,T_-}$ (violet dashed line) are seen to be on the same order of magnitude as $\Gamma^{\textrm{ine}}$ (gray dotted line) close to the level crossing. (c) The rate $\Gamma_{\sigma,T_\PP}$ (black full line) for tunneling out of the DQD from $T_\PP$ is reduced close to the crossing, since the DNP decreases [see Fig.~\ref{fig:ST-cross}(d)]. Thus, $\Gamma_{\sigma,T_\PP}$ becomes on the order of $\Gamma^{\textrm{ine}}$ (dotted gray line) close to the level crossing (see inset). In (a), (b) and (c), all rates are in units of $\Gamma_\tl$. The parameters are the same as in Fig.~\ref{fig:ST-cross}(d-e). For $T>\Tcs$ qualitatively similar curves are found.}
\label{fig:rates-ST-crossing}
\end{figure}

The leakage current is found from Eq.(\ref{eq:current-def}) by inserting the stable DNP found from the implicit equation (\ref{eq:implicit-eq-ST-cross}). To this end, we use $\sum_\nu n_\nu=1$ to specify all the occupations. In Fig.~\ref{fig:ST-cross}, we investigate the DNP, leakage current and energy levels in the regime, where the HF singlet-triplet rates dominate in magnitude over the triplet-triplet and inelastic rates close to the level crossing,\cite{footnote-compare-two-figs} see Fig.~\ref{fig:rates-ST-crossing}. 

Current hysteresis is found as a natural consequence of two stable DNP solutions for $T<\Tcs$ --- just as for the crossing of the triplets. For instance, if one increases the magnetic field from, say, $g\mu_BB=120\mu$eV for $T<\Tcs$ [Fig.~\ref{fig:ST-cross}(a)], then the DNP will remain on the lower DNP solution until the critical field of about $g\mu_BB\simeq140.9\mu$eV, where the lower branch cease to exist. At this point, the DNP jumps discontinuously to the upper stable branch, such that the current  also changes discontinuously as seen in Fig.~\ref{fig:ST-cross}(b). Likewise, when sweeping the field backwards from a high value of $g\mu_BB$, then a discontinues jump is found in the current at the point, where the upper stable DNP cease to exist. 

The stability of the DNP solution does not follow directly from the solution of the implicit equation (\ref{eq:implicit-eq-ST-cross}). To determine the stability of the DNP against small fluctuations, we numerically propagate the rate equations (\ref{eq:rate-eqs-ST-crossing}-\ref{eq:polarization-eq-ST-cross}) in time until a stationary solution is reached.\cite{Strogatz-BOOK} 

\begin{figure}
\includegraphics[width=0.39\textwidth,angle=0]{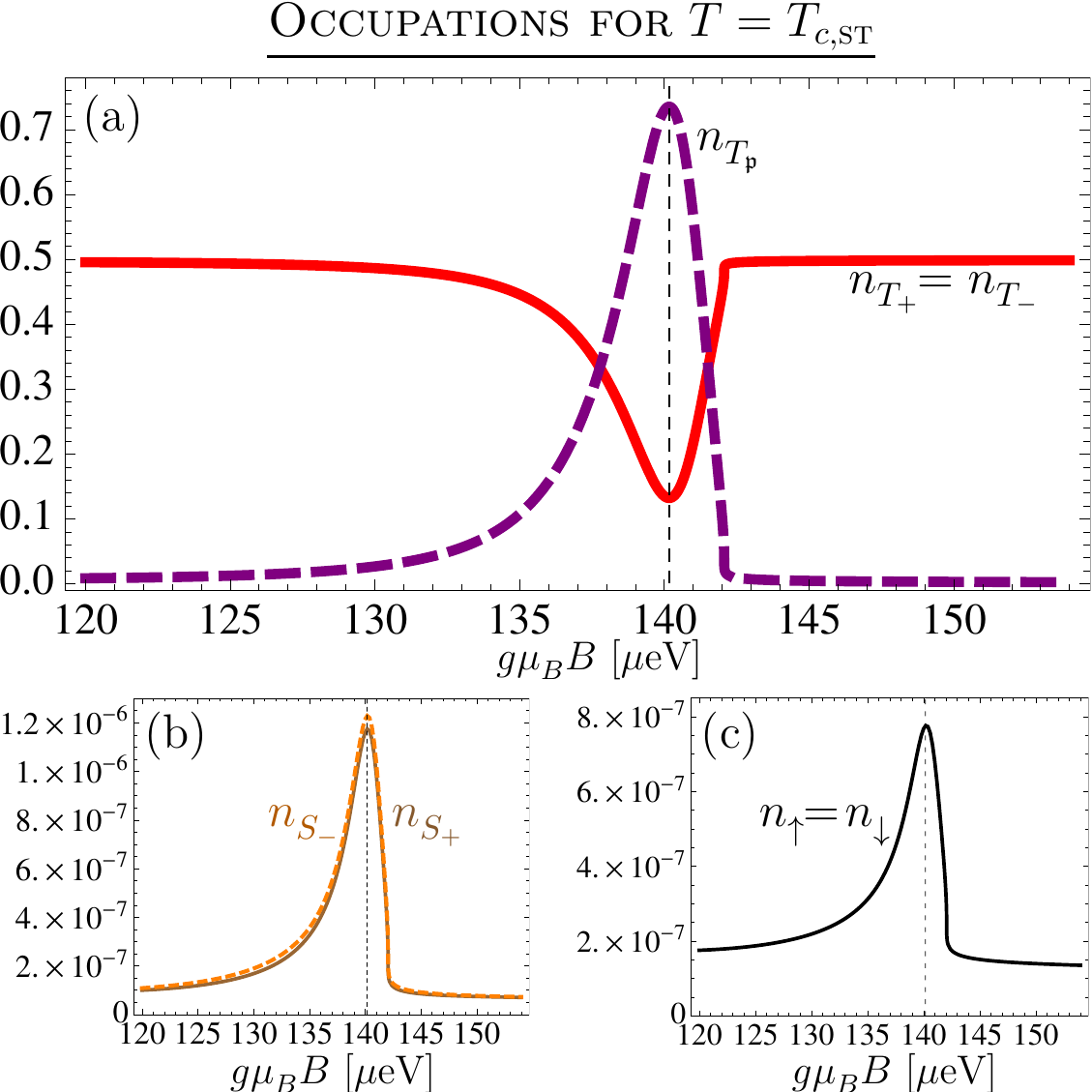}
\caption{(Color online) (a) Occupations of the triplets $n_{T_-}=n_{T_+}$ (red full line) and $n_{T_\PP}$ (purple dashed line). (b) Occupations of the singlets, $n_{S_-}$ (orange dotted line) and $n_{S_+}$ (brown full line), are seen to differ slightly as anticipated in Eq.(\ref{eq:occ-singlets-ST-cross}). (c) Occupations of the one-electron states $n_{\op}=n_{\ned}$ (black full line). The vertical dashed black line indicate the position of the singlet-triplet level crossing. We observe a decrease in the occupations of the pure triplets $T_{\pm}$ close to the level crossing due to the enhanced HF singlet-triplet rates in this region, see Fig.~\ref{fig:rates-ST-crossing}(a). In contrast, the occupation of the triplet $T_{\PP}$ increases heavily due to the DNP decrease [Fig.~\ref{fig:ST-cross}(d)], which reduces the escape rate form $T_\PP$ [see Eq.(\ref{eq:tunnel-out-from-Tx}) and Fig.~\ref{fig:rates-ST-crossing}(c)]. The parameters are the same as in Fig.~\ref{fig:ST-cross}(d-e) and the similar behavior is found for $T>\Tcs$.}
\label{fig:occupations-close-to-ST-crossing}
\end{figure}

The solution of the simplified rate equations (\ref{eq:rate-eqs-ST-crossing}-\ref{eq:polarization-eq-ST-cross}) and the numerical solution of the full rate equations (\ref{eq:full-rate-eqs-appendix},\ref{eq:full-polarization-eq}) with all rates match extremely well. The results in Fig.~\ref{fig:ST-cross}-\ref{fig:occupations-close-to-ST-crossing} calculated in the two ways fit perfectly. 

Next, we consider the regime of $T\geq \Tcs$, Fig.~\ref{fig:ST-cross}(d-i). By increasing the $B$-field away from the crossing of the triplets at $B=0$, the triplet-triplet rates decrease, while the singlet-triplet rates increase, since the triplet energy $E_{T_+}$ ($E_{T_-}$) approaches the singlet energy $E_{S_+}$ ($E_{S_-}$) from below (above) [Fig.~\ref{fig:ST-cross}(f,i)]. Therefore, two new processes come into play to lift the SB, namely $T_+\rightarrow S_+$ and $T_-\rightarrow S_-$, see Fig.~\ref{fig:ST-rate-eq-illustration}. The closer the singlet and triplet levels are, the more effective are these two new processes, which in turn produce a leakage current peak at the singlet-triplet level crossing as seen in Fig.~\ref{fig:ST-cross}(e,h). Moreover, the pure triplet occupations $n_{T_\pm}$ decrease close to the singlet-triplet level crossing as a consequence of the enhanced triplet-singlet processes as seen in Fig.~\ref{fig:occupations-close-to-ST-crossing}(a). Simultaneously, the occupation of the mixed triplet $T_\PP$ peaks at the level crossing.  
The reason is that the DNP decreases such that the escape rate $\Gamma_{\sigma,T_\PP}\sim \PP^2$ Eq.(\ref{eq:tunnel-out-from-Tx}) becomes heavily suppressed as seen in Fig.~\ref{fig:rates-ST-crossing}(c).

The two dominant HF processes close to the singlet-triplet crossing, $T_+\rightarrow S_+$ and $T_-\rightarrow S_-$, polarize the nuclei in opposite directions. When approaching the singlet-triplet crossing from below ($E_{T_+}<E_{S_+}$), the DNP decreases [Fig.~\ref{fig:ST-cross}(d,g)]. This is consistent with the fact that the negatively-polarizing phonon-emission process $T_-\rightarrow S_-$ is larger than the positively-polarizing phonon-absorption process $T_+\rightarrow S_+$  as seen in Fig.~\ref{fig:rates-ST-crossing}(a). The DNP is seen to increase again, once the magnetic field $g\mu_BB$ is tuned beyond the singlet-triplet crossing (indicated by the dashed vertical line in Fig.~\ref{fig:ST-cross}). 

Interesting, a very sharp -- yet continuous -- increase in the DNP is seen for $T=\Tcs$ at a \emph{higher} magnetic field than the one at which the singlet and triplet levels cross, see Fig.~\ref{fig:ST-cross}(d). This behavior is qualitatively different from the one observed for the crossing of the triplets. In that case, the sharp increase in DNP is found at the \emph{same} magnetic field as the one where the triplets cross; compare Fig.~\ref{fig:cross-triplets}(d) to Fig.~\ref{fig:ST-cross}(d). In both cases, the sharp DNP increase is a precursor of the DNP bistability. The sharp DNP increase at a $g\mu_BB$ beyond the singlet-triplet level crossing, is also reflected in the sudden increase of level splitting just \emph{after} the level crossing as seen in the inset of Fig.~\ref{fig:ST-cross}(f). The mismatch between the level crossing and the sharp DNP increase indicates that triplet-singlet processes are not the only important ingredient close to the singlet-triplet crossing -- although their rates dominate in magnitude. The triplet-triplet transitions also play a role. In fact, it is the inclusion of the triplet-triplet rates in the simplified rates (\ref{eq:rate-eqs-ST-crossing}-\ref{eq:polarization-eq-ST-cross}) that leads to an enhancement of the transition temperature.

\section{Monte Carlo simulations and the breakdown of the rate equation approach}\label{sec:rate-eq-breakdown-and-MC}

Now the rate equation approach is shown to be consistent with Monte Carlo simulation including an inelastic escape mechanism. We pay special attention to the case \emph{without} the inelastic escape mechanism, where the rate equation approach is shown to break down. In this case, the Monte Carlo simulations show that no polarization can be induced by the leakage current as expected.\cite{Jouravlev-Nazarov-PRL-2006,Rudner-Levitov-PRL-2007}   

\subsection{Breakdown of the rate equation description without the inelastic escape mechanism}

If HFI is the \emph{only} mechanism lifting SB, then the average DNP does not change.\cite{Jouravlev-Nazarov-PRL-2006,Rudner-Levitov-PRL-2007} Now, we show that this situation \emph{cannot} be described by the rate equations (\ref{eq:full-rate-eqs}-\ref{eq:full-polarization-eq}). To see this, we use rate equations (\ref{eq:full-rate-eqs-appendix},\ref{eq:full-polarization-eq}) to obtain 
\begin{align}
\dot{n}^{}_{T_+}-\dot{n}^{}_{T_-}&+\frac{N}{2}\dot{P}
+\frac{1}{2}\big[\dot{n}^{}_\op-\dot{n}^{}_\ned\big]
=
\nonumber\\
&\frac{1}{2}
(\Gamma_{S_+,\ned} + \Gamma_{S_+,\op} + \Gamma_{T_\PP,\op} - \Gamma_{T_+,\op}) 
[n^{}_\ned - n^{}_\op]
\nonumber\\
&+2 \Gamma^{\textrm{ine}} [n_{T_-}^{} - n_{T_+}^{}]
\end{align}
by utilizing the index invariances (\ref{eq:sym-for-tunnel-rates},\ref{eq:sym-for-HF-rates}) and equal inelastic escape rates (\ref{eq:equal-Gamma-ine}). The expressions for the rates (\ref{eq:tunnel-in-rates}) lead to $(\Gamma_{S_+,\ned} + \Gamma_{S_+,\op} + \Gamma_{T_\PP,\op} - \Gamma_{T_+,\op})=0$ such that
\begin{align}
\dot{n}^{}_{T_+} \!\!- \dot{n}^{}_{T_-}&\!+\!\frac{N}{2}\dot{P}
\!+\!\frac{1}{2}\big[\dot{n}^{}_\op\!-\!\dot{n}^{}_\ned\big]
=
2 \Gamma^{\textrm{ine}} [n_{T_-}^{}\!- n_{T_+}^{}].
\label{eq:conserved-quantity}
\end{align}
This shows that if $\Gamma^{\textrm{ine}}=0$, then the quantity $n^{}_{T_+} - n^{}_{T_-}+(N/2)P+(1/2)[n^{}_\op-n^{}_\ned]$ is \emph{conserved in the time evolution} of the rate equations. In other words, for $\Gamma^{\textrm{ine}}=0$, the stationary state of the rate equations \emph{depends} on the \emph{initial occupations} $n^{}_{T_\pm}$ and $n^{}_{\sigma}$, which is \emph{unphysical}. Thus, the rate equation description (\ref{eq:full-rate-eqs}-\ref{eq:full-polarization-eq}) breaks down for $\Gamma^{\textrm{ine}}=0$. This is the basic problem with the dynamics presented in Ref.~[\onlinecite{Lopez-Monis-et-al-NJP-2011}]. However, for $\Gamma^{\textrm{ine}}\neq0$ as used in this paper, the quantity $n^{}_{T_+} - n^{}_{T_-}+(N/2)P+(1/2)[n^{}_\op-n^{}_\ned]$ is not conserved, which is evident from Eq.(\ref{eq:conserved-quantity}).

\subsection{Monte Carlo simulations}

For $\Gamma^{\textrm{ine}}\neq0$  -- even if it is very small -- the rate equations (\ref{eq:full-rate-eqs}-\ref{eq:full-polarization-eq}) gives a reliable description of the DNP in the SB regime. To validate this, we have performed Monte Carlo simulations leading to the same results. 

The idea of the Monte Carlo simulation is -- in some sense -- to carry out a numerical experiment. The simulation is begun by placing the system in some initial state, say $|T_\PP\rangle$, with some initial polarization $P(t=0)$. Thereafter, the system is updated in discrete time steps. From each state $|i\rangle$, there is a certain probability $p_{f,i}$ to go to another state $|f\rangle$ of the system within a single time step. We use a computer-generated random number to decide, if the system goes to another state or simply remains in the same state in a time step. The probability for a certain transition in a time step is proportional to its rate. A HF transition changes the nuclear polarization for the next time step, and, in turn, also the transition probabilities. Thus, the polarization dynamically changes in time along with the probabilities during the simulation. At some point in time, the polarization is such that the system has found a stationary state on the average. 
In order to get average properties, that can be compared to the results of the rate equation approach, we need to time average over the fluctuations of the simulation.  

\begin{figure}
\includegraphics[width=0.38\textwidth,angle=0]{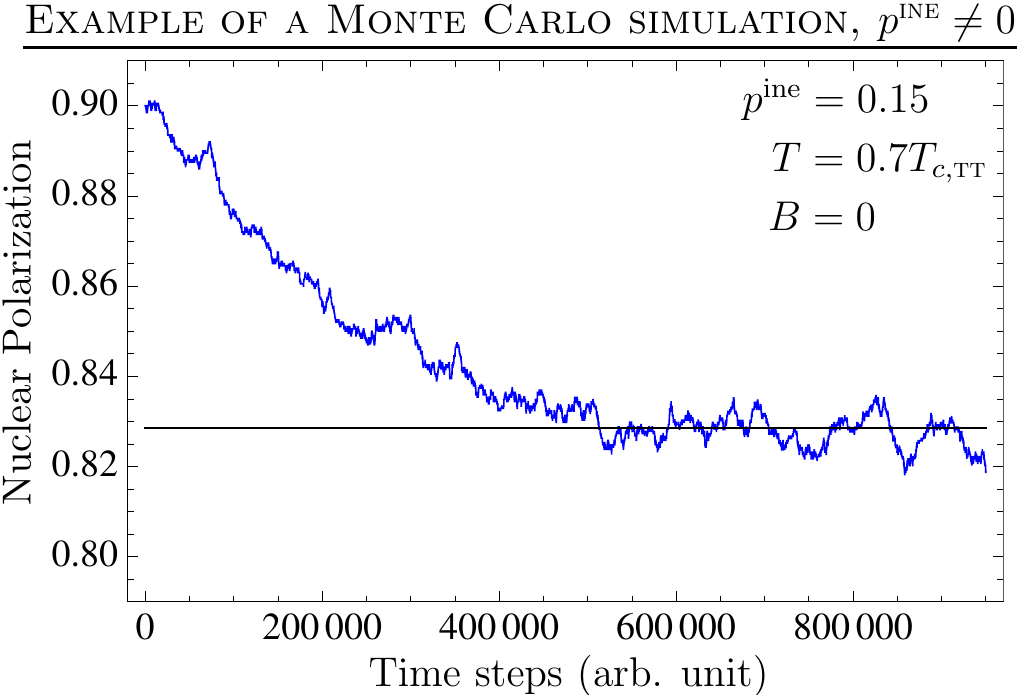}
\caption{(Color online) An example of a Monte Carlo simulation of the DNP (blue full line) versus time. The DNP is seen to level off to a stationary value of $P\simeq0.829$ (black horizontal line) from an initial DNP of $P(t=0)=0.9$. This is in perfect agreement with the rate equation result. As expected, the DNP is seen to fluctuate due to the randomness of the electron transport. The parameters are: $A_\tr=50\mu$eV, $A_\tl=30\mu$eV, $t=310\mu$eV and $\gamma_{ph}=5\mu$eV such that singlet and triplet levels are far apart for $B=0$. The probability for tunneling into $T_\pm$ is set to $p_\tl=p_\tr=0.45$ and the inelastic probability is chosen to be $p^{\textrm{ine}}=0.15$. Moreover, the change of the DNP due to a HFI is set to $dP=0.0005$ and an overall prefactor of $\eta^{}_{HF}=0.1$ is used in the HF probabilities (see Appendix C for details).}
\label{fig:MC-example-run}
\end{figure}

\begin{figure}
\includegraphics[width=0.38\textwidth,angle=0]{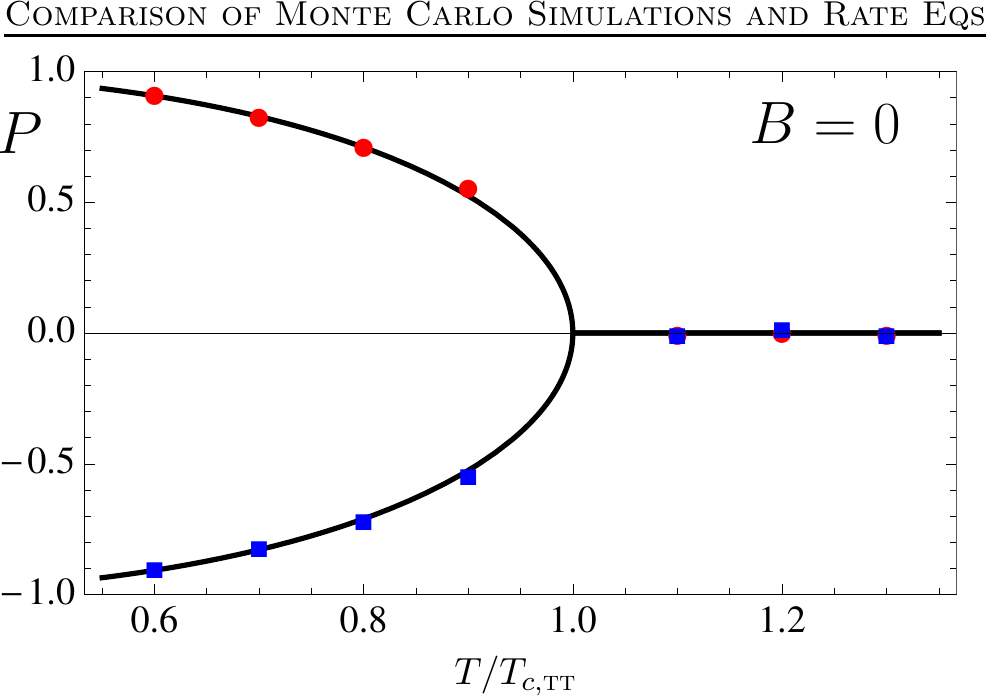}
\caption{(Color online) The stable polarization versus temperature $T/\Tct$ close to the crossing of the triplets from the rate equations (black full lines) and the Monte Carlo simulations with positive (red circles) and negative (blue squares) initial polarizations. The two methods agree very well. (We average over 25 simulations and use the same parameters as in Fig.~\ref{fig:MC-example-run} (except $dP=0.005$), see Appendix \ref{Appendix:Monte-carlo} for details).}
\label{fig:MC-comparison-to-Rate-eqs}
\end{figure}

\begin{figure*}
\includegraphics[width=0.88\textwidth,angle=0]{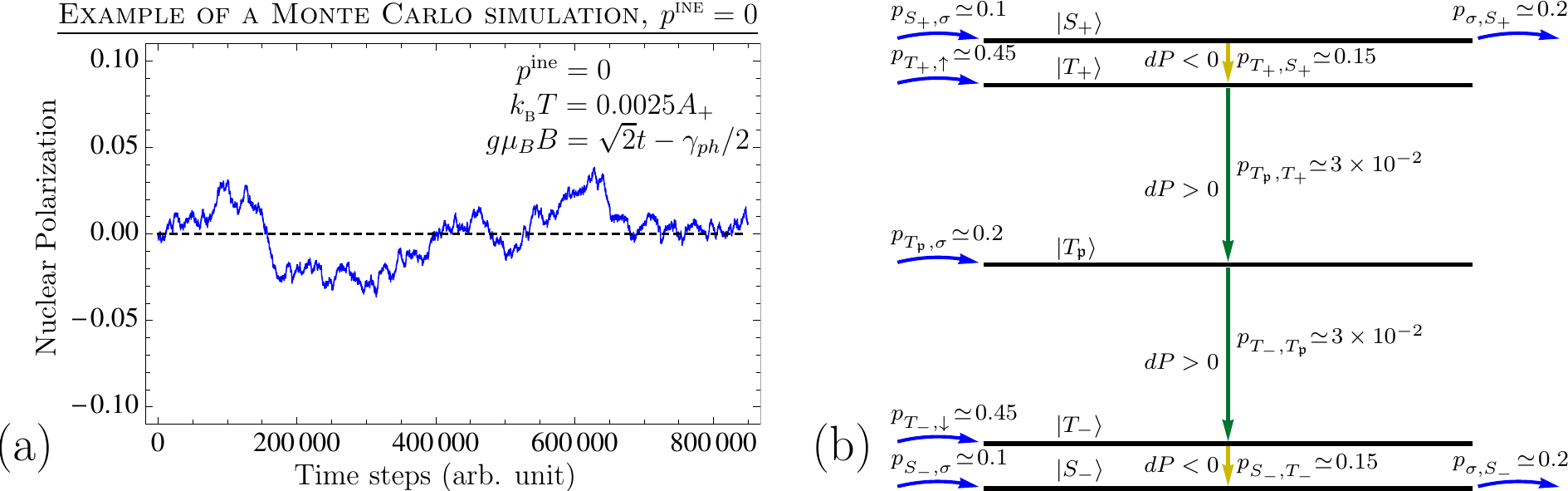}
\caption{(Color online) (a) An example of a Monte Carlo simulation showing that the nuclear polarization (blue full line) is not changed on the average due to the leakage current when \emph{only} HFIs lift the SB, i.e. $p^{\textrm{ine}}=0$. The magnetic field is tuned close to the singlet-triplet crossing leading to a level ordering as seen in (b). Moreover, we choose a very low temperature such that finite DNP could appear for $p^{\textrm{ine}}\neq0$. The initial polarization, $P(t=0)=0$ (black dashed line), is found to be equal to the time-averaged polarization, $\overline{P}\simeq0.002$, within the uncertainty. Similarly, the average occupations in the simulation are found to be $\overline{n}^{}_{T_+}\simeq0.26$, $\overline{n}^{}_{T_-}\simeq0.15$, $\overline{n}^{}_{T_\PP}\simeq0.48$, $\overline{n}^{}_{S_+}\simeq0.0063$, $\overline{n}^{}_{S_-}\simeq0.063$ and $\overline{n}^{}_{\op}\simeq\overline{n}^{}_{\ned}\simeq0.017$. Thus, in contrast to the $p^{\textrm{ine}}\neq0$ case, we find $\overline{n}^{}_{T_+}\neq\overline{n}^{}_{T_-}$. The parameters are: $A_\tr=50\mu$eV, $A_\tl=30\mu$eV, $t=50\mu$eV and $\gamma_{ph}=5\mu$eV such that the singlet and triplet energy levels are well separated, i.e. $t\gg\gamma_{ph}$. Moreover, we use $p_\tl=p_\tr=0.45$,  $dP=0.0005$ and $\eta^{}_{HF}=2$.\\
(b) An illustration of why no nuclear polarization is expected to be induced, when HFI is the only mechanism lifting SB. In the figure, we include all transition probabilities $p$ larger than $10^{-4}$ for the numerical example shown in (a). The HF spin-flip processes either increase (green arrows) or decrease (yellow arrows) the polarization on the average. We begin by noting that escape from neither $S_-$ nor $S_+$ change the overall polarization. For instance, the escape path $S_+\rightarrow T_+\rightarrow T_\PP\rightarrow T_-\rightarrow S_-\rightarrow \sigma$ consist of an equal amount of positive and negative nuclear spin-flips: $-|dP|+|dP|+|dP|-|dP|=0$. Similarly, escape from $T_\PP$ also leave the DNP polarization unchanged. In contrast, escape from $T_\pm$ polarize the nuclei by $\pm|dP|$, respectively. However, since $T_\pm$ also load with the same probability, $p_{T_+,\op}=p_{T_-,\ned}$,  no average DNP can be build up, even though escape from $T_+$ is less probable than from $T_-$ (as reflected in $\overline{n}^{}_{T_+}>\overline{n}^{}_{T_-}$) in the case considered here.}
\label{fig:MC-example-run-no-inelastic}
\end{figure*}

Appendix \ref{Appendix:Monte-carlo} gives more details on implementing the Monte Carlo simulations and shows examples of the DNP in single Monte Carlo simulations with and without the inelastic escape mechanism, respectively, in Figs.~\ref{fig:MC-example-run}-\ref{fig:MC-example-run-no-inelastic}.

The main difference between the rate equation approach and the Monte Carlo approach is that the rate equations solely deal with average quantities. Therefore, the rate equations allow in some sense many processes to take place on average side by side. In contrast, the system is in a specific state in each instant of time during a Monte Carlo simulation. Both approaches neglect all quantum mechanical coherences in the description. 

In Fig.~\ref{fig:MC-comparison-to-Rate-eqs}, we see that the rate equation description and the Monte Carlo simulations agree for the DNP versus $T$ close to the crossing of the triplets.\cite{footnote-MC-close-to-Tc} In the same way, we find excellent agreement between the two methods for finite $B$-fields close to the crossing of the triplets, and for $B$-fields around the singlet-triplet crossings. 

Furthermore, we have performed Monte Carlo simulations in the case of zero inelastic escape probability, $p^{\textrm{ine}}=0$, where the rate equation approach for the DNP breaks down. We find that if $p^{\textrm{ine}}=0$, then the time-averaged polarization is simply equal to the initial polarization of the simulation. See Fig.~\ref{fig:MC-example-run-no-inelastic} for an example and it caption for a discussion. These simulations therefore confirm that no finite DNP is built up on average for HFIs being the only mechanism lifting SB, as expected.\cite{Jouravlev-Nazarov-PRL-2006,Rudner-Levitov-PRL-2007}

\section{Summary, discussion and outlook}

In summary, we have analyzed the DNP and leakage current through a DQD in the SB regime due to a competition between HFIs and another inelastic escape mechanism from the triplets. We have demonstrated in detail how the DNP becomes bistable for temperatures $T$ below the transition temperature around both (i) the crossing of the three triplet levels and (ii)  the crossing of the triplets $T_\pm$ with the singlets $S_\pm$. The bistable DNP leads naturally to hysteresis in the leakage current. We have found that the transition temperature for the crossing of the triplet levels, $\Tct$, is generally different from the transition temperature for the singlet-triplet crossing, $\Tcs$. Moreover, $\Tct<\Tcs$ for experimentally relevant parameters and the difference can be sizable, e.g.~an order of magnitude. This enhancement of $\Tcs$ stems from an interplay between the triplet-triplet and singlet-triplet HF rates, even though the latter often dominates by at least an order of magnitude. For $\Tct<T<\Tcs$, current hysteresis appears around the singlet-triplet crossings at finite magnetic field, but is absent close to $B=0$. Moreover, we found analytically $\Tct=(A_\tl+A_\tr)/(8\kb)$, where $A_{\tl(\tr)}$ is the effective HF constant of the left (right) dot. In contrast, $\Tcs$ depends on various parameters, e.g. the inhomogeneity of the Overhauser field. Realistic HF constants of about\cite{Pfund-Shorubalko-Ensslin-Leturcq-PRL-2007,Churchill-Marcos-Marcus-et-al-nature-phys-2009} $100\mu$eV gives $\Tct\sim 300$mK, which is within experimental reach. Due to the enhancement of $\Tcs$ compared to $\Tct$, it might be harder to observe due to the broadening of the Coulomb blockade peaks. However, this depends heavily on the actual parameters and experimental setup (see Tables \ref{Table:Tc-ST-cross-inelastic-rate-vs-t}-\ref{Table:Tc-ST-cross-Am-og-N}). Furthermore, we have analyzed the details of the leakage current versus magnetic field and given various analytical limits in the case of the crossing of the triplet levels. 

Importantly, we have identified that the asymmetry between energy emission and absorption in the HF spin-flip transitions is the crucial ingredient for the existence of the transition temperatures at zero energy-detuning. Such an asymmetry can appear for many types of energy exchange mechanisms with an external bath due to detailed balance. 
Here we have considered phonons. In contrast, energy emission and absorption in Refs.[\onlinecite{Rudner-Levitov-PRL-2007,Rudner-Levitov-Nanotechnology-2010,Rudner-Rashba-PRB-2011}] is equally likely, such that \emph{no} DNP is found in these works for \emph{zero} energy-detuning. Nevertheless, they find bistabilities and current hysteresis for \emph{finite} energy-detuning. 

We have observed that our rate equation approach is consistent with the results produced by Monte Carlo simulations, if the inelastic escape mechanism is included. We discussed how the rate equation approach for the DNP becomes invalid without the inelastic escape rate. 

Through out the paper, we have neglected depolarizing processes such as nuclear spin-diffusion, since these are typically much slower than the HF spin-flip processes. Nevertheless, such processes might affect our results slightly in the case of large DNP, where the depolarization is stronger. On the other hand, very large DNP has also been reported experimentally.\cite{Baugh-Kitamura-Ono-Tarucha-PRL-2007}   

Furthermore, we have modeled the DNP of the nuclear spins as a single valued quantity, $P$, as in e.g. Refs.[\onlinecite{Rudner-Levitov-PRL-2007,Rudner-Levitov-Nanotechnology-2010,Rudner-Rashba-PRB-2011,Lopez-Monis-et-al-NJP-2011}]. In reality, the polarization will vary in space leading to a more complex behavior, which is more involved to model.\cite{Dominguez-Platero-PRB-2009} An extension of the model could be to use different DNPs for each dot.\cite{Inarrea-Platero-MacDonald-PRB-2007,Rudner-Koppens-Folk-Vandersypen-Levitov-PRB-2011} In such an approach, it is an open question, if the two DNPs would become bistable at the same transition temperature or not. Moreover, as emphasized in the paper, the difference in the Overhauser field between the dots is important to produce an escape path from the triplet with zero angular moment. Here, we have included this effect by having slightly different effective HF constants in the dots.   

We have studied in detail the case of a single constant inelastic escape rate from the triplets, which compete with the HF rates.\cite{Rudner-Levitov-PRL-2007,Rudner-Levitov-Nanotechnology-2010,Rudner-Rashba-PRB-2011,Rudner-Koppens-Folk-Vandersypen-Levitov-PRB-2011} Neglecting the energy dependence of the inelastic rates is a valid approach as long as the inelastic rate is smaller than the dominant HF rate close to the level crossing, as studied here. Nevertheless, the inelastic rates can be increased experimentally, e.g. by choosing a material with strong spin-orbit coupling\cite{Pfund-Shorubalko-Ensslin-Leturcq-PRL-2007} or by tuning the levels compared to the chemical potentials of the leads, so co-tunneling becomes more probable.\cite{Johnson-Petta-Marcus-PRB-2005} In such cases, it could be interesting to repeat our analysis including the energy dependences of the co-tunneling rates and/or spin-orbit rates. Future work could also analyze the effects of including a more detailed description of the phonons. However, we believe that the essential physics is captured by including the asymmetry between energy emission and absorption in the HF rates. 

\section{Acknowledgments}
 
We are especially thankful to J\'{a}nos Asb\'{o}th and Andras P\'alyi for insightful discussions on the nature of non-zero nuclear polarization in the SB setup. Moreover, we thank Andrea Donarini, Sigmund Kohler, Rafael S\'{a}nchez, Gerold Kiesslich, Marta Prada, Jeroen Danon, Jes\'us I\~{n}arrea and Fernando Dom\'inguez for useful discussions. AML acknowledges the Juan de la Cierva program (MICINN) and the Carlsberg Foundation. AML, CLM and GP acknowledge Grant No.~MAT2011-24331 and the ITN Grant 234970(EU). LLB and IV acknowledge Grant No.~FIS2011-28838-C02-01. We acknowledge FIS2010-22438-E (Spanish National Network Physics of Out-of-Equilibrium Systems).

\appendix 

\section{Expressions for the full rate equations of the model}\label{appendix:full-rate-eqs}

In the main paper, we give the full set of rate equations including all non-zero rates in a compact form in Eq.(\ref{eq:full-rate-eqs}). For completeness, we provide here the detailed equations  
\begin{widetext}
\begin{subequations}
\label{eq:full-rate-eqs-appendix}
\begin{align} 
\dot{n}^{}_{T_+}
&=  
W^{}_{T_+,S_+}n^{}_{S_+} 
{+}W^{}_{T_+,S_-} n^{}_{S_-} 
{+}W^{}_{T_+,T_\PP}n^{}_{T_\PP}
{+}\Gamma^{}_{T_+,\op} n^{}_\op 
-\big[W^{}_{S_+,T_+} + W^{}_{S_-,T_+} + W^{}_{T_\PP,T_+} 
+\Gamma_{\op,T_+}^{\textrm{ine}}+\Gamma_{\ned,T_+}^{\textrm{ine}}\big]n^{}_{T_+}, 
\\  
\dot{n}^{}_{T_-}&=
W^{}_{T_-,S_+}n^{}_{S_+}
{+}W^{}_{T_-,S_-}n^{}_{S_-} 
{+}W^{}_{T_-,T_\PP}n^{}_{T_\PP}
{+}\Gamma^{}_{T_-,\ned} n^{}_\ned 
\!-\!\big[W_{S_+,T_-}^{} + W_{S_-,T_-}^{} 
+W^{}_{T_\PP,T_-} 
+\Gamma_{\ned,T_-}^{\textrm{ine}}+\Gamma_{\op,T_-}^{\textrm{ine}}\big] n^{}_{T_-}, 
\\  
\dot{n}_{T_\PP}&=  
W_{T_\PP,T_+}n^{}_{T_+}
{+}W_{T_\PP,T_-}n^{}_{T_-} 
{+}\Gamma_{T_\PP,\op}n^{}_{\op}
+\Gamma_{T_\PP,\ned}n^{}_{\ned} 
-\big[W_{T_+,T_\PP} {+} W_{T_-,T_\PP}+\Gamma_{\op,T_\PP} 
+\Gamma_{\ned,T_\PP} +\Gamma_{\op,T_\PP}^{\textrm{ine}} 
+\Gamma_{\ned,T_\PP}^{\textrm{ine}}\big] n^{}_{T_\PP},
\\  
\dot{n}^{}_{S_+}&=
W^{}_{S_+,T_+}n^{}_{T_+}+W^{}_{S_+,T_-}n^{}_{T_-} 
+\Gamma^{}_{S_+,\op}n^{}_{\op}+\Gamma^{}_{S_+,\ned}n^{}_{\ned}
-\big[W^{}_{T_+,S_+} + W^{}_{T_-,S_+} 
+\Gamma^{}_{\op,S_+} + \Gamma^{}_{\ned,S_+}\big]n^{}_{S_+},
\\  
\dot{n}^{}_{S_-}&=
W^{}_{S_-,T_+}n^{}_{T_+}+W^{}_{S_-,T_-}n^{}_{T_-}
+\Gamma^{}_{S_-,\op}n^{}_{\op}+\Gamma^{}_{S_-,\ned}n^{}_{\ned}
-\big[W^{}_{T_+,S_-}+W^{}_{T_-,S_-}+\Gamma^{}_{\op,S_-}+\Gamma^{}_{\ned,S_-}\big]n^{}_{S_-},
\\
\dot{n}^{}_\op&=
\Gamma^{}_{\op,S_+}n^{}_{S_+}
{+}\Gamma^{}_{\op,S_-}n^{}_{S_-} 
+(\Gamma^{}_{\op,T_\PP}{+}\Gamma_{\op,T_\PP}^{\textrm{ine}})n^{}_{T_\PP} 
+\Gamma_{\op,T_+}^{\textrm{ine}}n^{}_{T_+}+\Gamma_{\op,T_-}^{\textrm{ine}}n^{}_{T_-} 
{-}\big[\Gamma^{}_{S_+,\op}+\Gamma^{}_{S_-,\op}+\Gamma^{}_{T_\PP,\op}+\Gamma^{}_{T_+,\op}\big]n^{}_\op,
\\
\dot{n}^{}_\ned&=
\Gamma^{}_{\ned,S_+}n^{}_{S_+}
{+}\Gamma^{}_{\ned,S_-}n^{}_{S_-} 
+(\Gamma^{}_{\ned,T_\PP}{+}\Gamma_{\ned,T_\PP}^{\textrm{ine}})n^{}_{T_\PP} 
+\Gamma_{\ned,T_-}^{\textrm{ine}}n^{}_{T_-} +\Gamma_{\ned,T_+}^{\textrm{ine}}n^{}_{T_+} 
{-}\big[\Gamma^{}_{S_+,\ned}\!+\Gamma^{}_{S_-,\ned}
+\Gamma^{}_{T_\PP,\ned}+\Gamma^{}_{T_-,\ned}\big]n^{}_\ned, 
\end{align}
\end{subequations}
\end{widetext}
as illustrated in Fig.~\ref{fig:full-rate-eq-illustration}.

\section{Current expressions close to the crossing of the triplet levels for non-equal inelastic escape rates}\label{appendix:non-equal-inelastic-rates}

Throughout the paper, we have considered the case of \emph{equal} and constant inelastic escape rates $\Gamma_{\sigma,T}^{\textrm{ine}}$ from the triplet states $T=T_\pm,T_\PP$, see Eq.(\ref{eq:equal-Gamma-ine}).\cite{Rudner-Levitov-PRL-2007,Rudner-Levitov-Nanotechnology-2010,Rudner-Koppens-Folk-Vandersypen-Levitov-PRB-2011,Rudner-Rashba-PRB-2011} However, depending on the inelastic escape mechanism, the rates might be different. For completeness, we discuss this briefly in this appendix in a particularly simple case.

Here we consider the case, where the inelastic escape rates are invariant under the same exchange of indices as the tunneling rates Eq.(\ref{eq:sym-for-tunnel-rates}), i.e.
\begin{align}
\Gamma^{\textrm{ine}}_{\op,T_\PP}=
\Gamma^{\textrm{ine}}_{\ned,T_\PP},
\quad
\Gamma^{\textrm{ine}}_{\op,T_+}=
\Gamma^{\textrm{ine}}_{\ned,T_-},
\quad
\Gamma^{\textrm{ine}}_{\ned,T_+}=
\Gamma^{\textrm{ine}}_{\op,T_-}.
\label{eq:Gamma-ine-with-tunnel-rate-symmetry}
\end{align}
Following the same steps leading to the polarization equation in Sec.~\ref{subsec:pol-eq-TT}, we find that $n_{\op}^{}=n_{\ned}^{}$ and $n_{T_+}^{}=n_{T_-}^{}$ still hold true such that the implicit equation (\ref{eq:pol-condition-for-rates}) for the polarization,
\begin{align} 
W^{}_{T_\PP,T_+}=W^{}_{T_+,T_\PP}
=W_{T_\PP,T_-}=W_{T_-,T_\PP},
\end{align}
and the transition temperature, $\kb \Tct=A_+/4$ (\ref{eq:trans-temp-TT}), remain unchanged compared to the main text. Furthermore, $n_{S_+}=n_{S_-}$ and Eqs.(\ref{eq:nSp-propto-np}) and (\ref{eq:np-via-norm-condition}) also still hold true, whereas the explicit expressions for the occupations (\ref{eq:triplet-occupations}) and the current (\ref{eq:current-general-for-TTT-crossing}) are changed slightly. In the numerator of $n_{T_+}$ (\ref{eq:Tp-occupation}), the rate $\Gamma^{\textrm{ine}}$ is replaced by $\Gamma^{\textrm{ine}}_{\op,T_\PP}$. Similarly, in the numerator of $n_{T_\PP}$ (\ref{eq:Tx-occupation}), one has to make the replacement $2\Gamma^{\textrm{ine}}\rightarrow \Gamma^{\textrm{ine}}_{\op,T_+}+\Gamma^{\textrm{ine}}_{\op,T_-}$. The common denominator of the occupations (\ref{eq:triplet-occupations}) changes to $\tilde{\Lambda}=
2 \Gamma_{T_+,\op} \Gamma_{\op,S_+} (\Gamma^{\textrm{ine}}_{\op,T_\PP} + \Gamma_{\op,T_\PP})
+\big[3 \Gamma_{T_+,\op} \Gamma_{\op,S_+} + (\Gamma^{\textrm{ine}}_{\op,T_\PP} + \Gamma_{\op,T_\PP}) \Upsilon + 3 \Gamma_{\op,S_+} \Gamma_{T_\PP,\op}\big] W_{T_\PP,T_+}
+(\Gamma^{\textrm{ine}}_{\op,T_-} + \Gamma^{\textrm{ine}}_{\op,T_+}) 
\big[\Gamma^{\textrm{ine}}_{\op,T_\PP} \Upsilon + (\Gamma_{S_+,\ned} + \Gamma_{S_+,\op}) (\Gamma_{\op,T_\PP} + W_{T_\PP,T_+}) + 
   \Gamma_{\op,S_+} (2 \Gamma_{\op,T_\PP} + \Gamma_{T_\PP,\op} + 2 W_{T_\PP,T_+})\big]$. Therefore the current expression becomes
\begin{align}
I&=
\frac{2\Gamma_{\op,S_+}}{\tilde{\zeta}}  
(\Gamma_{T_+,\op} + \Gamma_{S_+,\ned} + \Gamma_{S_+,\op} + \Gamma_{T_\PP,\op})
\label{eq:current-non-equal-ine-rates} \\
&\times
\Big[(\Gamma_{\op,T_-}^{\textrm{ine}}+\Gamma_{\op,T_+}^{\textrm{ine}}) 
(\Gamma_{\op,T_\PP}^{\textrm{ine}} + \Gamma_{\op,T_\PP})
\nonumber\\ 
&\hspace{5mm}+ W_{T_+,T_\PP}
(\Gamma_{\op,T_-}^{\textrm{ine}} +\Gamma_{\op,T_+}^{\textrm{ine}} 
+\Gamma_{\op,T_\PP}^{\textrm{ine}} + \Gamma_{\op,T_\PP})\Big]
\nonumber 
\end{align}
where
\begin{align}
\tilde{\zeta}\equiv
&
(\Gamma_{\op,T_\PP}^{\textrm{ine}} {+} \Gamma_{\op,T_\PP}) 
\big[2\Gamma_{T_+,\op} \Gamma_{\op,S_+} 
+ (\Gamma_{\op,T_+}^{\textrm{ine}} {+} \Gamma_{\op,T_-}^{\textrm{ine}}) \Upsilon\big]
\nonumber\\
&+
(\Gamma_{\op,T_+}^{\textrm{ine}}+\Gamma_{\op,T_-}^{\textrm{ine}})
\Gamma_{\op,S_+} \Gamma_{T_\PP,\op}
\nonumber\\
&+
W_{T_+,T_\PP}
\Big[3 \Gamma_{T_+,\op} \Gamma_{\op,S_+} 
+ 3 \Gamma_{\op,S_+} \Gamma_{T_\PP,\op}
\nonumber\\
&\hspace{15mm}+ \Upsilon
(\Gamma_{\op,T_\PP}^{\textrm{ine}}+\Gamma_{\op,T_+}^{\textrm{ine}}
+\Gamma_{\op,T_-}^{\textrm{ine}}+ \Gamma_{\op,T_\PP}) 
\Big].
\nonumber
\end{align}
Note the similarity to the simpler case of equal inelastic rates in Eq.(\ref{eq:current-general-for-TTT-crossing}). 

In passing, we note that even though the current expressions (\ref{eq:current-non-equal-ine-rates}) and (\ref{eq:current-general-for-TTT-crossing}) are similar, they have an interesting difference: \emph{Without} singlet-triplet mixing $\PP=0$ the current (\ref{eq:current-non-equal-ine-rates}) still depends on the HF rate in contrast to the simpler case of the main text, see Eq.(\ref{eq:current-for-zero-TS-mixing-parameter}). In order to see this explicitly, we insert $\PP=0$ and the tunneling rates (\ref{eq:tunnel-in-rates}-\ref{eq:tunnel-out-rates}) into the current expression (\ref{eq:current-non-equal-ine-rates}), i.e.
\begin{subequations}
\label{eq:current-non-equal-ine-rates-without-ST-mixing}
\begin{align}
I(\PP=0)=
\frac{8\Gamma_{\tl}}{\beth} 
\Big[&
W_{T_+,T_\PP}^{} 
(\Gamma_{\op,T_-}^{\textrm{ine}} + \Gamma_{\op,T_+}^{\textrm{ine}} + \Gamma_{\op,T_\PP}^{\textrm{ine}}) 
\nonumber\\
&+
\Gamma_{\op,T_\PP}^{\textrm{ine}} 
(\Gamma_{\op,T_-}^{\textrm{ine}} + \Gamma_{\op,T_+}^{\textrm{ine}}) 
\Big]
\end{align}
where $\Gamma_\tl=\Gamma_\tr$ was used for simplicity and 
\begin{align}
\beth\equiv&
W_{T_+,T_\PP}^{}  
\big[ 
6 \Gamma_{\op,T_\PP}^{\textrm{ine}} 
+ 6 \Gamma_{\op,T_-}^{\textrm{ine}} 
+6 \Gamma_{\op,T_+}^{\textrm{ine}}
+ 9 \Gamma_{\tl}
\big]
\nonumber\\
&+(6 \Gamma_{\op,T_\PP}^{\textrm{ine}} + \Gamma_{\tl} ) 
(\Gamma_{\op,T_-}^{\textrm{ine}} + \Gamma_{\op,T_+}^{\textrm{ine}})
+4 \Gamma_{\op,T_\PP}^{\textrm{ine}} \Gamma_{\tl}.
\end{align}
\end{subequations}
Thus, the current for $\PP=0$ still depends on the HF rate $W_{T_+,T_\PP}$ in general. 

In summary, when the inelastic rates have the same invariances under interchange of indices as the tunneling rates, then the polarization equation (\ref{eq:pol-condition-for-rates}) does not change and the current expression changes only slightly. However, if the inelastic rates are invariant under other interchange of indices, then the polarization condition (\ref{eq:pol-condition-for-rates}) and the transition temperature might change.

\section{Details on the implementation of the Monte Carlo simulations}\label{Appendix:Monte-carlo}

We implement all the possible transitions between the states $\{\op,\ned,T_+,T_-,T_\PP,S_+,S_-\}$ as shown graphically on Fig.~\ref{fig:full-rate-eq-illustration}. Therefore, we are not limited to simulate a specific level crossing. For the transition probabilities in a single time step, we use the same functional dependences as for the rate expressions (\ref{eq:equal-Gamma-ine},\ref{eq:tunnel-in-rates},\ref{eq:tunnel-out-rates},\ref{eq:TT-HF-rates},\ref{eq:TS-HF-rates}), since rates and probabilities are proportional. In the formulas (\ref{eq:equal-Gamma-ine},\ref{eq:tunnel-in-rates},\ref{eq:tunnel-out-rates}), we exchange the rates $\Gamma^{}_{\tr (\tl)}$ by $p^{}_{\tr (\tl)}$ and $\Gamma_{}^{\textrm{ine}}$ by $p^{\textrm{ine}}_{}$. For instance, $p^{}_{T_\PP,\op}=p_{\tl}^{}/(2\mathcal{N}^2)$ is the probability for going to $T_\PP$ given that the system is in the one-electron state $\op$. Likewise, we exchange the factor $1/(2N)$ in the HF rates (\ref{eq:TT-HF-rates},\ref{eq:TS-HF-rates}) by the parameter $\eta_{HF}$ in the HF probabilities. Thereby we can tune the magnitude of the HF transition probabilities compared to the inelastic transition probabilities. Thus, we can easily study the same physical situation as in the rate equation approach. For instance, Figs.~\ref{fig:P-vs-T-triplet-crossing} and \ref{fig:MC-comparison-to-Rate-eqs} both study large singlet-triplet energy separation and zero magnetic field.  

To minimize the computational load, we choose the transition probabilities within a single time step as \emph{high} as possible, such that the system does not remain in the same state over too many time steps. This can be understood as a long physical time duration for each time step. Nevertheless, we have to choose numbers such that the sum of all probabilities for leaving a specific state is always \emph{smaller} than one in each time step, e.g.~for $T_\PP$ this amounts to 
\begin{align}
2p_{}^{\textrm{ine}}+
p_{\ned,T_\PP}^{}+
p_{\op,T_\PP}^{}
+p^{HF}_{T_+,T_\PP}(t)+
p^{HF}_{T_-,T_\PP}(t)
<1.
\end{align}
In this way, the possibility of staying in the same state (here $T_\PP$) within a time step remains in the simulation.

In the real experiment, the polarization change by $dP=\pm 2/N$ due to a \emph{single} HF transition. In the simulation, however, $dP$ is increased substantially in order to obtain faster convergence to a stationary polarization. We emphasize that the choice of $dP$ does \emph{not} affect the value of the stationary polarization, but it does indeed affect the typical fluctuations around this value. Thus, an artifact of choosing $dP$ larger than $2/N$ is the artificially increased fluctuations around the stationary polarization -- as seen Figs.~\ref{fig:MC-example-run} and \ref{fig:MC-example-run-no-inelastic} -- compared to the experimental situation. However, since we are only interested in average values, this is not a concern here. Choosing $dP$ is therefore a compromise between maximizing convergence time and minimizing fluctuations.  

In order to find the stationary DNP, we choose a $dP$, perform the Monte Carlo simulation a number of times for a given initial DNP and then average over the results. The averaging makes it easier to decide in a computationally cheap way, if convergence is reached. To make sure that the found stationary DNP $\bar{P}$ is stable, we show that an initial DNP $P(t=0)>\bar{P}$ decreases versus time and that an initial DNP with $P(t=0)<\bar{P}$ increases versus time. We stress that the stationary DNP can also be found from doing the single Monte Carlo simulation as seen in Fig.~\ref{fig:MC-example-run}, but it requires a much smaller $dP$. Moreover, the fluctuations in DNP increase with temperature, since higher $T$ increases the phonon-absorption HF transition probabilities, which increases the number of likely transitions.


\begin{thebibliography}{135}
\bibitem[{foo({\natexlab{c}})}]{footnote-dE-nuclear-levels-is-negligible}
\bibinfo{note}{Here the Zeeman energy splitting between the nuclear spins
  states is negligible compared to the one of the electronic states. This is
  because the nuclear magneton is about 2000 times smaller then the Bohr
  magneton due to the mass ratio of the electron and the proton.}
\bibitem[{foo({\natexlab{d}})}]{footnote-degenerate-per}
\bibinfo{note}{Formally, this is equivalent to including the inter-dot
  tunneling and the Overhauser field in usual degenerate perturbation theory
  for the chosen states.}
\bibitem[{foo({\natexlab{e}})}]{footnote-depol}
\bibinfo{note}{The depolarization can be described by including a
  phenomenological term $-W_{{\rm rel}} P$ in the DNP rate
  Eq.(\ref{eq:full-polarization-eq}).}
\bibitem[{foo({\natexlab{g}})}]{footnote-general-TcTT}
\bibinfo{note}{Note for instance that if $E_{T_+}>E_{T_\PP}$ and we take
  $W_{T_\PP,T_+}\propto n_{B}(E_{T_+}-E_{T_\PP})+1$ and $W_{T_+,T_\PP}\propto
  n_{B}(E_{T_+}-E_{T_\PP})$, then we get the same ratio on the right-hand side
  of Eq.(\ref{eq:pol-condition-for-rates-mellem}) as if we use our model for
  the rates in Eq.(\ref{eq:ph-factor}). Thus, in this respect, the DNP becomes
  model independent. Here $n_{B}(E)=[e^{E/\kb T}-1]^{-1}$ is the Bose
  function.}
\bibitem[{foo({\natexlab{h}})}]{footnote-way-to-get-TcTT}
\bibinfo{note}{This can, for instance, be done by using that
  Eq.(\ref{eq:pol-condition-tanh}) for $B=0$ goes from having one to three
  solutions for $P$, when the derivative of $\tanh[A_+P/(4\kb T)]$ at $P=0$ is
  one.}
\bibitem[{foo({\natexlab{i}})}]{footnote-w-universal}
\bibinfo{note}{Remember that all the non-zero HF rates between the triplets
  become equal in steady state due to the implicit polarization equation
  (\ref{eq:pol-condition-for-rates}), and therefore the final and initial state
  is not emphasized in the notation of $w$.}
\bibitem[{foo({\natexlab{j}})}]{footnote-number-discussion}
\bibinfo{note}{Note that there is no need to choose the rate $\Gamma_\tl$
  explicitly in these plots. However, to specify the dimensionless hyperfine
  rate $w=W_{T_+,T_\PP}/\Gamma_{\tl}$ Eq.(\ref{eq:def-w-gRL-gine}) one needs to
  give a number for $\h\Gamma_\tl N$, where $N$ is the total number of nuclear
  spins in the DQD, around $10^4-10^7$ depending on the material etc. Choosing
  $\h\Gamma_\tl N=10^7\mu$eV corresponds to -- for instance -- having
  $\h\Gamma_\tl \sim 100\mu$eV and $N\sim 10^5$.}
\bibitem[{foo({\natexlab{l}})}]{footnote-negative-B}
\bibinfo{note}{For $g\mu_BB<0$, the energy levels that cross are reversed, i.e.
  $E_{T_+}$ cross $E_{S_-}$ and $E_{T_-}$ cross $E_{S_+}$. This means that one
  has to neglect other HF rates in order to construct a simple model describing
  the level crossing for $g\mu_BB<0$. However, this is straight forward
  following the same principles as outlined in the main text.}
\bibitem[{foo({\natexlab{m}})}]{footnote-remark-on-approx}
\bibinfo{note}{We note that this approach fails for the parameters considered
  here, e.g. $A_{\tl}=80\mu$eV, $A_{\tr}=70\mu$eV, $t=100\mu$eV,
  $\gamma_{ph}=1\mu$eV, $\Gamma_\tl=\Gamma_\tr$, $\h\Gamma_\tl N=10^6\mu$eV and
  $\gamma_{\textrm{ine}}=10^{-7}$. Nevertheless, the approximation becomes
  better (but far from perfect) very close to the singlet-triplet crossing by
  increasing $t$ to $t=10^4\mu$eV, since this decreases the influence of the
  triplet-triplet HF rates. However, we are not interested in such extreme
  cases of very large $t$.}
\bibitem[{foo({\natexlab{n}})}]{footnote-G-details}
\bibinfo{note}{To see that $G(P)>0$ and $G(P)\neq0$, we begin by noting that
  $G(P)$ is a sum of products of rates, which are all greater than or equal to
  zero. Thus, $G(P)\geq0$. However, at least one of these terms is always
  strictly positive, namely $\Gamma_{T_+,\op}\Gamma^{\textrm{ine}}>0$. Hence,
  $G(P)>0$.}
\bibitem[{foo({\natexlab{o}})}]{footnote-compare-two-figs}
\bibinfo{note}{We choose slightly different parameters in
  Fig.~\ref{fig:ST-cross} compared to Fig.~\ref{fig:cross-triplets}. Using the
  parameters of Fig.~\ref{fig:cross-triplets}, a very similar transition
  temperature is found, $\Tcs\simeq2.81A_+/\kb$, -- even though the inelastic
  rate dominates for magnetic fields close to the singlet-triplet crossing.}
\bibitem[{foo({\natexlab{p}})}]{footnote-MC-close-to-Tc}
\bibinfo{note}{We remark that close to the critical temperature in
  Fig.~\ref{fig:MC-comparison-to-Rate-eqs}, it is hard to obtain conclusive
  average polarization values of the Monte Carlo simulations due to
  fluctuations.}
\end{thebibliography}

\end{document}